\documentclass[fleqn,usenatbib]{mnras}

\usepackage{newtxtext,newtxmath}
\usepackage[T1]{fontenc}
\usepackage{multirow}

\DeclareRobustCommand{\VAN}[3]{#2}
\let\VANthebibliography\thebibliography
\def\thebibliography{\DeclareRobustCommand{\VAN}[3]{##3}\VANthebibliography}

\usepackage{graphicx}	
\usepackage{amsmath}	

\usepackage{tikz}
\definecolor{blue1}{HTML}{1f77b4}
\definecolor{orange1}{HTML}{ff7f0e}
\definecolor{green1}{HTML}{2ca02c}

\def \cii {C{\sc ii}}
\def \Tkin {$T_{\rm kin}$}

\def \twC  {$^{12}$C}
\def \stO  {$^{16}$O}

\def \oi  {O{\sc i}}
\def \mum  {$\mu$m}

\def\purple#1 {{\textcolor{purple}{#1}}\ }
\def\red#1 {\textcolor{red}{#1}}
\def\blue#1 {{\textcolor{blue}{#1}}\ }
\definecolor{forestgreen}{rgb}{0.13, 0.55, 0.13}
\def\zy#1 {\textcolor{forestgreen}{ zy: #1}} 
\def\eg#1 {\textcolor{magenta}{ eg: #1}} 

\title[$\alpha$-enhanced ISM]{$\alpha$-enhanced Astrochemistry: The Carbon cycle in extreme galactic conditions}

\author[T. G. Bisbas et al.]{
Thomas G. Bisbas$^1$\thanks{E-mail: tbisbas@zhejianglab.com (TGB)},
Zhi-Yu Zhang$^{2,3}$\thanks{E-mail: zzhang@nju.edu.cn (ZYZ)},
Eda Gjergo$^{2,3}$,
Ying-He Zhao$^{4,5}$,
Gan Luo$^{6,2}$,
Donghui Quan$^{1,7}$,
\newauthor
Xue-Jian Jiang$^{1}$, 
Yichen Sun$^{2,3}$,
Theodoros Topkaras$^8$,
Di Li$^{9,1,10}$ and
Ziyi Guo$^{2,3}$ 
\\
$^1$Research Center for Intelligent Computing Platforms, Zhejiang Lab, Hangzhou 311100, China\\
$^2$School of Astronomy and Space Science, Nanjing University, Nanjing, China\\
$^3$Key Laboratory of Modern Astronomy and Astrophysics, Nanjing University, Ministry of Education, Nanjing, China\\
$^4$Yunnan Observatories, Chinese Academy of Sciences, Kunming 650011, China\\
$^5$Key Laboratory of Radio Astronomy and Technology, Chinese Academy of Sciences, A20 Datun Road, Chaoyang District, Beijing, 100101, P. R. China \\
$^6$Institut de Radioastronomie Millimetrique, 300 rue de la Piscine, Domaine Universitaire de Grenoble, 38406, Saint-Martin d'H\'eres, France\\
$^7$Xinjiang Astronomical Observatory, Chinese Academy of Sciences, No. 150 Science 1-Street, Urumqi 830011, People's Republic of China\\
$^8$I. Physikalisches Institut, Universit\"at zu K\"oln, Z\"ulpicher Stra\ss e 77, 50937 K\"oln, Germany\\
$^9$CAS Key Laboratory of FAST, National Astronomical Observatories, Chinese Academy of Sciences, Beijing 100101, China\\
$^{10}$NAOC-UKZN Computational Astrophysics Centre, University of KwaZulu-Natal, Durban 4000, South Africa
}

\date{Accepted XXX. Received YYY; in original form ZZZ}

\pubyear{2023}

\begin{document}
\label{firstpage}
\pagerange{\pageref{firstpage}--\pageref{lastpage}}
\maketitle


\begin{abstract} 
Astrochemistry has been widely developed as a power tool to probe physical
properties of the interstellar medium (ISM) in various conditions of the Milky Way (MW)
Galaxy, and in near and distant galaxies. Most current studies conventionally
apply linear scaling to all elemental abundances based on the gas-phase
metallicity. However, these elements, including carbon and oxygen, are enriched
differentially by stellar nucleosynthesis and the overall galactic chemical
evolution, evident from $\alpha$-enhancement in multiple galactic observations
such as starbursts, high-redshift star-forming galaxies, and low-metallicity
dwarfs. We perform astrochemical modeling to simulate the impact of
an $\alpha$-enhanced ISM gas cloud on the abundances of the three phases of
carbon (C$^+$, C, CO) dubbed as `the carbon cycle'. 
The ISM environmental parameters considered include two cosmic-ray
ionization rates ($\zeta_{\rm CR}=10^{-17}$ and $10^{-15}\,{\rm s}^{-1}$), two 
isotropic FUV radiation field strengths ($\chi/\chi_0=1$ and $10^2$),
and (sub-)linear dust-to-gas relations against metallicity, mimicking the ISM
conditions of different galaxy types. In galaxies with [C/O] $<$ 0, CO, 
C and C$^+$ all decrease in both abundances and emission, 
though with differential biases.  
The low-$J$ CO emission is found to be the most stable
tracer for the molecular gas, while C and C$^+$ trace H$_2$ gas 
only under limited conditions, in line with recent discoveries of [C{\sc i}]-dark galaxies.  
We call for caution when using [C{\sc ii}]~$158\mu$m and [C{\sc i}](1-0) as alternative
H$_2$-gas tracers for both diffuse and dense gas with non-zero [C/O] ratios.
\end{abstract}




\section{Introduction}
\label{sec:intro}

To study the evolution of the multi-phase interstellar medium (ISM) and the
star-formation process in galaxies through cosmic time, it is required to
understand the astrochemistry that occurs under the different environmental
conditions. These conditions act as an energy source that powers the ISM, which
in turn compensates with a number of cooling functions leading to its thermal
balance. Various computational works model these
processes, either in static but chemically evolved clouds \citep[e.g.][]{LePetit06, Roellig07, Bisbas12, Ferland17, Roellig22} or by evolving them together with (magneto-)hydrodynamics \citep[e.g.][]{Glover10, Walch15, Girichidis16, Richings16, Gong17, Bate19, Seifried20, Hu21}. These help to understand the physical properties and the line emission
of the ISM both in Galactic local clouds and distant galaxies, especially 
nowadays since the James Webb Space Telescope (JWST) in combination with the
Atacama Large Millimeter/sub-millimeter Array (ALMA) offer an unprecedented
view of our Universe.

Metallicity is one of the most important environmental parameters of the ISM, 
as it plays a significant role in controlling its chemistry \citep[see][for a review]{Maiolino19}. Metals
are the key coolants in the cold phase (\Tkin\ $< 10^4$ K) of the ISM, such as the [\cii]\ $158\,\mu$m and [\oi]\ $63\,\mu$m fine-structure lines, molecular lines such as the CO ladder, and dust emission \citep{Tielens05, Draine11, Goldsmith12}.  Denoted as $Z$, metallicity refers to the mass fraction of elements heavier than hydrogen ($X$) and helium
($Y$) present in a medium, so that $X  + Y  + Z = 1$. The galactic metallicity is often presented by the
abundance measurement of a single element; either by Oxygen ([O/H], frequently observed for
gas), or by Iron ([Fe/H], frequently observed for stars). In the MW and also in other galaxies, $Z$ may change 
as a function of the galactocentric radius with the general trend to decrease towards the outer 
galactic regions \citep[e.g.][]{Wolfire03,Wuyts16,Kreckel19,Esteban22,Matsunaga23}.

Throughout this work, we will refer to the oxygen and carbon abundances using the square bracket notation \citep{Aller60,Pagel09,Maiolino19}, as was also used above.
The notation [A/B] represents the logarithmic ratio of the number of atoms of element A to element B, normalized to a reference abundance, typically the Sun:
\begin{eqnarray}
\mathrm{[A/B]} 
= \log_{10}\left(\frac{N_{\mathrm{A}}}{N_{\mathrm{B}}}\right) - \log_{10}\left(\frac{N_{\mathrm{A}}}{N_{\mathrm{B}}}\right)_{\odot}.
\end{eqnarray}
For the purposes of the above equation, we adopt solar abundances \citep{Asplund2009, Asplund21} in which $12+\log_{10}(\rm O/H)=8.69$\footnote{The most recent measurements by \citet{Pietrow23} set the solar oxygen abundance to $12+\log_{10}(\rm O/H)=8.73\pm0.03$} and $12+\log_{10}(\rm C/H)=8.43$. Consequently, for $\rm [O/H]=0$ and $\rm [C/O]=0$ it is understood that solar values are assumed. However, throughout the paper, gas-phase elemental abundances will be considered which are lower than the solar values to account for effects related to depletion on grains (see \S\ref{ssec:initialabund}).

\subsection{Chemical evolution of elements in galaxies}
\label{ssec:Intro-chemicalevolution}

Oxygen and carbon are the most abundant metals in the
Universe and they, therefore, constitute the core of chemical evolution of
galaxies. Similar to all other elements with atomic mass $A>12$, i.e. from
carbon to iron, C and O are continuously enriched via stellar nucleosynthesis
throughout the galactic evolution \citep[see][for a review]{Romano22}, 
while elements heavier than $^{56}$Fe are mainly produced via neutron capture \citep{B2FH} by Asymptotic Giant Branch stars (AGB) \citep{Gallino98, Cescutti22} and core collapse supernovae \citep{Limongi18}, or by more catastrophic events such as binary-neutron star mergers or collapsars \citep{Kasen17,Kajino19}. 
However, carbon and oxygen originate from different processes, making their abundances 
vary both from galaxy-to-galaxy and also within a galaxy.

The \stO\ (O-16) isotope is mostly synthesised from \twC\ and $\alpha$
particles (i.e., the Helium nucleus), through the reaction of
$^{12}$C($\alpha,\gamma$)$^{16}$O, which has a low reaction rate in
low-temperature hydrostatic He-burning. The major producer of \stO\ is the 
short-lived massive stars with mass $\gtrsim8\,{\rm M}_{\odot}$. 
Through type-II supernovae explosions, oxygen is almost always released back to the ISM within in a 
short period of time \citep{Romano22}. 

The \twC\ (C-12) isotope, on the other hand, is synthesised through the
triple-$\alpha$ capture process during the He-burning stage, in which three
$\alpha$ particles  are synthesised to the Hoyle state of \twC\ \citep{Hoyle54,Freer14}. Apart from massive stars, 
carbon is also and mainly produced in low- and intermediate-mass stars ($1<M<8\,{\rm M}_{\odot}$) which live much longer than
the first ones. Carbon is released into the ISM through type Ia supernovae and through 
the Asymptotic Giant Branch (AGB) stars \citep[e.g.][]{Berg19}. As discussed below, this delay in the release compared to oxygen, makes
the C/O ratio to decrease throughout the chemical evolution period, allowing also astronomers to use it as a clock. 

Although the abundance evolution of C and O elements involves various processes, including stellar
physics, the stellar Initial Mass Function (IMF), and the evolutionary history
of the host galaxy, the relative evolutionary patterns of C and O started to be
known through observational and theoretical studies \citep{Skillman1998}. In the MW 
and in nearby galaxies, both C and O abundances as well as their relative ratio C/O 
have been systematically measured from the surface of stars of various ages 
and metallicities, which reflects the conditions of the ISM throughout history \citep[e.g.,][]{Surez2018}. 

In particular, throughout the galactic evolution and starting from metallicities as low as
$\rm [O/H]{\approx}-3$ (or $\rm [Fe/H]\approx-2$) and until $\rm [O/H]{\approx}-1.5$, several 
observations \citep[e.g.][]{Akerman04,Fabbian09,Berg16,Berg19} show that the [C/O] ratio drops from 
solar values, reaching a minimum of $\rm [C/O]{\approx}-0.7$ to $-1$ 
\citep{Trainor16, Cooke17, Maiolino19}. This trend (which decreases also as a function
of [Fe/H]), is measured from Carbon-Enhanced Metal-Poor (CEMP) stars. CEMP stars are possibly 
enriched by the first generation (Population~III) stars or by binary stars whose one member is an AGB resulting in mass transfer \citep{Bonifacio15}. 
At $\rm [O/H]{\approx}-1.5$ (or $\rm [Fe/H]>-2$), a turnover occurs bringing the [C/O] ratio back 
to solar values ($\rm [C/O]{\approx}0$) \citep{Spitoni19, Romano20, Delgado21}. This trend is 
observed in both MW stars and metal-poor gas of local dwarf galaxies \citep{Berg19,Romano22}.

For [C/O] ratios less than the expected ones from solar scaling, it means that the relative abundance of oxygen (the most typical $\alpha$-element) is enhanced relative to the expected value. Therefore, the aforementioned conditions can be considered as `$\alpha$-enhanced'. 

\subsection{Treatment of elemental abundances and dust in astrochemical models}

In astrochemical models concerning small-scale objects, such as the circumstellar disks
of AGB stars \citep[e.g.][]{Li16} and protoplanetary discs
\citep[e.g.][]{Oberg11,vDishoeck23}, non-Solar values of the C/O ratio are considered due 
to their particular importance in icy chemistry. For example, the Gas phase Elemental abundances in Molecular 
cloudS \citep[`GEMS';][]{Fuente19} is a 
dedicated IRAM program to study the depletion of key elements, such as S, C, N, and O, 
in star-forming filaments. However, such systematic studies in molecular clouds and on galactic scales,
especially considering extra-galactic conditions, are still lacking. 

In most astrochemical models of the MW and external galaxies, it is
frequently assumed that the CNO abundances have a linear correlation with
metallicity. This assumption means that, for example, the abundances of C and N will decrease by the same amount with [O/H],
leading to constant [C/O] and [N/O] as a function of [O/H]
\citep[e.g.][]{Kaufman99,Glover16,Gong17,Gong20,Bisbas19,Bisbas21}. It was only recently since the impact of negative [C/O] and [N/O] ratios on the larger-scale ISM and star-formation process have
been considered in numerical and analytical works (see \S\ref{ssec:impact}). 

Similar to the abundance scaling, the dust-to-gas ratio is
frequently assumed to decrease linearly with metallicity \citep[see
e.g.][]{Bate14,Bate19,Tanaka14,Tanaka18,Bisbas21}. However, observations
\citep{Galametz11,Herrera12,RemyRuyer14} show that for low values of [O/H]
(e.g. for the $\rm [O/H]=-1.21$ modelled here), the dust-to-gas ratio decreases
even further, favouring the photodissociation processes of both H$_2$ and CO
molecules. This behavior is important for the C/O ratio all the more since the 
carbon abundance and the dust-to-gas relation are related \citep{Mathis90, Dwek98}.
The assumption of a linear correlation leads to an
overestimation of the dust opacity having consequences in 
the formation of H$_2$ on grains. 
Apart from the fact that the grain abundance controls the FUV attenuation along a column, it is also directly proportional to the rate of photoelectric heating and, thus, important for the thermal balance \citep{Wolfire08}.

\subsection{Impact of a non-linear [C/O] with [O/H] in the ISM of galaxies}
\label{ssec:impact}

The impact of a negative [C/O] in the ISM has been recently considered in observations 
and numerical simulations of high-redshift galaxies. Using ALMA, \cite{Harikane20} 
studied galaxies at a redshift of $z>6$ and found that low [C/O] partially explain 
the enhanced $L_{\rm [OIII]}/L_{\rm [CII]}$ and possibly the $L_{\rm [CII]}/{\rm SFR}$ 
(Star Formation Rate) ratios observed in these systems. Cosmological radiation 
hydrodynamics simulations of \citet{Arata20} find that the $L_{\rm [OIII]}/L_{\rm [CII]}$ 
luminosity ratio decreases with increasing [C/O], which is eventually connected with the 
metal enrichment during the galaxy evolution. Similar conclusions were addressed in the models of
\citet{Katz22} who showed that low C/O ratios are needed to reproduce the observed relation 
of $\rm [C\textsc{ii}]_{\rm 158\mu m}$--SFR and $\rm [O\textsc{iii}]_{\rm 88\mu m}$--SFR in $z>6$
galaxies, as well as the observed higher $\rm [O\textsc{iii}]_{\rm 88\mu
m}/[C\textsc{ii}]_{\rm 158\mu m}$ line ratio in the Epoch of Reionization. 
\citet{Katz22} further suggest that this results in a top-heavy IMF in early-Universe star-forming galaxies. This finding is in agreement
with the \citet{Zhang18} $^{13}$C/$^{18}$O isotope ratio observations in
high-$z$ galaxies, implying an increased population of massive stars. 

In studying the IMF, the analytical work of \citet{Sharda23} showed that negative [C/O]
values in metal-poor ISM environments impact the transition point from
top-heavy to bottom-heavy IMF, shifting it upwards by $\sim0.5-1.0\,\rm dex$ in
metallicity and that the characteristic mass \citep[which sets the peak of the stellar IMF depending on the ISM environmental conditions;][]{Sharda22} also increases by a factor of
$\sim7$. This effect is a consequence of the cooling processes in the ISM,
since both carbon and oxygen are major cooling components in the total cooling
function. It is, therefore, expected that such negative [C/O] values
-regardless of the metallicity- will have a direct impact on the observables,
especially concerning the carbon cycle as described below.

\subsection{Carbon cycle: Background}
\label{sec:background}

Photodissociation regions (PDR) are essential for understanding the astrochemistry of molecular clouds and the ISM at large. They are the sites where the
atomic-to-molecular transition occurs, as well as the transition between the
three carbon phases (ionized, atomic, and molecular in the form of carbon
monoxide) known as the `carbon cycle' \citep[hereafter `C-cycle'; see reviews
by][]{Hollenbach99,Wolfire22}. However, the exact location of these phase
transitions depends on a variety of environmental parameters
\citep{Jura74,Black77,vDishoeck86,Offner13,Sternberg14,Bialy15,Bisbas23}. The most
important of those include the radiation due to far-UV photons emitted by
massive stars which photodissociate the molecules of H$_2$ and CO
\citep{vDishoeck88}, the ionization rate due to the interaction of ISM gas with
charged particles carrying high-energies at high column densities known as
`cosmic-rays' \citep[see reviews by][]{Strong07, Grenier15}, and the
metallicity which is described above and consists the main focus of this work.
Other environmental parameters, which will not be considered here, include
X-rays \citep{Maloney96,Meijerink06,Mackey19}, turbulence \citep{Xie95} and
shocks \citep{Meijeirink11,Kelly17,Cosentino19,James20}.

Ionized carbon (`C{\sc ii}' when referring to its emission or `C$^+$' when referring to its abundance) is an ion with significant importance to 
the study of the ISM. Its fine-structure transition, at $157.7\mu$m, is one of the brightest 
cooling lines and is commonly used to trace warm neutral gas and ionized gas. 
[C{\sc ii}] is excited due to inelastic collisions mainly with e$^-$, H{\sc i}, and H$_2$ \citep{Goldsmith12,Lique13}, and can be enhanced by interstellar shocks \citep{Draine93,Appleton13}.
It is emitted by both diffuse and dense gas in star-forming regions \citep{Stacey91, Brauher08, Accurso17, Franeck18, Cormier19}. The [C{\sc ii}] 158 \mum\ line is also often adopted as a tracer of 
Star Formation Rate (SFR) of galaxies \citep{DeLooze11, Pineda14, Herrera15, Sutter19, Bisbas22, Liang23} and as a tracer of the molecular gas mass \citep{Accurso17, Combes18, Zanella18, Madden20}, including the dynamical evolution of colliding clouds \citep{Bisbas18,Schneider23}.

Atomic carbon (`C{\sc i}' when referring to its emission or `C' when referring to its abundance) is another important tracer of the ISM. It is emitted at frequencies of $492.2$ and $809.3,{\rm GHz}$ through the [C{\sc i}] $^{3}P_1\rightarrow {^3}P_0$ (hereafter `[C{\sc i}]~(1-0)') and $^3P_2\rightarrow {^3}P_1$ (hereafter `[C{\sc i}]~(2-1)') fine-structure lines, respectively. While the C layer is believed to be thin and located between C$^+$ and carbon monoxide (CO) in classical one-dimensional PDRs \citep{Draine11}, it has been found to accurately trace the molecular mass content of ISM gas \citep{Papadopoulos04, Papadopoulos18, Bell07, Lo14, Offner14, Zhang14, Glover15, Jiao17, Jiao19, Gaches19b, Bisbas21, Dunne22}. This suggests that C emission originates from a wider range of densities than those where the C abundance peaks. The ratio of the two [C{\sc i}] lines is often used to investigate the properties of the observed gas \citep{Bothwell17, Valentino20}, including the environmental parameters of the ISM \citep{Bisbas21, Bisbas23}.

$^{12}$CO (hereafter referred to as CO) is the most abundant molecule (except for H$_2$) in the ISM and it is the most commonly used tracer of the molecular gas in the MW and extragalactic objects \citep{Tacconi08, Genzel12, Narayanan12, Bolatto13, Gong20, Luo20, FriasCastillo23, MontoyaArroyave23}. It emits at different wavelengths depending on the $J\rightarrow J-1$ transition. The collection of these transitions consists  the Spectral Line Energy Distribution (SLED), which is used as an important diagnostics for studying the chemical and dynamical states of the ISM \citep{Papadopoulos10b, Narayanan14, Mashian15, Rosenberg15, Vallini18, Klitsch22, Stanley23} and a discriminant between the X-ray and FUV heating \citep{Meijerink07,Vallini19,Esposito22}.

\subsection{This work: C-cycle in extreme galactic environments}

The CO molecule has a very high binding energy of $\sim$ 7.7 eV, making it stable against various chemical conditions. Intuitively speaking, molecular clouds with $\rm  [C/O]<0$ would first 
bond most carbon in the CO molecule. Only dissociation processes, such as UV radiation 
fields, cosmic rays, and shocks, would release carbon into free atoms and produce C 
and C$^+$. Therefore, to quantitatively evaluate such impacts, we perform detailed chemical modeling to simulate various galactic conditions. 

This work aims to evaluate how an ISM component with a $\rm [C/O]<0$ ratio affects the abundances and the line emission of the gas-phase C-cycle as well as the location of the H{\sc i}-to-H$_2$ transition. To our knowledge, the impact of a negative [C/O] on the C-cycle observables and the H{\sc i}-to-H$_2$ transition has not been studied in the past. Particular attention is drawn on how an $\rm [C/O]<0$ environment affects the C layer and the emission of both its fine-structure lines, and its impact on extragalactic observations.

A total of nineteen three-dimensional PDR models have been performed in this regard using the density distribution of a molecular cloud that has been studied in previous works \citep{Bisbas17b,Bisbas21,Gaches22,Gaches22b} and considering various combinations of cosmic-ray ionization rates, dust-to-gas ratios and FUV intensities to best represent different galactic environments. We examine the response in the abundances of C$^+$, C and CO, and we also perform radiative transfer to produce velocity integrated emission maps of [C{\sc ii}]~$158\,\mu$m, [C{\sc i}](1-0) and (2-1), and the first ten CO transitions ($J=1-0$ to $10-9$).

This paper is organized as follows. In Section~\ref{sec:methods} we present the numerical approach followed. Section~\ref{sec:results} shows the results of our calculations including velocity integrated emission maps, column densities, and line ratios. Section~\ref{sec:discussion} discusses the impact of our results on observations. We conclude in Section~\ref{sec:conclusions}.

\section{Numerical method}
\label{sec:methods}

The PDR calculations in this study have been performed using the {\sc 3d-pdr}\footnote{https://uclchem.github.io/3dpdr/} code \citep{Bisbas12}. {\sc 3d-pdr} is a publicly available astrochemical code that treats one- and three-dimensional photodissociation regions. By taking into account several heating and cooling processes, the code performs thermal balance calculations and terminates when the total heating is approximately equal to the total cooling. Once the code is terminated, it outputs the abundances of species, gas and dust temperatures, heating and cooling functions as well as the level populations of various coolants, all versus the column depth. 

A subset of the UMIST2012 chemical network \citep{McElroy13} consisting of 33 species and 330 reactions is used. For the purposes of this work, we only consider the elements of carbon, oxygen, helium, and hydrogen in the gas-phase. All abundances are normalized to the total hydrogen. The intensity of the incident FUV radiation field, $\chi/\chi_0$, is normalized according to the spectral shape of \citet{Draine78}. Two cosmic-ray ionization rates per H$_2$ molecule are adopted; a low one with $\zeta_{\rm CR}=10^{-17}\,{\rm s}^{-1}$ and a high one with $\zeta_{\rm CR}=10^{-15}\,{\rm s}^{-1}$. Such higher values of $\zeta_{\rm CR}$ are found near supernovae \citep{Indriolo23} and are expected in galaxies with high star-formation rates \citep{Papadopoulos10a, Indriolo18, Gaches19, Yang23}. In all simulations, the isotropic FUV radiation field impinges radially and incident to each \textsc{HEALPix} ray used in the ray-tracing scheme \citep{Bisbas12}. In most of the models, the FUV intensity is taken to be $\chi/\chi_0=1$, however we also consider an additional and stronger intensity of $\chi/\chi_0=10^2$ for the $\zeta_{\rm CR}=10^{-15}\,{\rm s}^{-1}$ models. At all times, the microturbulence velocity is set to $v_{\rm turb}=2.0\,{\rm km}\,{\rm s}^{-1}$. The formation of H$_2$ molecule is modelled using the treatment of \citet{Cazaux02,Cazaux04,Cazaux10}. The H$_2$ self-shielding is calculated along each \textsc{HEALPix} ray emanated from each computational cell and using the results of \citet{Lee96}.

For the three-dimensional models, the `Dense' cloud of \citet{Bisbas21} is used, which is a sub-region of the binary collision simulation of giant molecular clouds presented in \citet{Wu17}. The selected sub-region has a resolution of $112^3$ uniform cells and contains a dense filamentary structure that remains molecular under most ISM conditions explored and that undergoes star-formation. It has a size of $L=13.88\,{\rm pc}$ and a total mass of $M=5.9\times10^4\,{\rm M}_{\odot}$ with a mean density of $\langle n_{\rm H}\rangle = 640\,{\rm cm}^{-3}$. In a full hydro-chemical simulation, it is expected that different ISM environmental parameters would affect hydrodynamical pressures resulting in different density and velocity distributions. These would, in turn, impact the abundances and the corresponding line emission. In the present work, however, the astrochemical models have been performed in a static cloud; this provides a different -and perhaps more educational- view of how negative [C/O] ratios affect the observables.

\begin{figure*}
    \centering
    \includegraphics[width=0.488\textwidth]{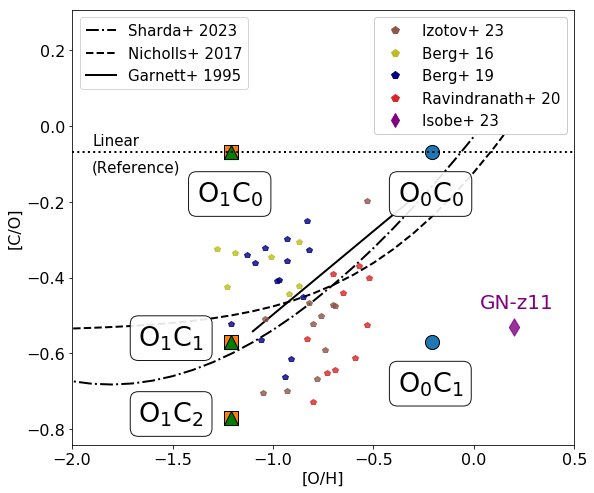}
    \includegraphics[width=0.494\textwidth]{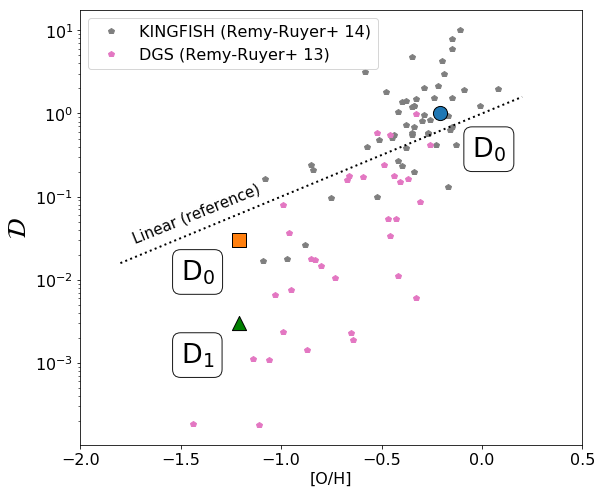}
    \caption{\textit{Left panel}: Pairs of [O/H] and [C/O] used in models. Blue circles correspond to models O$_0$C$_0$ and O$_0$C$_1$. Orange squares and green triangles correspond to models O$_1$C$_0$, O$_1$C$_1$, O$_1$C$_2$ according to the assumed $\cal D$, which is the dust-to-gas ratio normalized to the solar value (see right panel). The horizontal dotted black line (`Linear') corresponds to the assumption of a constant [C/O] (equal to $-0.07$) and independent on [O/H].  The \citet{Sharda23,Sharda23cor} (Eqn.~\ref{eqn:sharda}) cubic-fit, the \citet{Nicholls2017} (Eqn.\ref{eqn:nicholls}) best-fit, and the \citet{Garnett95} (Eqn.\ref{eqn:garnett}) best-fit are shown in dot-dashed, dashed, and solid lines, respectively. Small brown diamonds refer to Lyman-continuum leaking galaxies at $z\sim0.3-0.4$ of \citet{Izotov23}, yellow and dark blue to metal-poor dwarf galaxies of \citet{Berg16} and \citet{Berg19}, respectively and red to green pea galaxies of \citet{Ravindranath20}. The purple diamond refers to the \citet{Isobe23} JWST observations of GN-z11. \textit{Right panel}: Relation of [O/H] versus $\cal D$. The big blue circle, orange square and green triangle correspond to the values used in our models while their colours are in tandem with those of the left panel. Blue and orange models follow an approximately linear decrease of $\cal D$ as a function of [O/H] and are labeled as `D$_0$'. The green triangle has a sub-linear $\cal D$ and is labeled as `D$_1$'. Black dotted line corresponds to a linear connection between [O/H] and $\cal D$ to guide the eye. Gray diamonds correspond to the KINGFISH observations of \citet{RemyRuyer14}, while magenta diamonds to the Dwarf Galaxy Survey data of \citet{RemyRuyer13}.}
    \label{fig:panel}
\end{figure*}

Numerical instabilities may arise in PDR models as a result of the very non-linear nature of the ordinary differential equations that are solved until thermal balance is reached. In order to minimize bi-stabilities \citep{LeBourlot93,Viti01,Roueff20,Dufour21} which result in a significant noise in the calculations, a mixture of carbon phase abundances is preferred in which the total abundance of carbon is always divided in three parts for C$^+$, C, and CO, respectively.

\subsection{Initial abundances}
\label{ssec:initialabund}

The fiducial (MW) gas-phase carbon abundance\footnote{Hereafter, $x_{\rm sp}$ is to denote the relative abundance of species `sp'.} is taken to be $x_{\rm C, MW}=1.4\times10^{-4}$ and refers to the interstellar carbon abundance within the 600~pc solar neighbourhood \citep{Cardelli96,Roellig07,Draine11}. The fiducial gas-phase oxygen abundance, is taken to be $x_{\rm O, MW}=3\times10^{-4}$ \citep{Cartledge04,Draine11}. Both aforementioned abundances are lower than the reported Solar ones \citep{Asplund2009} due to depletion on grains and have been measured by optical/UV absorption lines in diffuse clouds \citep{Cardelli96,Cartledge04}. In addition, we assume a helium abundance of $x_{\rm He}=1.00\times10^{-1}$. The fiducial model has, thus, an $\rm [O/H]=-0.21$ and a $\rm [C/O]=-0.07$.

\begin{table}
    \centering
    \caption{Initial conditions and PDR parameters of the presented simulations. The first column gives the identifier of each simulation. To ease the reader, we represent the key models with the colours and shapes shown in Fig.~\ref{fig:panel} and elsewhere. The `x' notation refers to the negative logarithm of the cosmic-ray ionization rate and it is either 17 (for $\zeta_{\rm CR}=10^{-17}\,{\rm s}^{-1}$) or 15 (for $\zeta_{\rm CR}=10^{-15}\,{\rm s}^{-1}$). Models including the `y' notation refer to the additional calculations with a higher FUV intensity (2 for $\chi/\chi_0=10^2$, otherwise no number noted). The second and third columns show the [O/H] and [C/O] ratios, respectively. The fourth and fifth column refers to the fractional gas-phase abundances with respect to hydrogen nuclei used in the {\sc 3d-pdr} simulations. Simulations O$_0$C$_0$D$_0$, O$_1$C$_0$D$_0$ and O$_1$C$_0$D$_1$ follow a linear decrease of the C/O ratio (they have all C$_0$). The sixth column shows $\cal D$, which is the dust-to-gas ratio normalized to the solar one of $10^{-2}$. D$_0$ refers to an approximately linear decrease of $\cal D$, whereas D$_1$ to a sub-linear. In all cases the bracketed numbers indicate the order of magnitude.}
    \begin{tabular}{lccccc}
    \hline
    ID & [O/H] & [C/O] & $x_{\rm C}$ & $x_{\rm O}$  & $\cal D$ \\
    \hline
    \tikz\draw[black,fill=blue1] (0,0) circle (.6ex); O$_0$C$_0$D$_0$-x & -0.21 & -0.07 & $1.40(-4)$ & $3.00(-4)$ & 1 \\
    \tikz\draw[black,fill=blue1] (0,0) circle (.6ex); O$_0$C$_1$D$_0$-x-y & -0.21 & -0.57 & $4.43(-5)$ & $3.00(-4)$ & 1\\
    \tikz\draw[black,fill=orange1] (0,0) rectangle (1ex,1ex); O$_1$C$_0$D$_0$-x & -1.21 & -0.07 & $1.40(-5)$ & $3.00(-5)$ & $3(-2)$ \\
    \tikz\draw[black,fill=orange1] (0,0) rectangle (1ex,1ex); O$_1$C$_1$D$_0$-x & -1.21 & -0.57 & $4.43(-6)$ & $3.00(-5)$ & $3(-2)$\\
    \tikz\draw[black,fill=orange1] (0,0) rectangle (1ex,1ex); O$_1$C$_2$D$_0$-x-y & -1.21 & -0.77 & $2.80(-6)$ & $3.00(-5)$ & $3(-2)$ \\
    \tikz\draw[black,fill=green1] (0,0) -- (.7ex,1.2ex) -- (1.4ex,0) -- cycle; O$_1$C$_0$D$_1$-x & -1.21 & -0.07 & $1.40(-5)$ & $3.00(-5)$ & $3(-3)$ \\
    \tikz\draw[black,fill=green1] (0,0) -- (.7ex,1.2ex) -- (1.4ex,0) -- cycle; O$_1$C$_1$D$_1$-x & -1.21 & -0.57 & $4.43(-6)$ & $3.00(-5)$ & $3(-3)$\\
    \tikz\draw[black,fill=green1] (0,0) -- (.7ex,1.2ex) -- (1.4ex,0) -- cycle; O$_1$C$_2$D$_1$-x-y & -1.21 & -0.77 & $2.80(-6)$ & $3.00(-5)$ & $3(-3)$ \\
      \hline
    \end{tabular}
    \label{tab:ids}
\end{table}

\begin{table*}
    \centering
    \caption{Average abundances and velocity integrated emissions of all simulations. Column 1 shows the IDs of all models. Column 2 refers to the object type each simulation best represents. Columns 3-7 refer to the average (density-weighted) abundances of H{\sc i}, H$_2$, C$^+$, C, and CO, respectively. The numbers in the parenthesis denote the order of magnitude. Column 8 refers to the average (density-weighted) gas temperature (K). Columns 9-11 refer to the area average emissions of [C{\sc ii}]~$158\mu$m, [C{\sc i}](1-0), and CO $J=1-0$, respectively. The average emission is given in units of $\rm K\,km\,s^{-1}$.}
    \label{tab:all}
    \begin{tabular}{llccccccccc}
        \hline
        ID & Representative type$^{\star}$ & $\langle x_{\rm HI}\rangle$ & $\langle x_{\rm H2}\rangle$ & $\langle x_{\rm C^+}\rangle$ & $\langle x_{\rm C}\rangle$ & $\langle x_{\rm CO}\rangle$ & $\langle T_{\rm gas}\rangle$ & [C{\sc ii}] & [C{\sc i}](1-0) & CO(1-0) \\
        \hline
        \tikz\draw[black,fill=blue1] (0,0) circle (.6ex); O$_0$C$_0$D$_0$-17                                 & Solar Neighbourhood                   & 0.27(-2) & 0.49(0) & 1.07(-5) & 5.42(-6) & 1.21(-4) & 11.8 & 0.21 & 2.53  & 45.32 \\
        \tikz\draw[black,fill=blue1] (0,0) circle (.6ex); O$_0$C$_1$D$_0$-17                                & Local post-starbursts                  & 0.25(-2) & 0.49(0) & 4.03(-6) & 1.12(-6) & 3.87(-5) & 14.7 & 0.14 & 1.01  & 51.11 \\
        \tikz\draw[black,fill=orange1] (0,0) rectangle (1ex,1ex); O$_1$C$_0$D$_0$-17                         & Dwarfs (C/O$_{\rm L-S}$)              & 0.88(-1) & 0.45(0) & 4.47(-6) & 3.55(-6) & 5.84(-6) & 17.6 & 0.23 & 4.59  & 12.64 \\
        \tikz\draw[black,fill=orange1] (0,0) rectangle (1ex,1ex); O$_1$C$_1$D$_0$-17                        & Dwarfs (C/O$_{\rm Obs}$)              & 0.82(-1) & 0.45(0) & 1.46(-6) & 1.11(-6) & 1.80(-6) & 24.7 & 0.20 & 2.23  & 10.82 \\
        \tikz\draw[black,fill=orange1] (0,0) rectangle (1ex,1ex); O$_1$C$_2$D$_0$-17                        & Dwarfs (C/O$_{\rm Ext}$)              & 0.80(-1) & 0.46(0) & 9.32(-7) & 7.27(-7) & 1.13(-6) & 28.2 & 0.18 & 1.62  & 9.49 \\
        \tikz\draw[black,fill=green1] (0,0) -- (.7ex,1.2ex) -- (1.4ex,0) -- cycle; O$_1$C$_0$D$_1$-17     & Dwarfs (C/O$_{\rm L-S}$, oD)          & 0.27(0)  & 0.36(0) & 5.61(-6) & 5.01(-6) & 3.26(-6) & 15.2 & 0.17 & 4.72  & 5.36 \\
        \tikz\draw[black,fill=green1] (0,0) -- (.7ex,1.2ex) -- (1.4ex,0) -- cycle; O$_1$C$_1$D$_1$-17    & Dwarfs (C/O$_{\rm Ext}$, oD)          & 0.25(0)  & 0.37(0) & 1.85(-6) & 1.54(-6) & 9.90(-7) & 20.8 & 0.16 & 2.73  & 4.89 \\
        \tikz\draw[black,fill=green1] (0,0) -- (.7ex,1.2ex) -- (1.4ex,0) -- cycle; O$_1$C$_2$D$_1$-17    & Dwarfs (C/O$_{\rm L-S}$, oD)          & 0.25(0)  & 0.37(0) & 1.18(-6) & 9.83(-7) & 6.20(-7) & 23.1 & 0.15 & 1.97  & 4.31 \\
        \tikz\draw[black,fill=blue1] (0,0) circle (.6ex); O$_0$C$_0$D$_0$-15                                 & Solar Neighbourhood (hCR)             & 0.20(-1) & 0.49(0) & 2.27(-5) & 3.29(-5) & 8.22(-5) & 28.3 & 2.77 & 41.42 & 67.63 \\
        \tikz\draw[black,fill=blue1] (0,0) circle (.6ex); O$_0$C$_1$D$_0$-15                                & Local ULIRGs (hCR)                    & 0.20(-1) & 0.49(0) & 6.26(-6) & 2.30(-6) & 3.53(-5) & 41.2 & 1.26 & 4.57  & 124.63 \\
        \tikz\draw[black,fill=blue1] (0,0) circle (.6ex); O$_0$C$_1$D$_0$-15-2                             & Local ULIRGs (hCR, hUV)               & 0.50(-1) & 0.47(0) & 9.68(-6) & 2.87(-6) & 3.13(-5) & 86.8 & 3.93 & 5.93  & 91.01 \\
        \tikz\draw[black,fill=orange1] (0,0) rectangle (1ex,1ex); O$_1$C$_0$D$_0$-15                         & S-B Dwarfs (C/O$_{\rm L-S}$,hCR)      & 0.16(0)  & 0.41(0) & 4.39(-6) & 5.09(-6) & 4.39(-6) & 46.1 & 2.65 & 10.24 & 24.75 \\
        \tikz\draw[black,fill=orange1] (0,0) rectangle (1ex,1ex); O$_1$C$_1$D$_0$-15                        & S-B Dwarfs (C/O$_{\rm Obs}$,hCR)      & 0.17(0)  & 0.41(0) & 1.40(-6) & 1.06(-6) & 1.91(-6) & 61.3 & 1.36 & 2.42  & 17.79 \\
        \tikz\draw[black,fill=orange1] (0,0) rectangle (1ex,1ex); O$_1$C$_2$D$_0$-15                        & S-B Dwarfs (C/O$_{\rm Ext}$,hCR)      & 0.17(0)  & 0.41(0) & 8.86(-7) & 6.24(-7) & 1.28(-6) & 66.0 & 0.93 & 1.48  & 13.80 \\
        \tikz\draw[black,fill=orange1] (0,0) rectangle (1ex,1ex); O$_1$C$_2$D$_0$-15-2                     & S-B Dwarfs (C/O$_{\rm Obs}$,hCR,hUV)  & 0.38(0)  & 0.30(0) & 1.70(-6) & 6.95(-7) & 3.83(-7) & 93.6 & 3.19 & 1.56  & 3.00  \\
        \tikz\draw[black,fill=green1] (0,0) -- (.7ex,1.2ex) -- (1.4ex,0) -- cycle; O$_1$C$_0$D$_1$-15     & Dwarfs (C/O$_{\rm L-S}$, oD,hCR)      & 0.40(0)  & 0.29(0) & 5.19(-6) & 6.45(-6) & 2.24(-6) & 34.4 & 1.50 & 11.31 & 10.08 \\
        \tikz\draw[black,fill=green1] (0,0) -- (.7ex,1.2ex) -- (1.4ex,0) -- cycle; O$_1$C$_1$D$_1$-15    & Dwarfs (C/O$_{\rm Obs}$, oD,hCR)      & 0.40(0)  & 0.29(0) & 1.72(-6) & 1.74(-6) & 9.22(-7) & 41.8 & 0.99 & 3.67  & 7.88 \\
        \tikz\draw[black,fill=green1] (0,0) -- (.7ex,1.2ex) -- (1.4ex,0) -- cycle; O$_1$C$_2$D$_1$-15    & Dwarfs (C/O$_{\rm Ext}$, oD,hCR )     & 0.41(0)  & 0.29(0) & 1.10(-6) & 1.06(-6) & 6.21(-7) & 44.9 & 0.77 & 2.33  & 6.09 \\
        \tikz\draw[black,fill=green1] (0,0) -- (.7ex,1.2ex) -- (1.4ex,0) -- cycle; O$_1$C$_2$D$_1$-15-2 & Dwarfs (C/O$_{\rm Ext}$, oD,hCR, hUV) & 0.73(0)  & 0.13(0) & 2.13(-6) & 5.59(-7) & 1.00(-7) & 46.7 & 2.01 & 1.13  & 0.64 \\
        
        \hline
    \end{tabular}
      \\\small
      {$^{\star}$ ULIRGs: Ultra Luminous Infrared galaxies, S-B: Starburst, L-S: Linearly-scaled, Obs: Observed, Ext: Extreme, oD: observed dust-to-gas ratio, hCR: high cosmic-ray ionization rate, hUV: high FUV intensity}
\end{table*}

In both panels of Fig.~\ref{fig:panel}, the larger shapes of circles, boxes, and triangles, show the combinations of [O/H] and [C/O] (left panel), and of [O/H] and $\cal D$ (right panel; see \ref{ssec:d2g} for the definition of $\cal D$) considered here. 
In all our models, the notation `O$_i$C$_j$D$_k$-x-y' is followed, where $i$ is the identifier of the oxygen abundance, $j$ of the carbon abundance and $k$ of the adopted $\cal D$. The terms `x' and `y' correspond to the logarithm of the cosmic-ray ionization rate (`17' for $\zeta_{\rm CR}=10^{-17}$ or `15' for $10^{-15}\,{\rm s}^{-1}$) and to the logarithm of the FUV intensity field (no number for $\chi/\chi_0=1$ or `2' for $\chi/\chi_0=10^2$), respectively. Table~\ref{tab:ids} shows a summary of all presented models. 

In the left panel of Fig.~\ref{fig:panel}, individual observations of metal-poor dwarf galaxies \citep{Berg16, Berg19}, star-forming galaxies of low-mass and low-metallicity at a redshift of $z\sim0.1-0.3$ known as `green pea galaxies' \citep{Ravindranath20} as well as Lyman-continuum leaking galaxies at a redshift of $z\sim0.3-0.4$ \citep{Izotov23} are shown with small diamonds. The purple diamond refers to the recent JWST observations of GN-z11 \citep[][]{Isobe23}. The horizontal dotted black line refers to the constant [C/O] (equal to $-0.07$), which corresponds to the commonly assumed linear relationship between carbon and oxygen. In addition, three best-fit relations are illustrated. The dot-dashed line shows the \citet{Sharda23} expression\footnote{The \citet{Sharda23} expression is given by: 
\begin{equation}
\label{eqn:sharda}
{\rm [C/O]_{MW}} = a_S{\rm [O/H]_{MW}}^3+b_S{\rm [O/H]_{MW}}^2 + c_S{\rm [O/H]_{MW}}+d_S
\end{equation}
where $a_S=-0.02$, $b_S=0.14$, $c_S=0.6$ and $d_S=-0.09$. Note that the above presented in \citet{Sharda23} uses the MW abundances as a reference value \citep[see][for correction]{Sharda23cor}. The corresponding curve of Fig.~\ref{fig:panel} is plotted after some further calculations considering the Solar abundances adopted here as the reference value.}. This expression is a cubic-fit of the \citet{Amarsi19c} 3D non-LTE calculations in a sample of 187 stars that exist in the thin and thick discs and the metal-poor halo of the MW.
The dashed line shows the \citet{Nicholls2017} expression\footnote{The \citet{Nicholls2017} expression is given by:
\begin{equation}
\label{eqn:nicholls}
    {\rm [C/O]} = \log_{10}\left(10^{a_N}+10^{{\rm [O/H]}+\log_{10}\left(\frac{x_{\rm O}}{x_{\rm H}}\right)_{\odot}+b_N}\right) 
    -\log_{10}\left(\frac{x_{\rm C}}{x_{\rm O}}\right)_{\odot}
\end{equation}
where $a_N=-0.8$ and $b_N=2.72$.}. 
This expression is a best-fit of various observations including disk and halo MW stars and damped Ly$\alpha$ systems \citep[see][and references therein]{Berg16,Berg19}.
The thin solid line corresponds to the \citet{Garnett95} best-fit of least squares\footnote{The \citet{Garnett95} expression is given by
\begin{equation}
\label{eqn:garnett}
    {\rm [C/O]} = 0.44\left\{{\rm [O/H]}+\log_{10}\left(\frac{x_{\rm O}}{x_{\rm H}}\right)_{\odot}\right\}+1.14-\log_{10}\left(\frac{x_{\rm C}}{x_{\rm O}}\right)_{\odot}.
\end{equation}
}.

\subsection{Dust-to-gas mass ratio}
\label{ssec:d2g}

The reference value of the dust-to-gas ratio is taken to be $10^{-2}$ \citep{Sandstrom13}. We further define with $\cal D$ the dust-to-gas ratio normalized to the above value, thus ${\cal D}=1$ is the fiducial one. This ratio affects both the opacity and the grain photoelectric heating. The right panel of Fig.~\ref{fig:panel} shows the observed relationship between $\cal D$ and [O/H]. The dotted black line shows the linear connection between [O/H] and $\cal D$ that is frequently adopted. As described in the \hyperref[sec:intro]{Introduction} this linear correlation of the dust-to-gas ratio with metallicity, may not be valid for very metal-poor galaxies \citep{RemyRuyer14}. This is further shown with the individual small triangles which correspond to observations from the Dwarf Galaxy Survey \citep{RemyRuyer13} and the KINGFISH survey of nearby galaxies \citep{RemyRuyer14}. To accommodate this observed behavior and explore its imprint on the C-cycle, we consider two $\cal D$ values: ${\cal D}=3\times10^{-2}$ (which is a $30\times$ decrease as opposed to the $10\times$ decrease in abundances) to represent an approximately linear/sub-linear scaling, as well as a much lower $\cal D$ of $3\times10^{-3}$. Our calculations assume a standard MRN distribution \citep{Mathis97} as described in \citet{Bisbas12}.

Models O$_0$C$_0$D$_0$ refer to the standard MW carbon and oxygen abundances. In model O$_0$C$_1$D$_0$, the carbon abundance is reduced by half an order of magnitude while keeping $\cal D$ constant. Such a combination between [O/H] and [C/O] has been observed in the GN-z11 galaxy using JWST \citep[][]{Isobe23}, which is located at a redshift of $z\simeq10.6$ \citep[][]{Oesch16}. In particular, the O$_0$C$_1$D$_0$-15 model is to represent conditions that are met in metal-rich starburst galaxies while O$_0$C$_1$D$_0$-15-2 is to explore also the effect of an FUV enhancement. 

Models O$_1$C$_0$D$_0$ use carbon and oxygen abundances that are both reduced by one order of magnitude (linear decrease). This model is used as a benchmark for the O$_1$C$_1$D$_0$ and O$_1$C$_2$D$_0$ which use the same (reduced) oxygen abundance but a further suppressed carbon abundance by $\sim3.15$ (C$_1$) and $5$ (C$_2$) times corresponding to $\rm [C/O]=-0.5$ and $\rm [C/O]=-0.7$, respectively. In all these cases, the normalized dust-to-gas ratio is ${\cal D}=3\times10^{-2}$ (D$_0$). The effect of a lower, sub-linear $\cal D$ is also explored and indicated with `D$_1$' in the ID of the corresponding model. Recently, \citet{Arellano23} reported JWST observations (not shown in the left panel of Fig.~\ref{fig:panel}) of the $z>7$ galaxy cluster SMACS J0723.3-7327 having $\rm [O/H]\simeq-1.57$ and $\rm [C/O]\simeq-0.57$ (see also Figure~4 of \citealt{Jones23} for a more complete picture).

\subsection{Synthetic observations and radiative transfer} 

To calculate the line emission of the C-cycle and thus obtain the corresponding synthetic observations \citep[see][for a review]{Haworth18}, the equation of radiative transfer is solved along the line-of-sight. For each C-cycle coolant, the methodology described in \citet{Bisbas17b} is followed, including the updates described in \citet{Bisbas21} to account for the dust contribution. We redirect the reader to these works for further details. 

\section{Results}
\label{sec:results}

\subsection{General behaviour}
\label{ssec:general}

\begin{figure}
    \centering
    \includegraphics[width=0.48\textwidth]{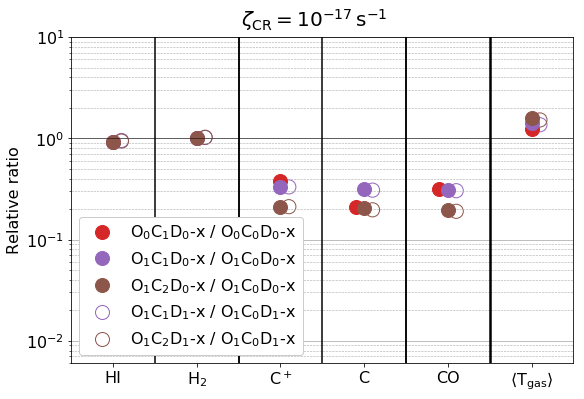}
    \includegraphics[width=0.48\textwidth]{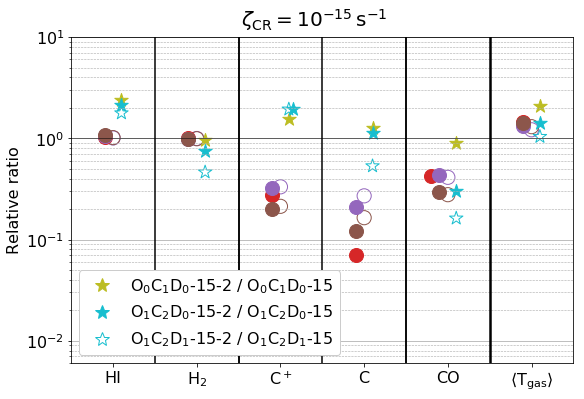}
    \caption{Comparative predictions of abundances for H{\sc i}, H$_2$, C$^+$, C, CO and the average gas temperature. The top panel presents models for $\zeta_{\rm CR}=10^{-17}\,{\rm s}^{-1}$, while the bottom panel depicts the $\zeta_{\rm CR}=10^{-15}\,{\rm s}^{-1}$ models. Consequently, `x' in the legend identifies the `17' and `15' models for the top and bottom panels, respectively. Abundances are normalized according to their designated reference models. In particular, the top panel reference models are characterized by $\rm [C/O]=-0.07$ for a constant [O/H], while the bottom panel reference models align with the corresponding lower FUV intensity in the ISM. Models marked with open circles adopt a sub-linear ${\cal D}=3\times10^{-3}$.}
    \label{fig:rabund}
\end{figure}

\begin{figure}
    \centering
    \includegraphics[width=0.48\textwidth]{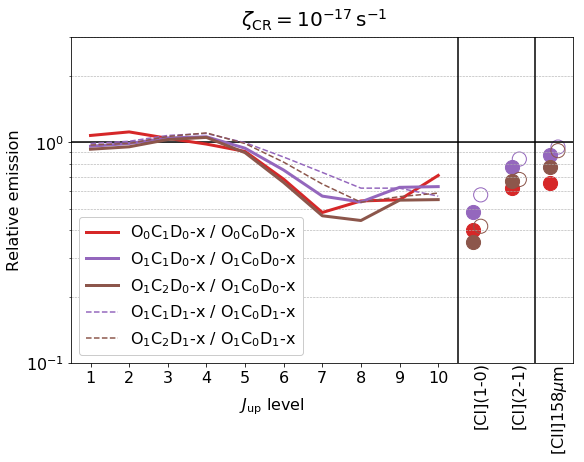}
    \includegraphics[width=0.48\textwidth]{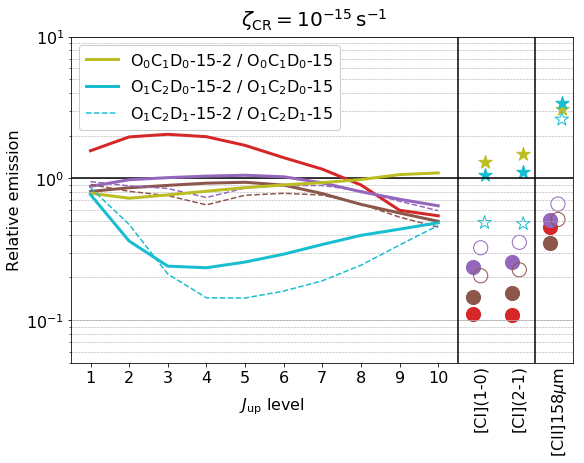}
    \caption{Relative emission lines of the CO SLED, both [C{\sc i}] lines and the [C{\sc ii}]~$158\,\mu$m line. Dashed lines indicate models with ${\cal D}=3\times10^{-3}$. The rest of the notation follows the one of Fig.~\ref{fig:rabund}.}
    \label{fig:remission}
\end{figure}

Table~\ref{tab:all} shows a summary of all simulation results. Column~2 refers to the most representative type of objects based on the ISM environmental parameters of each simulation. These types include local clouds in the solar neighbourhood, post-starburst, starburst, and dwarf galaxies with different C/O and dust-to-gas ratios. From column~3 onward, the density-weighted average abundances of H{\sc i}, H$_2$, C$^+$, C, and CO are shown, as well as the average gas temperature and the average velocity integrated emission of [C{\sc ii}]~158\,$\mu$m, [C{\sc i}](1-0) and CO $J=1-0$. The latter quantity is defined as 
\begin{equation}
    \langle W\rangle = \frac{\int W {\rm d}S}{\int {\rm d}S},
\end{equation}
where $W$ is the velocity integrated emission of the line (in units of ${\rm K}\,{\rm km}\,{\rm s}^{-1}$) and $S$ is the area of the whole map, without adopting a lower observational limit. As can be seen from columns 3 and 4, the cloud remains molecular for most of the cases ($\langle x_{\rm HI}\rangle < \langle x_{\rm H2}\rangle$), except for the extreme conditions of O$_1$C$_2$D$_0$-15-2 and for all O$_1$C$_j$D$_1$-15-y models (last five rows of Table~\ref{tab:all}).

The above results are visualized in Figures~\ref{fig:rabund} and \ref{fig:remission} where the aforementioned average abundance ratios and emissions of the C-cycle are shown, respectively. In particular, Figure~\ref{fig:rabund} shows the relative behaviour of the above species for $\zeta_{\rm CR}=10^{-17}\,{\rm s}^{-1}$ (top panel) and $\zeta_{\rm CR}=10^{-15}\,{\rm s}^{-1}$ (bottom panel). This relative behaviour refers to the normalization of the $\rm [C/O]<-0.07$ models (C$_1$ and C$_2$) with the corresponding ones for $\rm [C/O]=-0.07$ at a fixed [O/H] (e.g. O$_0$C$_1$D$_0$-17 / O$_0$C$_0$D$_0$-17 etc.). For the higher-FUV runs (O$_0$C$_1$D$_0$-15-2 and O$_1$C$_2$D$_1$-15-2) these are normalized with their counterparts of $\chi/\chi_0=1$ (e.g. O$_0$C$_1$D$_0$-15 etc.). 

It is found that the atomic-to-molecular mass content remains unaffected as a function of [C/O] and [O/H] for fixed $\zeta_{\rm CR}$ (the models with higher FUV intensity are described below). In all cases, the C$^+$, C and CO abundances decrease by a factor between $3-5$ when compared to the reference $\rm [C/O]=-0.07$ value. Notably, in the higher $\zeta_{\rm CR}$ models the C abundance can be decreased up to approximately $15\times$ (simulation O$_0$C$_1$D$_0$-15). 

The stronger FUV intensity in the selected $\zeta_{\rm CR}=10^{-15}\,{\rm s}^{-1}$ simulations affect all aforementioned species. In particular, the stronger $\chi/\chi_0$ slightly increases $\langle x_{\rm HI}\rangle$ by approximately $2.5\times$ in the O$_0$C$_1$D$_0$-15-2 simulation while the cloud remains molecular. However, in the lower metallicity model of O$_1$C$_2$D$_0$-15-2, the stronger FUV radiation pushes the H{\sc i}-to-H$_2$ transition towards higher column densities increasing its $x_{\rm HI}$. 
This effect is more prominent in the O$_1$C$_2$D$_1$-15-2 model where the additionally lower dust-to-gas ratio allows the FUV photons to propagate even further, thereby reducing the abundance of H$_2$ even more. A similar behavior is also reflected in the C-cycle in the aforementioned models, where an increase in C$^+$ and a decrease in CO are observed due to the photodissociation of the latter. 

\begin{figure*}
    \centering
    \includegraphics[width=0.99\textwidth]{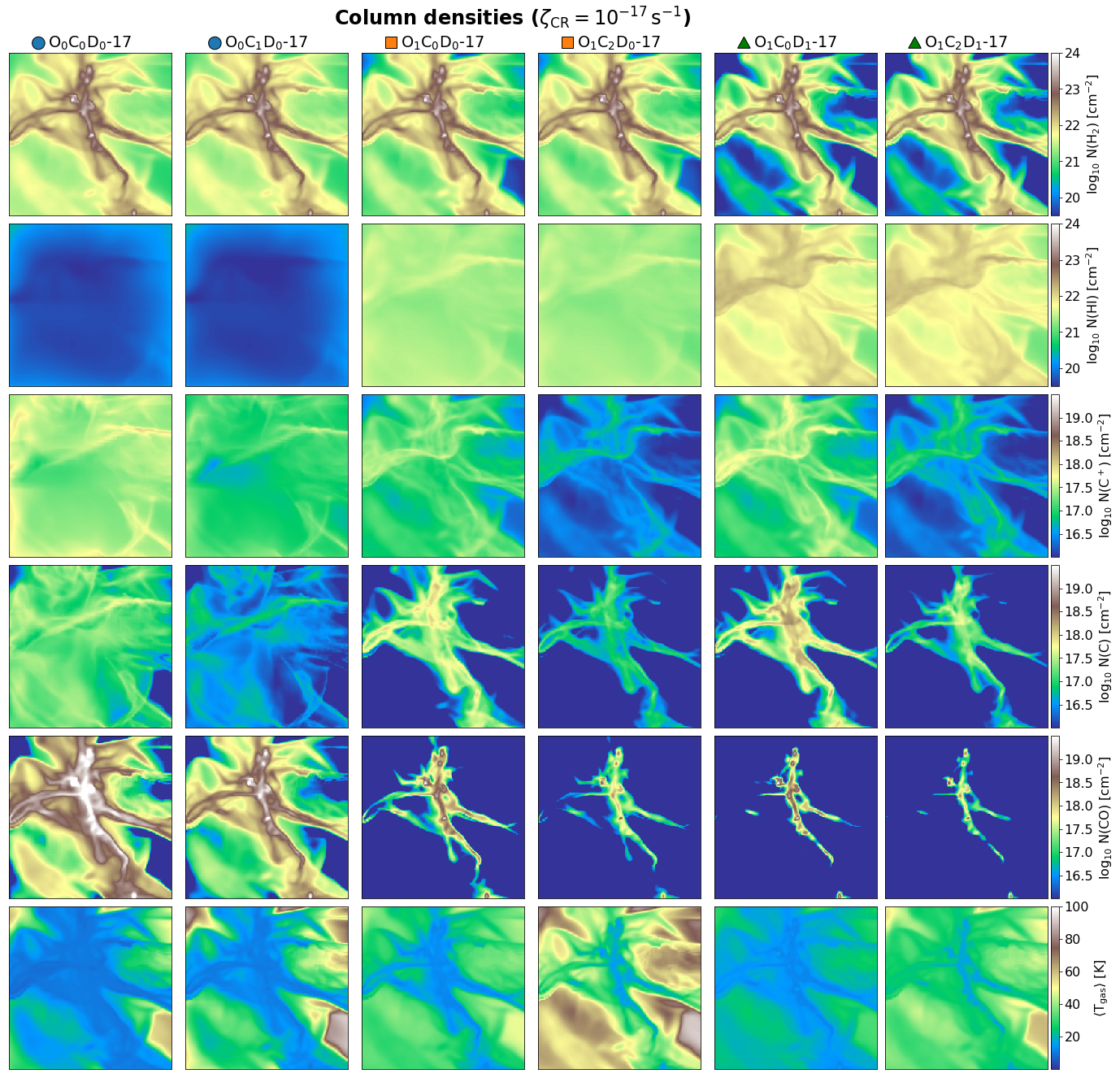}
    \caption{Column densities of species for $\zeta_{\rm CR}=10^{-17}\,{\rm s}^{-1}$. The first pair of columns correspond to simulations O$_0$C$_{0,1}$D$_0$-17, the second pair to O$_1$C$_{0,2}$D$_0$-17 and the third pair to O$_1$C$_{0,2}$D$_1$-17 in which ${\cal D}=3\times10^{-3}$. From top-to-bottom, the column densities of H$_2$, H{\sc i}, C$^+$, C, and CO are shown, respectively. In general, the column densities of all these species decrease with decreasing [C/O], except for the H{\sc i} and H$_2$ which depend more strongly on the [O/H] ratio. The bottom row shows the density-weighted gas temperature, which increases with decreasing [C/O].}
    \label{fig:cds1e17}
\end{figure*}

Figure~\ref{fig:remission} shows the relative emission lines from the C-cycle. These include the CO SLED for the first 10 transitions, both [C{\sc i}] and the [C{\sc ii}]~$158\,\mu$m fine-structure lines. For $\zeta_{\rm CR}=10^{-17}\,{\rm s}^{-1}$ (top panel), the high-$J$ CO lines become weaker for lower [C/O] ratios. For 
$J=6-5$ and above, the emission of CO is found to be approximately $2\times$ weaker. 
The [C{\sc i}]~(1-0) line becomes on average $2.5\times$ weaker as [C/O] decreases at a fixed [O/H]. The emission of both [C{\sc i}]~(2-1) and [C{\sc ii}] is also decreasing, albeit at a smaller factor. For $\zeta_{\rm CR}=10^{-15}\,{\rm s}^{-1}$ and low FUV intensity (bottom panel), it is found that the relative CO SLEDs of the models O$_0$C$_1$D$_0$-15 and O$_0$C$_0$D$_0$-15 are mostly affected, with the (low-to-mid)-$J$ lines to become twice as bright. This effect is largely connected with the efficient destruction of CO in the more diffuse gas and also with the increase of the local heating due to cosmic-rays \citep{Bisbas15,Bisbas17,Bisbas21}. This makes the emission of the line to be closely associated with a more dense and slightly warmer gas, hence its increase in the (low-to-mid)-$J$ transitions \citep[][]{Gaches22b}. All models with $\rm [C/O]<-0.07$ and $\chi/\chi_0=1$ appear to be dimmer for the CO $J=7-6$ transition and above. Both [C{\sc i}] emission lines are very suppressed ($\approx3-10\times$) and [C{\sc ii}]~$158\mu$m becomes also weaker by a factor of $2-3$. 

\begin{figure*}
    \centering
    \includegraphics[width=0.99\textwidth]{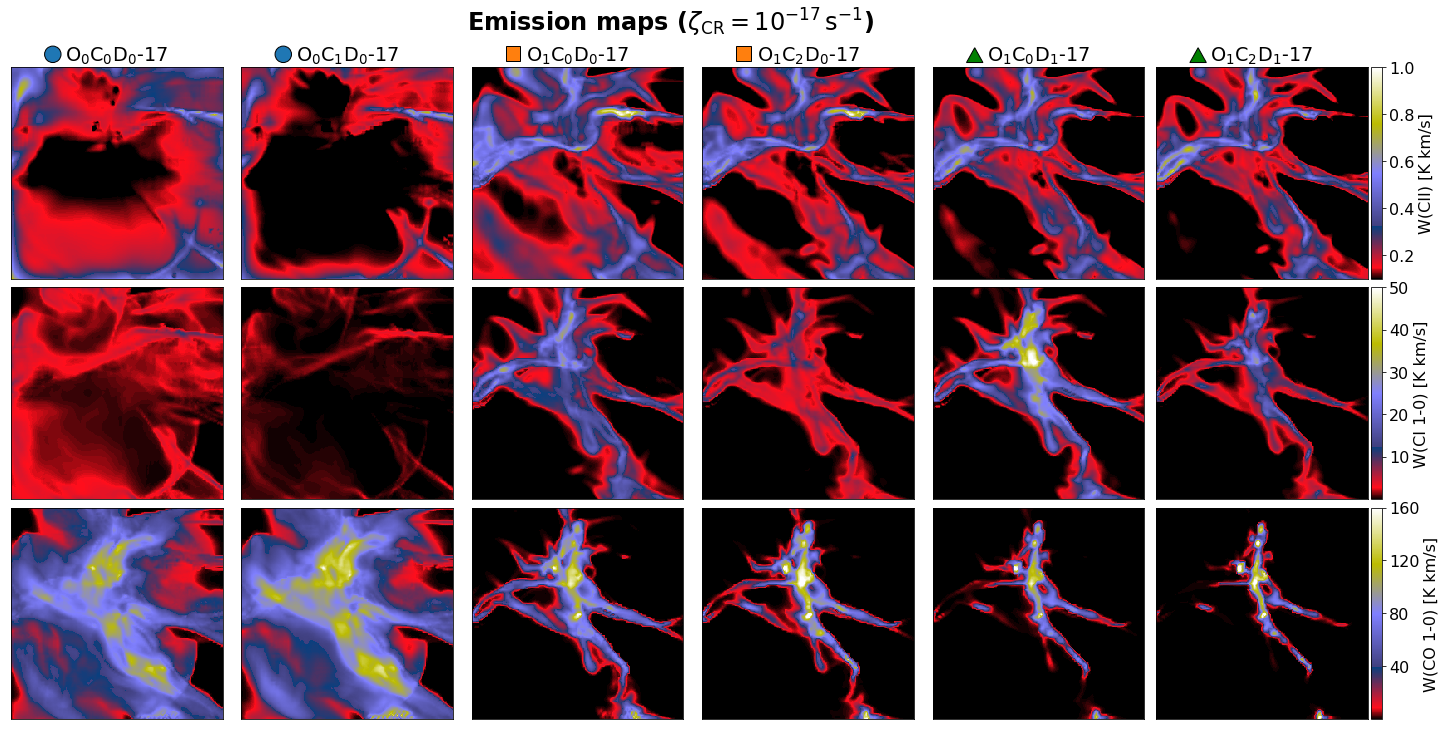}
    \caption{Emission maps for $\zeta_{\rm CR}=10^{-17}\,{\rm s}^{-1}$. The first pair of columns correspond to simulations O$_0$C$_{0,1}$D$_0$-17, the second pair to O$_1$C$_{0,2}$D$_0$-17 and the third pair to O$_1$C$_{0,2}$D$_1$-17. From top-to-bottom, the velocity integrated emissions of [C{\sc ii}]~$158\mu$m, [C{\sc i}](1-0) and CO $J=1-0$ are shown, respectively. As [C/O] decreases with respect to the reference value, the emission of [C{\sc i}](1-0) decreases substantially while CO $J=1-0$ remains bright. Note, however, that as [O/H] decreases, [C{\sc i}](1-0) increases. The effect is stronger for sub-linear $\cal D$ due to the additional CO photodissociation.}
    \label{fig:emissionz1e17}
\end{figure*}

In the O$_0$C$_1$D$_0$-15-2 model, the higher FUV radiation field does not affect much the CO SLED. However, the CO SLED is significantly affected in the models of lower $\rm [O/H]=-1.21$ values, especially for all transitions with $J>2-1$ (models O$_1$C$_2$D$_0$-15-2 and O$_1$C$_2$D$_1$-15-2). This is because the abundances of CO and H$_2$ molecules decrease in those models resulting in a weaker excitation of CO for mid/high-$J$ transitions. The emission of both [C{\sc i}] lines is slightly enhanced for the higher FUV simulations, except for the O$_1$C$_2$D$_1$-15-2 one, since the much lower dust-to-gas ratio contributes (through the propagation of radiation) to the ionization of atomic carbon. Contrary, the emission of [C{\sc ii}]~$158\mu$m line is enhanced by a factor of $\approx3$ at all cases of higher FUV intensity.

In Appendix~\ref{app:ratios}, the behaviour of the most commonly used line ratios between CO and [C{\sc i}] is discussed. 

\subsection{Low cosmic-ray ionization rate}
\label{ssec:lowCR}

Figure~\ref{fig:cds1e17} shows (from top to bottom) the column densities of 
H$_2$, H{\sc i}, C$^+$, C and CO of the most representative models. The bottom row shows the density-weighted gas temperature. The first pair of columns correspond to models O$_0$C$_{0,1}$D$_0$-17, the second pair to models O$_1$C$_{0,2}$D$_0$-17 (all O$_1$C$_1$ models are omitted in these plots) and the third pair to models O$_1$C$_{0,2}$D$_1$-17, respectively. 

The gas within the dense, star-forming filamentary structure remains always molecular as can be seen from the first row of Fig.~\ref{fig:cds1e17}. The H$_2$ column density of that structure does not change, remaining almost completely unaffected in all cases. However, the N(H$_2$) of the gas in the more diffuse medium surrounding the filament is reduced particularly in the O$_1$C$_0$D$_1$-17 and O$_1$C$_2$D$_1$-17 models due to the lower dust-to-gas ratio, $\cal D$, as the latter favours the H$_2$ photodissociation process. On the other hand, the H{\sc i} column density remains low in the O$_0$C$_{0,1}$D$_0$-17 models, increases in the O$_1$C$_{0,2}$D$_0$-17 models, and eventually increases further in the O$_1$C$_{0,2}$D$_1$-17 models. Although the column densities of H$_2$ and H{\sc i} change with [O/H] for fixed [C/O] (e.g. moving ``across" the left panel of Fig.~\ref{fig:panel}) and also with $\cal D$, they do not change as a function of [C/O] at fixed [O/H] (e.g. moving ``top-to-bottom" of that latter panel).

As expected, the column densities of the C-cycle species (C$^+$, C and CO) are reduced with decreasing the [C/O] ratio at fixed [O/H] (e.g. O$_0$C$_0$D$_0$-17 with O$_0$C$_1$D$_0$-17 etc.). This is also evident from the average abundances shown in Table~\ref{tab:all} and also in the top panel of Fig.~\ref{fig:rabund}. Interestingly, all these three species are reduced by the same amount and by observing their column density maps, it can be seen that no significant changes in the structure of those column density maps can be found. 

However, considering fixed $\rm [C/O]=-0.07$ and a decreasing [O/H] ratio (models O$_0$C$_0$D$_0$-17, O$_1$C$_0$D$_0$-17 and O$_0$C$_0$D$_1$-17), and focusing on the maps of N(C$^+$), N(C) and N(H$_2$), it can be seen that the column densities of C$^+$ and C become more closely related with the filamentary structure \citep[see also][for a relevant discussion]{Bisbas19}. This connection is more evident in the O$_1$C$_0$D$_1$-17 model in which ${\rm N(C\textsc{i})}\gtrsim10^{18.5}\,{\rm cm}^{-2}$ (the highest C column density in all $\zeta_{\rm CR}=10^{-17}\,{\rm s}^{-1}$ models) implying that in such conditions, atomic carbon may provide more accurate measurements of the molecular mass content in the ISM. On the other hand, the photodissociation of CO at lower [O/H] (due to the further penetration of FUV photons at higher column densities) reduces even more the abundance and, therefore, the column density of this molecule. It is, thus, found that CO is abundant only at the very dense parts of the filamentary structure where star-formation is likely to occur. 

\begin{figure*}
    \centering
    \includegraphics[width=0.99\textwidth]{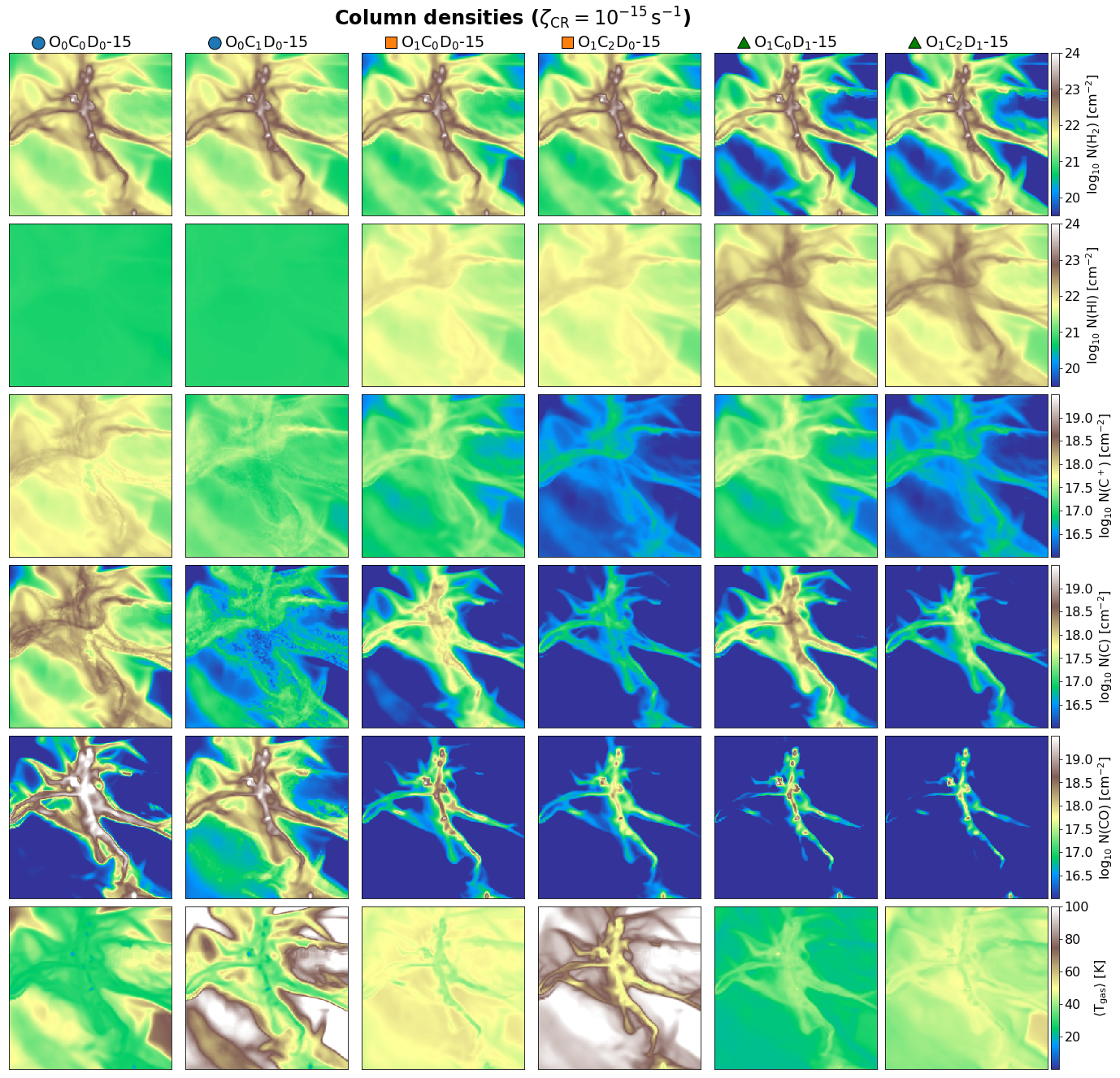}
    \caption{As in Fig.~\ref{fig:cds1e17} but for $\zeta_{\rm CR}=10^{-15}\,{\rm s}^{-1}$. The particular abundance of C decreases severely in model O$_0$C$_1$D$_0$-15 and substantially in models O$_1$C$_2$D$_0$-15 and O$_1$C$_2$D$_1$-15, even though a higher cosmic-ray ionization rate is present. It is therefore found that knowledge of the [C/O] ratio is highly important for a better understanding of the C-cycle emission under extreme environmental conditions. As expected, the column densities of CO are significantly reduced in comparison to the $\zeta_{\rm CR}=10^{-17}\,{\rm s}^{-1}$ ionization rate \citep{Bisbas15,Bisbas17}.}
    \label{fig:cds1e15}
\end{figure*}

Finally, the density-weighted gas temperature seen in the bottom panel of Fig.~\ref{fig:cds1e17} shows an interesting behavior. The average gas temperature increases with decreasing [C/O] at fixed [O/H] (see also column 8 in Table~\ref{tab:all}) since carbon is a major PDR coolant and therefore its suppression in abundance leads to a reduction in its cooling efficacy. However, the densest part of the filamentary structure remains always within $\langle T_{\rm gas}\rangle\sim10-20\,{\rm K}$. Perhaps the most striking feature is the gas temperature comparison between the pair of models O$_1$C$_0$D$_{0,1}$-17, and the pair of O$_1$C$_2$D$_{0,1}$-17. Although $T_{\rm gas}$ in both these models is higher than the corresponding one of the O$_0$C$_1$D$_0$-17 model, O$_1$C$_2$D$_0$-17 appears to have on average the warmest ISM with a $\langle T_{\rm gas}\rangle\simeq28.3\,{\rm K}$. The lower ${\cal D}$ ratio, although allows the FUV photons for reaching higher column densities, it increases $x_{\rm C^+}$ and $x_{\rm C}$ as described above, which in turn contribute positively to the total cooling. A lower $\cal D$ also reduces the efficiency of photoelectric heating in places. Notably, the spread of gas temperatures in both O$_1$C$_0$D$_1$-17 and O$_1$C$_2$D$_1$-17 models is smaller than in the other simulations, making overall a cloud of approximately uniform $T_{\rm gas}$. It is noted that our models show that the photoelectric heating and the cosmic-ray heating (the two most dominant heating mechanisms) are not affected by the decrease of [C/O] ratio but rather by the lower $\cal D$ ratio.

Figure~\ref{fig:emissionz1e17} shows (from top to bottom) the velocity integrated emission maps of [C{\sc ii}]~$158\mu$m, [C{\sc i}]~(1-0) and CO $J=1-0$ for the above models. As with the behaviour of the C-cycle column densities described earlier, the [C{\sc ii}]~$158\mu$m and [C{\sc i}]~(1-0) lines are better connected with the N(H$_2$) map for lower [O/H] values (O$_1$ models), therefore originating from denser gas than the one in the O$_0$C$_{0,1}$D$_0$-17 models. In the particular O$_1$C$_0$D$_{0,1}$-17 models, the additionally higher gas temperature results in the brightness increase of the [C{\sc ii}]~$158\mu$m and [C{\sc i}]~(1-0) emission lines.

\begin{figure*}
    \centering
    \includegraphics[width=0.99\textwidth]{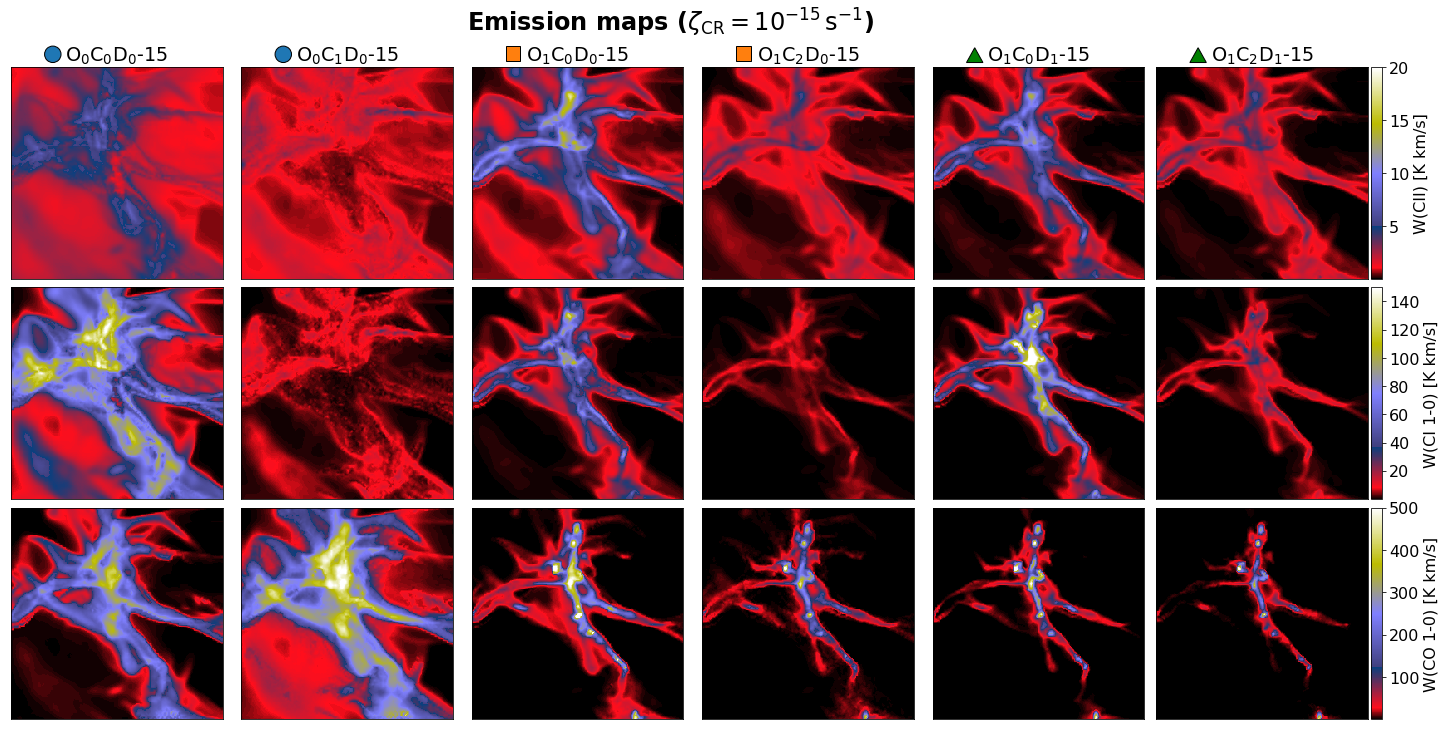}
    \caption{As in Fig.~\ref{fig:emissionz1e17} but for $\zeta_{\rm CR}=10^{-15}\,{\rm s}^{-1}$. The behavior of the C-cycle emission remains similar albeit all lines are much brighter due to the higher cosmic-ray heating.}
    \label{fig:emissionz1e15}
\end{figure*}

Focusing now on the changes observed in the line emission due to the decrease of [C/O] at fixed [O/H], it is found that [C{\sc ii}]~$158\mu$m although it is originated from different parts of the cloud, is not strongly influenced by the carbon abundance decrease and its brightness remains relatively unchanged as the abundance decrease compensates with the gas temperature increase (see top panel of Fig.~\ref{fig:remission}). However, [C{\sc i}]~(1-0) decreases with the decreasing [C/O] by an average factor of $\sim2-3$ as can be seen from Table~\ref{tab:all}. Finally, CO $J=1-0$ shows a different picture; although the abundance of CO molecule is lower, the emission of this line becomes stronger. This is due to the increase of gas temperature by a few K everywhere in the cloud ($\lesssim1.5\times$ on average) and particularly in the dense gas where this line originates from.

\subsection{High cosmic-ray ionization rate}
\label{ssec:highCR}

Figure~\ref{fig:cds1e15} shows the column densities of species as described in Fig.~\ref{fig:cds1e17} but for the case of higher cosmic-ray ionization rate of $\zeta_{\rm CR}=10^{-15}\,{\rm s}^{-1}$. The majority of the gas remains molecular in the O$_0$C$_{0,1}$D$_0$-15 and O$_1$C$_{0,1,2}$D$_0$-15 models (top row of Fig.~\ref{fig:cds1e15}), since the H{\sc i}-to-H$_2$ transition is not heavily affected by the cosmic-ray ionization rate \citep[][]{Bialy15,Bisbas15}. However, in the models of lower $\cal D$ (O$_1$C$_{0,1,2}$D$_1$-15) the gas remains molecular only in regions of higher densities where star-formation is likely to occur. As can be seen in Table~\ref{tab:all} for these models, $\langle \chi_{\rm HI}\rangle > \langle \chi_{\rm H2}\rangle$ thus the cloud is mainly atomic. 

The column densities of H{\sc i} (second row of Fig.~\ref{fig:cds1e15}) are enhanced by a factor between $\sim2\times$ (in models O$_1$C$_{0,2}$D$_1$-15) and  $\sim10\times$ (in models O$_0$C$_{0,1}$D$_0$-15). An interesting feature occurs, however, for both O$_0$C$_{0,1}$D$_0$-15. In these models, N(H{\sc i}) appears to be approximately constant ($\langle \rm N(H\textsc{i})\rangle\simeq8.1\times10^{20}\,{\rm cm}^{-2}$ with a standard deviation of $\sigma_{\rm N(H\textsc{i})}\simeq 9.6\times10^{19}\,{\rm cm}^{-2}$). Notably, this is in good agreement with H{\sc i} measurements in the Perseus cloud \citep[][]{Burkhart15} and the IC~348 cloud \citep{Luo23}. Interestingly, the latter work estimates a $\zeta_{\rm CR}\sim5\times10^{-16}\,{\rm s}^{-1}$ which is slightly lower than the one used.

The response of the C-cycle in an $\zeta_{\rm CR}$-enhanced environment has been extensively studied in previous works \citep[][]{Bisbas15,Bisbas17,Bisbas21}. Similarly to those studies, it is found here that CO is converted to C$^+$ and mainly to C through the interaction of He$^+$ resulting from cosmic-rays. However, the increase of $\chi_{\rm C}$ is not as effective for lower [C/O] values. In fact, we find that in the O$_0$C$_1$D$_0$-15 model, C$^+$ transitions to CO almost directly without having a C-dominated layer. This in turn lowers N(C), as evident from the corresponding panel in Fig.~\ref{fig:cds1e15} (fourth row, second column). In particular, it is estimated that in the O$_0$C$_0$D$_0$-15 model, $\sim57.7\%$ of the total carbon is in CO, $\sim23.1\%$ in C and $\sim19.1\%$ in C$^+$, whereas in the O$_0$C$_1$D$_0$-15 model these percentages are $\sim80.5\%$, $\sim5.2\%$ and $\sim14.3\%$, respectively. Similarly, although the cloud in O$_1$C$_0$D$_0$-15 is C-dominated, it becomes CO-dominated in O$_1$C$_2$D$_0$-15. The additionally lower dust-to-gas ratio in the O$_1$C$_0$D$_1$-15 and O$_1$C$_2$D$_1$-15 models favour both C$^+$ and C carbon phases. Indeed, the cloud remains poor in CO, which is found only in the very high density areas ($n_{\rm H}\gtrsim3\times10^4\,{\rm cm}^{-3}$).

By performing further analysis, it is found that for $\zeta_{\rm CR}=10^{-17}\,{\rm s}^{-1}$, C peaks for a narrow range of densities. The width of this range does not change for lower [C/O] values, so the column density of C decreases approximately linearly with the decreasing C abundance. However, for $\zeta_{\rm CR}=10^{-15}\,{\rm s}^{-1}$ and for $\rm [C/O]=-0.07$, a wider range of densities are C-rich \citep{Bisbas15}. For lower [C/O], the width of this range becomes significantly narrower. Chemical analysis show that this occurs because lower [C/O] values at high $\zeta_{\rm CR}$, influence the formation pathway of C. In particular, it is found that C is mainly formed via the reactions:
\begin{align}
    \rm H + CH \rightarrow &\ \rm C + H_2 \tag{R1}, \label{R1}\\
    \rm CH_3^+ + e^- \rightarrow &\ \rm C + H_2 + H \tag{R2}, \label{R2}
\end{align}
instead from the C$^+$ recombination with free electron which has a formation rate of $\gtrsim90\%$ for $\rm [C/O]\sim0$. For $\rm [C/O]<0$, reactions \ref{R1} and \ref{R2} can dominate the C formation pathway by $\sim50-70\%$, leading to a C-poor gas, since the abundances of CH and CH$_3^+$ are low.

The bottom row of Fig.~\ref{fig:cds1e15} shows the density-weighted gas temperature. The pattern discussed for the corresponding panels in Fig.~\ref{fig:cds1e17} is also repeated here; the average gas temperature increases with decreasing [O/H] and [C/O] (see also $\langle T_{\rm gas}\rangle$ in Fig.~\ref{fig:rabund} where the relative ratios are essentially the same for both $\zeta_{\rm CR}$). However, due to the enhanced $\zeta_{\rm CR}$ value, the higher cosmic-ray heating is consequently increasing the average temperatures $\gtrsim2-3\times$, affecting, in turn, all corresponding line emissions described below.

Figure~\ref{fig:emissionz1e15} shows the velocity integrated emission maps of [C{\sc ii}]~$158\mu$m, [C{\sc i}]~(1-0) and CO $J=1-0$. In general, all these emission lines are much stronger in the O$_0$C$_0$D$_0$-15 model when compared to the O$_0$C$_0$D$_0$-17 one, as was also discussed in \citet{Bisbas21, Bisbas23}. The most striking feature found in this figure, is that both [C{\sc ii}]~$158\mu$m and [C{\sc i}]~(1-0) become significantly dimmer when [C/O] decreases for fixed [O/H]. 
For instance, recent NOEMA observations of [C{\sc ii}] in the GN-z11 galaxy \citep{Fudamoto23} find a weak emission of this line, in agreement with our models for negative values of [C/O].
In regards to the [C{\sc i}]~(1-0) line and even though the cosmic-ray ionization rate is high, it is weak and remains much weaker than CO $J=1-0$. The latter emission line on the other hand becomes approximately twice brighter in the O$_0$C$_1$D$_0$-15 model than in O$_0$C$_0$D$_0$-15 and remains approximately unaffected by the lower [C/O] in the low-metallicity models (O$_1$) and even for sub-linear values of $\cal D$. This in turn favours the CO $J=1-0$ line as a molecular gas tracer in such an environment. The fact that under these conditions the molecular gas can remain bright in CO $J=1-0$ but dark in [C{\sc i}](1-0) may have important consequences in observations of high-redshift star-forming galaxies which can potentially have lower [C/O] than expected (see \S\ref{sec:discussion}).

\subsection{High cosmic-ray ionization rate and FUV intensity}
\label{ssec:highCRFUV}

\begin{figure}
    \centering
    \includegraphics[width=0.49\textwidth]{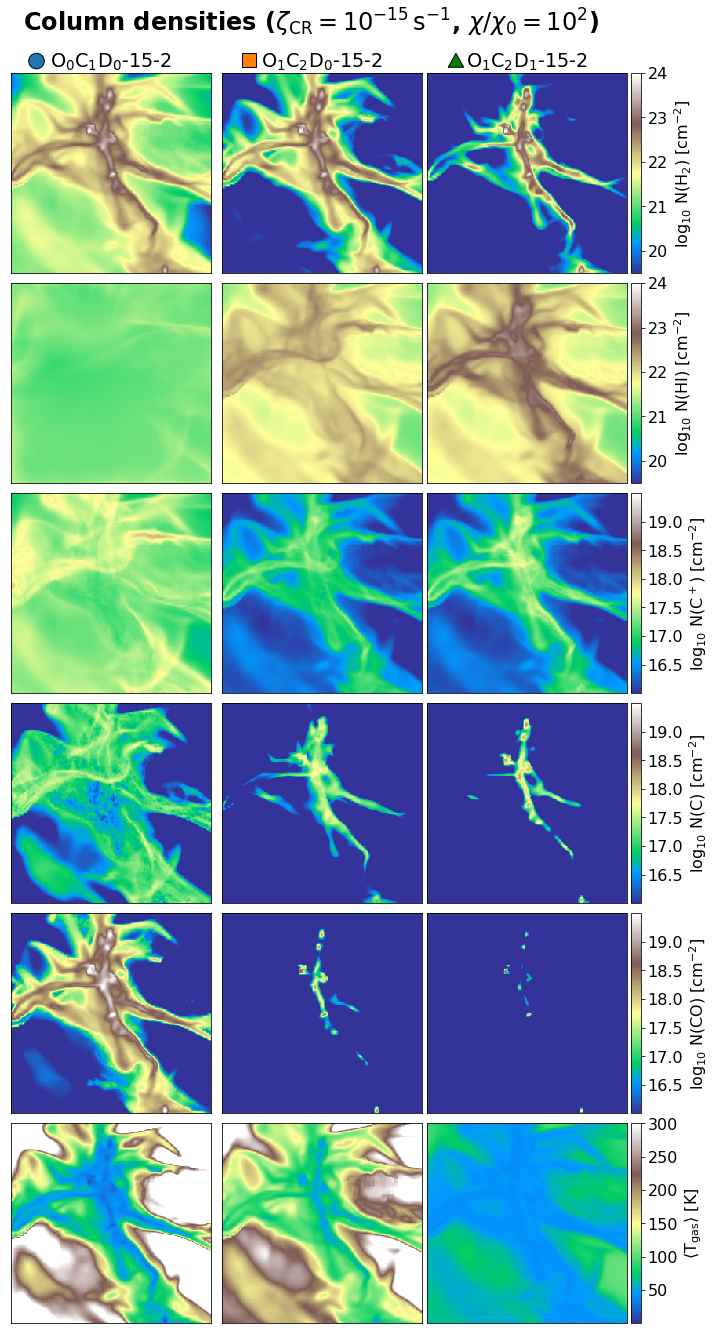}
    \caption{As in Fig.~\ref{fig:cds1e15} but with a higher FUV intensity of $\chi/\chi_0=10^2$. In the O$_1$C$_2$D$_0$-15 and O$_1$C$_2$D$_1$-15-2 models, N(CO) is severely suppressed even in the densest part of the molecular filament.}
    \label{fig:columndz1e15x1e2}
\end{figure}

We will now explore the case where both $\zeta_{\rm CR}$ and FUV intensity are enhanced. Figure~\ref{fig:columndz1e15x1e2} shows how the column density maps presented in \S\ref{ssec:highCR} change if an FUV field with intensity $\chi/\chi_0=10^2$ is applied (for $\zeta_{\rm CR}=10^{-15}\,{\rm s}^{-1}$). The models calculated consider the lower [C/O] values for the explored [O/H]. The combination of high-$\zeta_{\rm CR}$ and high-FUV represents better the conditions of star-forming metal-rich and metal-poor galaxies. In general, it is expected that a galaxy of higher star-formation rate, will also contain on average higher FUV radiation fields and higher cosmic-ray ionization rates. While the exact relationship between these quantities is not yet observationally determined, they are frequently assumed to scale in a linear fashion \citep[e.g.][]{Papadopoulos10a}.

It is found that lower [O/H] values lead to an increase in atomic fraction. More specifically, when the dust-to-gas ratio $\cal D$ is lower, the cloud becomes H{\sc i}-dominated.
In the particular case of O$_1$C$_2$D$_1$-15-2, 
the higher $\chi/\chi_0$ and the higher $\zeta_{\rm CR}$, dissociates the H$_2$ molecule almost everywhere in the cloud, except for the dense filamentary structure. As discussed in \S\ref{ssec:highCR}, the column density of H{\sc i} in the O$_0$C$_1$D$_0$-15-2 model is approximately uniform with $\langle \rm N(H\textsc{i}) \rangle \simeq 1.4\times10^{21}\,{\rm cm}^{-2}$ and with a standard deviation of $\sigma_{\rm N(H\textsc{i})}\simeq3.5\times10^{20}\,{\rm cm}^{-2}$ (see also O$_0$C$_0$D$_0$-15 and O$_0$C$_1$D$_0$-15 in Fig.~\ref{fig:cds1e15}). Considering all simulations presented in this work, the H{\sc i} column density obtains its highest average value of $\langle \rm N(H\textsc{i})\rangle\simeq2.1\times10^{22}\,{\rm cm}^{-2}$ in the O$_1$C$_2$D$_1$-15-2 model.

In regards to the C-cycle, it is found that the cloud is CO-rich in the O$_0$C$_1$D$_0$-15-2 model but becomes C$^+$-rich in the O$_1$C$_2$D$_0$-15-2 and O$_1$C$_2$D$_1$-15-2 models. However, the column density maps of all carbon phases agree well with the structure of the N(H$_2$) maps. It is interesting to note that in both O$_1$C$_2$D$_0$-15-2 and O$_1$C$_2$D$_1$-15-2 simulations, the molecular gas contains a small amount of CO, thus making the H$_2$-rich cloud to be severely CO-dark. 

\begin{figure}
    \centering
    \includegraphics[width=0.49\textwidth]{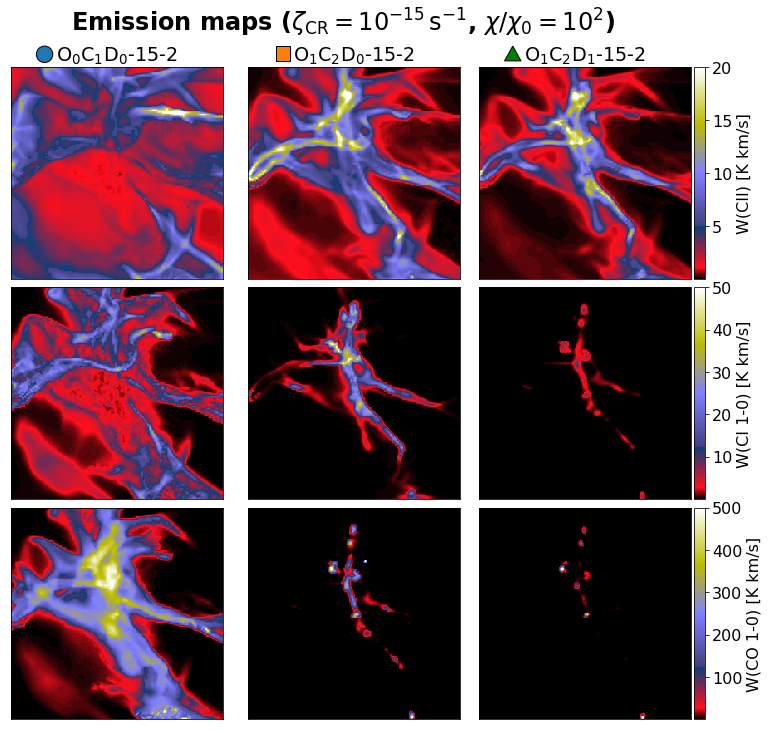}
    \caption{As in Fig.~\ref{fig:emissionz1e15} but with a higher FUV intensity of $\chi/\chi_0=10^2$. In the O$_0$C$_1$D$_0$-15-2 model, CO~(1-0) is very bright and well associated with N(H$_2$) (see Fig.~\ref{fig:columndz1e15x1e2}) whereas in the O$_1$C$_2$D$_0$-15-2 and O$_1$C$_2$D$_1$-15-2, the filament is bright in [C{\sc ii}]~$158\mu$m. In the O$_1$C$_2$D$_1$-15-2 simulation, the molecular cloud is dark in both [C{\sc i}](1-0) and CO(1-0). Note the change in the extent of the colour-bar for the [C{\sc i}](1-0) line between this figure and Fig.~\ref{fig:emissionz1e15}.}
    \label{fig:emissionz1e15x1e2}
\end{figure}

The gas temperatures are shown in the bottom row. The variations in $T_{\rm gas}$ in the different parts of the cloud and between the aforementioned models appear to be somehow counter-intuitive. In particular, the diffuse part of the cloud surrounding the dense filamentary structure obtains the highest $T_{\rm gas}$ value for the O$_0$C$_1$D$_0$-15-2 simulation. Although in O$_1$C$_2$D$_0$-15-2 and O$_1$C$_2$D$_1$-15-2 models both metallicity and $\cal D$ are lower, the gas temperature in the diffuse part is not higher as expected. This is better seen in the O$_1$C$_2$D$_1$-15-2 simulation. This effect is due to the increase of C$^+$ abundance, and consequently its line emission, which cools down the gas \citep[see also][]{Bisbas17}. On the other hand, the dense gas along the filament is warmer in the lower [O/H] models as a result of photoelectric heating due to the lower $\cal D$ compared to the O$_0$C$_1$D$_0$-15-2 model. A double behaviour is, therefore, seen as [O/H] decreases for negative [C/O]; a decrease in $T_{\rm gas}$ for densities corresponding to diffuse gas and an increase in $T_{\rm gas}$ for the higher density gas. 

\begin{figure*}
    \centering
    \includegraphics[width=0.87\textwidth]{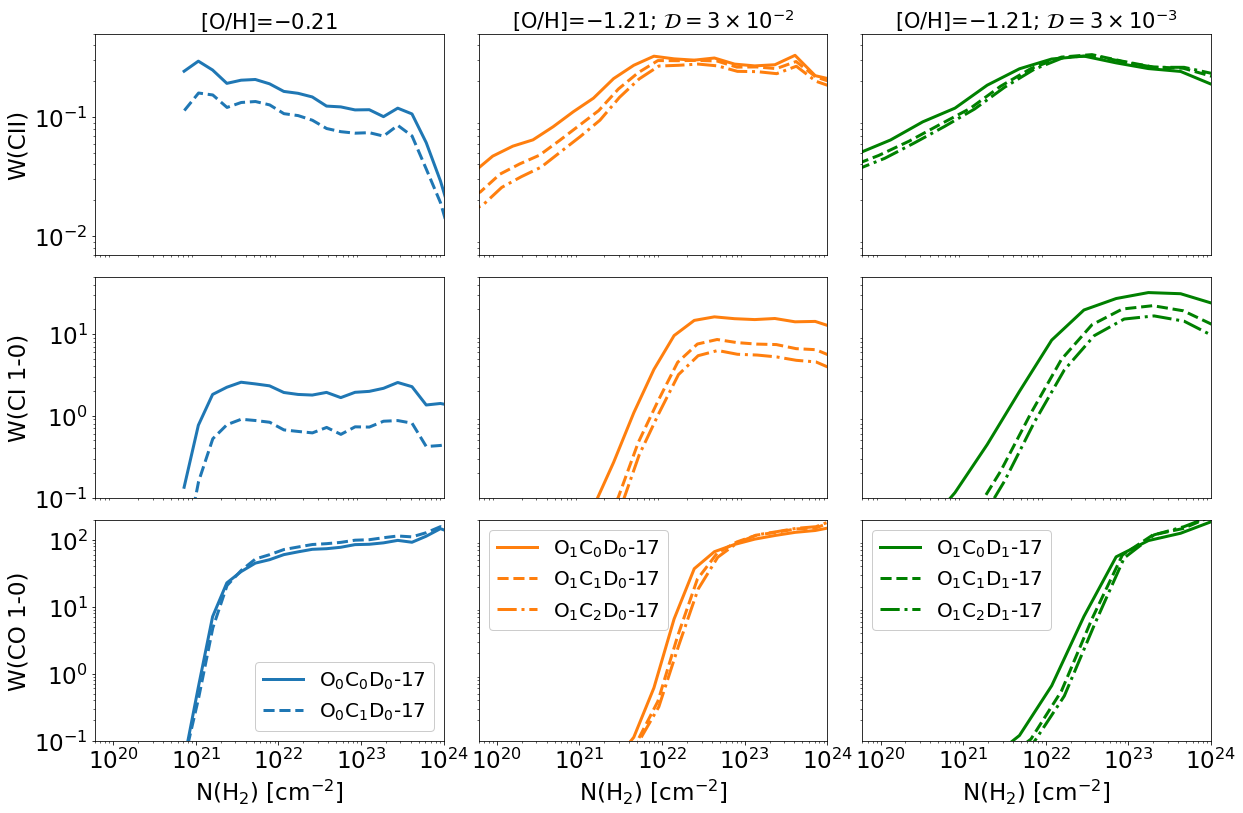}
    \caption{Mean values of the velocity integrated emission (in units of $\rm K\,km\,s^{-1}$) versus N(H$_2$) for $\zeta_{\rm CR}=10^{-17}\,{\rm s}^{-1}$. From top-to-bottom the emission of [C{\sc ii}], [C{\sc i}](1-0) and CO(1-0) are shown for all simulations considered, in units of ${\rm K}\,{\rm km}\,{\rm s}^{-1}$. Solid lines show the $\rm [C/O]=-0.07$ case whereas dashed and dot-dashed the $\rm [C/O]=-0.57$ and $-0.77$, respectively. The first column is for $\rm [O/H]=-0.21$, the second is for $\rm [O/H]=-1.21$ and ${\cal D}=3\times10^{-2}$ and the third for $\rm [O/H]=-1.21$ and ${\cal D}=3\times10^{-3}$.}
    \label{fig:I17}
\end{figure*}

\begin{figure*}
    \centering
    \includegraphics[width=0.87\textwidth]{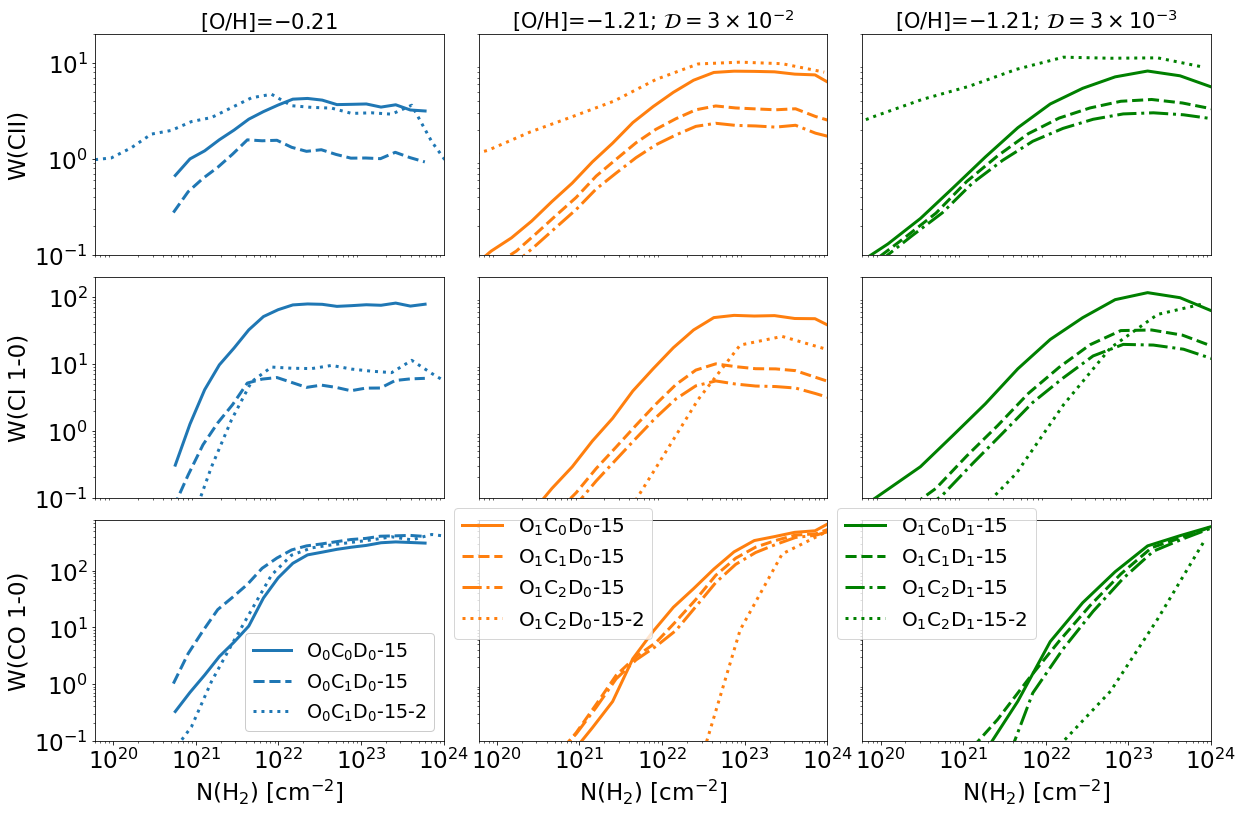}
    \caption{As in Fig.~\ref{fig:I17} for $\zeta_{\rm CR}=10^{-15}\,{\rm s}^{-1}$. The dotted lines are for the models with $\chi/\chi_0=10^2$. The W(CO 1-0) versus N(H$_2$) relationship does not depend on the [C/O] ratio.}
    \label{fig:I15}
\end{figure*}

Figure~\ref{fig:emissionz1e15x1e2} shows the corresponding velocity integrated emission maps. The emission lines follow closely the pattern of the corresponding column density maps of the C-cycle. CO $J=1-0$ becomes very bright and extended in the O$_0$C$_1$D$_0$-15-2 simulation and is much brighter than [C{\sc i}]~(1-0) and [C{\sc ii}]~$158\mu$m. The most interesting feature, however, can be seen in the O$_1$C$_2$D$_0$-15-2 and particularly in the O$_1$C$_2$D$_1$-15-2 models. In both these simulations, CO $J=1-0$ originates only from the very dense clumps of the cloud while it is very dark otherwise. This makes the H$_2$-rich part of the cloud to be essentially CO-dark. Although [C{\sc i}]~(1-0) shows a good correlation with the N(H$_2$) map in O$_1$C$_2$D$_0$-15-2, it also becomes very weak in the O$_1$C$_2$D$_1$-15-2 simulation making the molecular structure essentially [C{\sc i}]-dark. 

In both these models, [C{\sc ii}]~$158\mu$m appears to give the most reasonable and detailed correlation with the molecular cloud structure, thus favouring the observation of this line in low-metallicity galaxies with high star-formation rates \citep[][]{Cormier19, Madden20} and $\rm [C/O]<0$. Remarkably, its brightness remains approximately constant even for sub-linear $\cal D$ (O$_1$C$_2$D$_{0,1}$-15-2 simulations). For the above simulations of high $\zeta_{\rm CR}$ and high $\chi/\chi_0$, it is also found that [O{\sc i}]~$63\mu$m increases with decreasing [C/O] and decreasing $\cal D$, contributing significantly to the total cooling.

\subsection{Relation of line intensities with H$_2$ column density}
\label{ssec:NH2}

Figure~\ref{fig:I17} shows the mean values of the velocity integrated emission of [C{\sc ii}]~$158\mu$m (top row), [C{\sc i}]~(1-0) (middle row) and CO $J=1-0$ (bottom row) versus the H$_2$ column density for $\zeta_{\rm CR}=10^{-17}\,{\rm s}^{-1}$. [O/H] and $\cal D$ decrease from left to right. The different line styles correspond to different [C/O] values. 
With the exception of [C{\sc ii}]~$158\mu$m for $\rm [O/H]=-0.21$, it is found that the emission of all lines increases with N(H$_2$) and eventually saturates. The different behaviour of the [C{\sc ii}] emission in the top left panel is due to its different origin from the ISM, other than the molecular phase, and thus no correlation with N(H$_2$) can be found. Looking at all panels, perhaps the most significant feature is the decrease of the W(C{\sc i} 1-0) emission with [C/O] for fixed [O/H] as has been also discussed earlier. W(C{\sc ii}) remains weak for all [O/H] and W(C{\sc i} 1-0), although increases its brightness for $\rm [O/H]=-1.21$ and for an H$_2$ column density of a few $\times10^{22}\,{\rm cm}^{-2}$, it becomes weaker with the [C/O] decrease.

Interestingly, the correlation of W(CO 1-0) with N(H$_2$) remains nearly unchanged with [C/O]. However, some dependency on [O/H] can be seen. Even more remarkably, comparing the results between the second and third column (bottom row) of Fig.~\ref{fig:I17}, the W(CO 1-0) versus N(H$_2$) relation does not depend on the normalized dust-to-gas ratio, $\cal D$.  

Figure~\ref{fig:I15} shows the aforementioned correlation of emission lines with the H$_2$ column densities for $\zeta_{\rm CR}=10^{-15}\,{\rm s}^{-1}$. The models with $\chi/\chi_0=10^2$ are also plotted with dotted lines. With the exception of these higher-FUV simulations, the general behavior of [C{\sc ii}]~$158\mu$m, [C{\sc i}]~(1-0) and CO $J=1-0$ is similar to the one described above for $\zeta_{\rm CR}=10^{-17}\,{\rm s}^{-1}$. The models with $\chi/\chi_0=10^2$ show two important differences in the W(C{\sc ii}) and W(CO 1-0) emission when compared to the fiducial models of lower FUV intensity. Since the higher radiation photodissociates CO and increases the abundance of C$^+$, an increase in W(C{\sc ii}) at lower N(H$_2$) is observed. At the same time, W(CO 1-0) builds for much higher N(H$_2$) in the O$_1$C$_2$D$_0$-15-2 and O$_1$C$_2$D$_1$-15-2 models. A similar behavior, albeit at a lower scale, can be also observed for W(C{\sc i} 1-0). 

Like above with $\zeta_{\rm CR}=10^{-17}\,{\rm s}^{-1}$, it appears that CO $J=1-0$ does not depend on the [C/O] value for any [O/H] and $\cal D$ considered. This means that the W(CO 1-0) versus N(H$_2$) relation is not a function of [C/O] neither for the low $\zeta_{\rm CR}$ nor for the high $\zeta_{\rm CR}$ rate considered. We find that this is because the optical depth of CO (1-0) decreases with decreasing [C/O] resulting in a weak change in the W(CO 1-0)--N(H$_2$) relation as a function of [C/O]. However, it seems that it is affected mostly when a higher radiation field is applied. It is interesting to note that for $\rm [O/H]=-1.21$ and regardless of the dust-to-gas ratio, W(CO 1-0) builds in an approximately linear fashion. However, as the corresponding panels in Fig.~\ref{fig:I15} show, it originates from the very dense and clumpy medium leading to a CO-dark molecular gas (see also Figs.~\ref{fig:emissionz1e15} and \ref{fig:emissionz1e15x1e2}). 

Given that the W(CO 1-0) -- N(H$_2$) correlation can be considered independent on the [C/O] ratio for fixed [O/H], it is expected that the corresponding CO-to-H$_2$ conversion factor will behave in a similar fashion. This finding also favors the alternative methodology of \citet{Seifried20} in calculating the H$_2$ gas mass content for solar metallicity as their formulas (see their Eqns.~12--14) may hold also for $\rm [C/O]<0$.

\subsection{Distribution of C$^+$-, C-, and CO-rich gas masses across various galactic environments}
\label{ssec:ccycleenv}

How do the gas masses of C$^+$, C, and CO distribute in each chemical phase of the cloud? To answer this question, a definition of `pure atomic', `PDR' and `molecular' regions needs to be introduced, based on the relative abundances of H{\sc i} and H$_2$. It is defined, here, as `pure atomic' (or simply `atomic') the region that satisfies the relation $x_{\rm HI}\ge0.9$, as `PDR' the region that satisfies the relation $x_{\rm HI}<0.9$ and $x_{\rm H2}<0.49$, and as `molecular' the region with $x_{\rm H2}\ge0.49$. The choice of the latter criterion is taken after consideration of the H{\sc i} narrow-line self-absorption observations of Galactic infrared dark clouds \citep{Zuo2018}, as the early stage of high-column density, H$_2$-dominated clouds. However, a more relaxed criterion satisfying the relation of $x_{\rm H2}\ge0.45$ is also explored (see Appendix~\ref{app:origin}).

\begin{figure}
    \centering
    \includegraphics[width=0.5\textwidth]{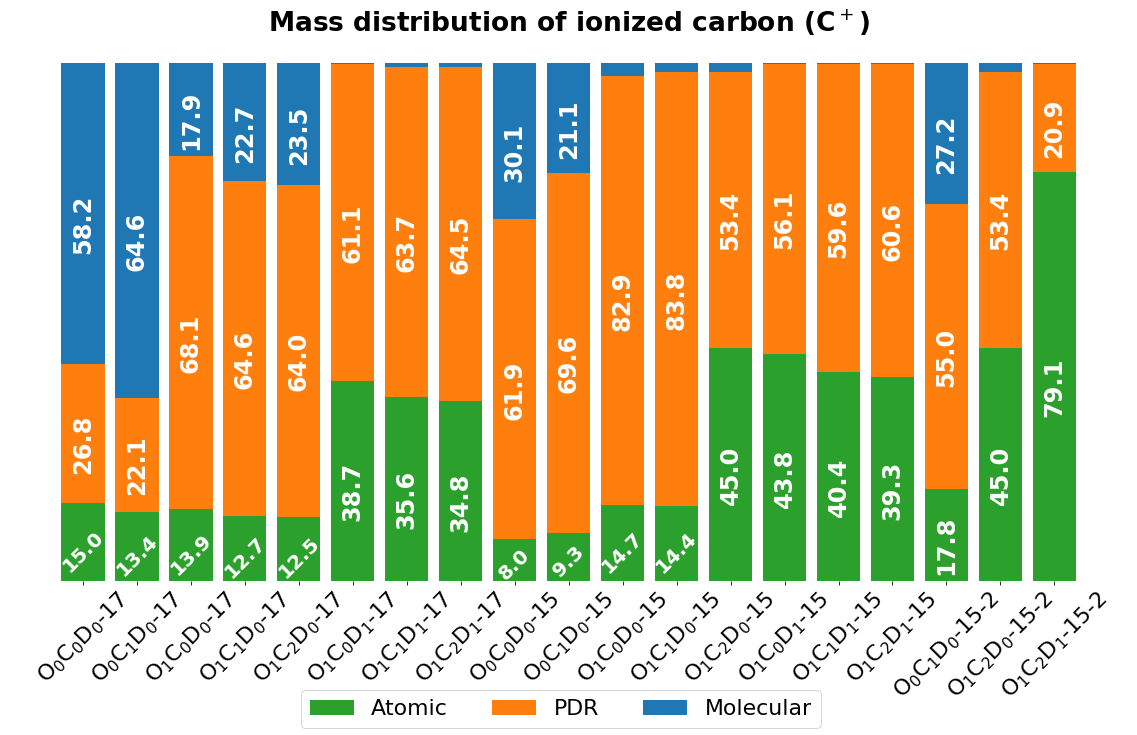}
    \includegraphics[width=0.5\textwidth]{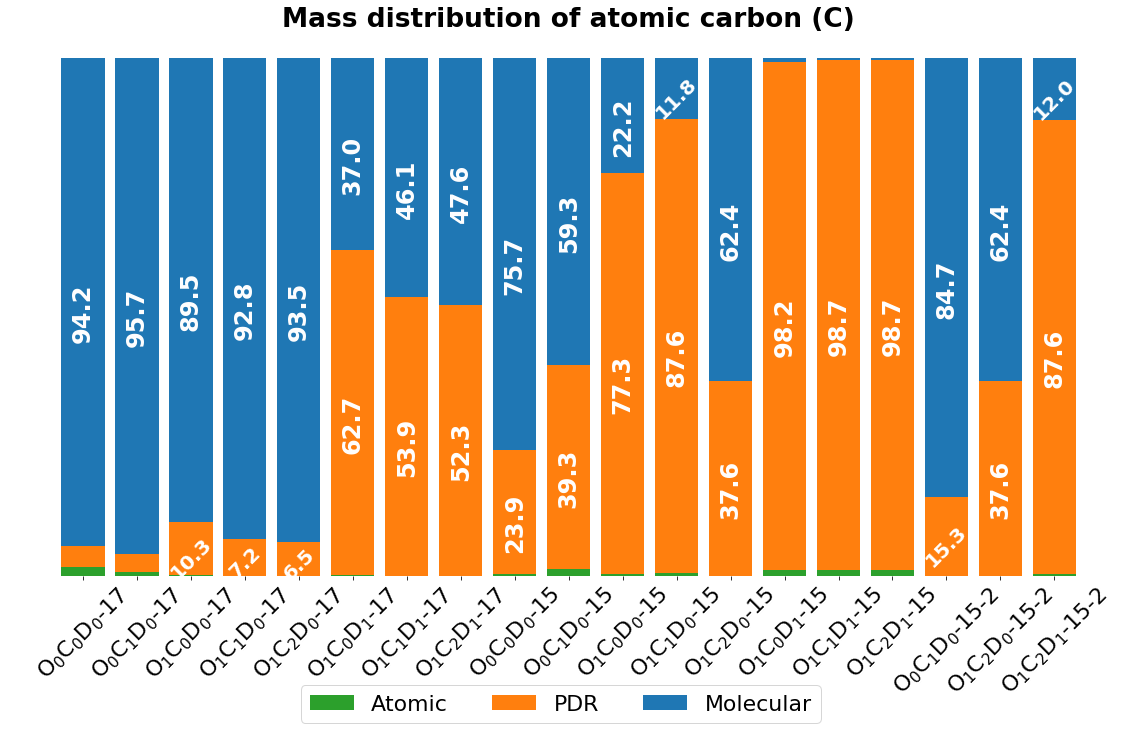}
    \includegraphics[width=0.5\textwidth]{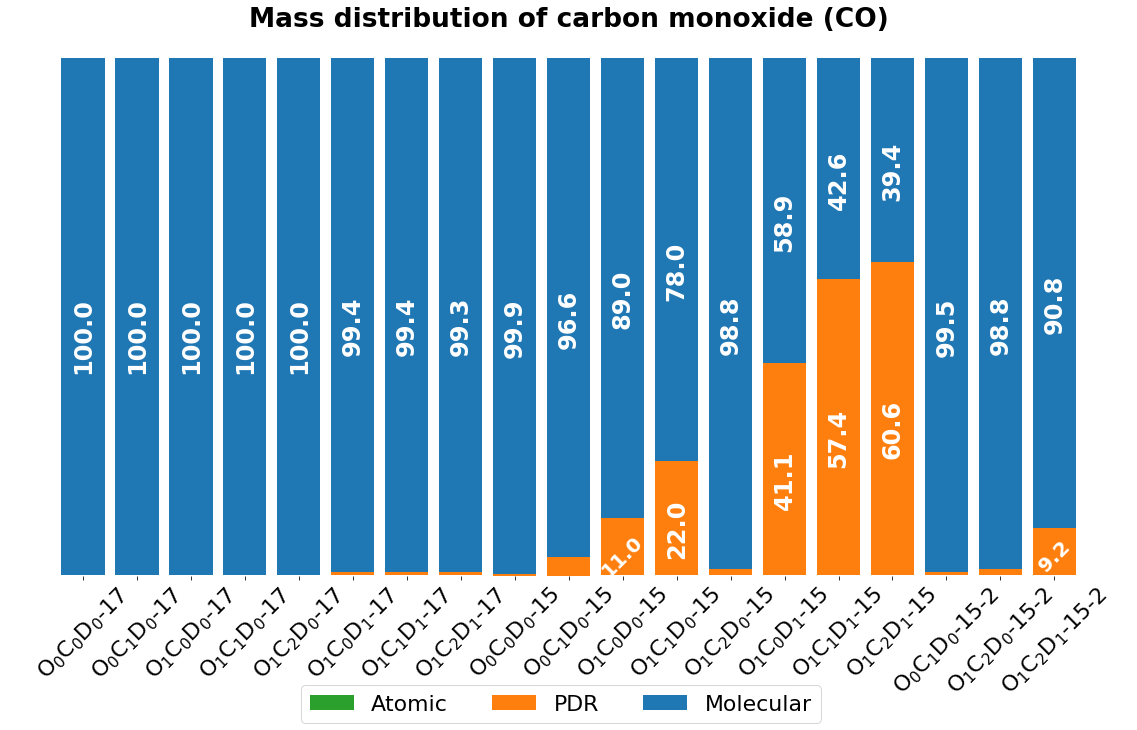}
    \caption{Distribution of C$^+$ mass (top), C mass (middle), and CO mass (bottom) in each model. Green colour refers to the atomic gas ($\chi_{\rm HI}\ge0.9$), orange colour to PDRs ($\chi_{\rm HI}<0.9$ and $\chi_{\rm H2}<0.49$), and blue colour to molecular gas ($\chi_{\rm H2}\ge0.49$). The numbers shown represent the percentage of the carbon mass associated with each different phase. Fractions less than $5\%$ are not labeled. See \S\ref{ssec:ccycleenv} for discussion.}
    \label{fig:origin2pcent}
\end{figure}

\subsubsection{Distribution of C$^+$}

The upper panel of Fig.~\ref{fig:origin2pcent} shows the percentage fraction of $M_{\rm C^+}$. Blue colour corresponds to the molecular region, orange to the PDR and green to the pure atomic. The general picture is that only two models (O$_0$C$_{0,1}$D$_0$-17) appear to contain a significant fraction of C$^+$ mass in the molecular phase of approximately $60\%$. For low metallicities at low $\zeta_{\rm CR}$ (O$_1$C$_{0,1,2}$D$_0$-17), it is found that approximately $20\%$ of the total C$^+$ mass is associated with the molecular gas phase. A similar-to-higher percentage ($20-30\%$) is also found when the higher $\zeta_{\rm CR}=10^{-15}\,{\rm s}^{-1}$ is applied in the two models of solar metallicities (so for O$_0$C$_{0,1}$D$_0$-15) and depending on the [C/O] ratio. Such a fraction is found also when the FUV intensity is increased (O$_0$C$_1$D$_0$-15-2). 

In all other cases, a trace amount of C$^+$ mass is located in the molecular gas. Instead, in such models, $M_{\rm C^+}$ is located in PDRs at a fraction of $\sim50-60\%$ with the remaining to be associated with the atomic phase. It is noted, here, that our simulations do not model the ionized phase; should that have occurred, the fraction corresponding to the `pure atomic' would also include $M_{\rm C^+}$ associated with that phase. For instance, the SHINING collaboration \citep{Herrera18} find that $\gtrsim60-90\%$ of [C{\sc ii}] may arise from neutral gas including gas of low-ionization.

\subsubsection{Distribution of C}

The middle panel of Fig.~\ref{fig:origin2pcent} shows the percentage fraction of $M_{\rm C}$ using the above criterion. It is found that $M_{\rm C}$ is associated either with the molecular phase or the PDR depending on the environmental condition. The atomic phase contains trace amount ($\lesssim1\%$) of C-rich gas. More specifically, for $\zeta_{\rm CR}=10^{-17}\,{\rm s}^{-1}$ and a linear decrease of $\cal D$ (models O$_0$C$_{0,1}$D$_0$-17, and O$_1$C$_{0,1,2}$D$_0$-17), a fraction of $M_{\rm C}$ exceeding $90\%$ of the total C is associated with the molecular phase. The corresponding fraction for the PDR phase is $\lesssim10\%$. Once a sub-linear $\cal D$ is adopted (models O$_1$C$_{0,1,2}$D$_1$-17), the photodissociation of CO at higher optical depths decreases the fraction of $M_{\rm C}$ associated with the molecular phase, giving an approximate fifty-fifty percent between this phase and the PDR phase. 

For the high cosmic-ray ionization rate of $\zeta_{\rm CR}=10^{-15}\,{\rm s}^{-1}$, it is found that for metallicities close to solar (model O$_0$C$_0$D$_0$-15) the majority of $M_{\rm C}$ originates from the molecular phase with a fraction of $\gtrsim75\%$. For $\rm [C/O]<0$ at this metallicity, this fraction can be reduced down to approximately $60\%$ (model 1A\_15). For the low-metallicity models (O$_1$C$_{0,1,2}$D$_0$-15), including the linear and sub-linear decrease of the dust-to-gas ratio $\cal D$, C is abundant on the surface of PDRs whereas the molecular gas is very C-poor. It can be, therefore, argued that in such systems the [C{\sc i}] emission may be associated with the PDR surface rather than directly with the H$_2$ rich gas.

Higher FUV intensities appear to increase the abundance of C inside the molecular gas. This is evident from the O$_0$C$_1$D$_0$-15-2, O$_1$C$_2$D$_0$-15-2 and O$_1$C$_2$D$_1$-15-2 models when compared to the corresponding ones of lower FUV intensity. In particular, it is found that the two orders of magnitude higher FUV intensity considered, results in an increase of $M_{\rm C}$ that is associated with the molecular phase by $\sim25\%$ for solar metallicities and by $\gtrsim10-50\%$ for low-metallicities, the latter depending on the adopted value of $\cal D$.

\subsubsection{Distribution of CO}

The bottom panel of Fig.~\ref{fig:origin2pcent} shows the corresponding mass distribution of CO. Overall, CO is associated with molecular gas to a very high percentage, as expected from PDR theory. However, for the models of low [O/H], low $\cal D$ and high $\zeta_{\rm CR}$, it is found that part of the total CO gas mass may be considerably connected with PDRs to $\gtrsim40\%$. This connection increases from $\sim40\%$ to $\sim60\%$ with decreasing [C/O] (models O$_1$C$_{0,1,2}$D$_1$-15). By relaxing the criterion for defining PDRs and molecular regions as introduced in the beginning of this section (i.e. from $x_{\rm H2}\ge0.49$ to $x_{\rm H2}\ge0.45$, see Appendix~\ref{app:origin}), we find that the aforementioned percentages are reduced to $\lesssim15\%$. This abrupt decrease implies that CO-rich gas may exist at the very edge of PDRs and that the transition from C$^+$ and/or C to CO is extremely sharp. For such extreme cases, fraction of the CO emission may originate from such regions.

\subsection{Model limitations}

The present models are to investigate the {\it trends} in the emission lines and the abundances of the carbon cycle as well as the location of the atomic-to-molecular transition resulting from $\rm [C/O]<0$ and sub-linear dust-to-gas relations as a function of metallicity ($\rm [O/H]$). It is expected that the dynamical evolution will differ under the various ISM environmental parameters explored \citep[e.g.][]{Gong20,Hu21}, affecting the aforementioned abundances and line emissions. However, it is not expected that the presented trends will be altered if dynamical evolution is included \citep[see][]{Offner13,Bisbas23}, and thus an understanding of how negative [C/O] ratios impact the observables can be reached. Numerical codes incorporating (magneto-)hydrodynamics coupled with chemistry \citep[such as][]{Glover10,Walch15,Gong17,Smith20,Hu21} will be needed to provide a more accurate insight into the presented outcomes.

\section{Discussion}
\label{sec:discussion}

The observed chemical evolution of galaxies shows that carbon and oxygen have been originated from different stellar processes, resulting in negative [C/O] ratios. While negative ratios have been considered in studies related to star-formation and the IMF, their impact on the observables using the carbon cycle emission lines and the location of the H{\sc i}-to-H$_2$ transition has not been addressed in the past. Through three-dimensional PDR and radiative transfer modelling, it has been demonstrated how such [C/O] values change the emission of [C{\sc ii}]~$158\mu$m, [C{\sc i}]~(1-0) and (2-1), and the first ten CO transitions in a star-forming cloud which is embedded in different cosmic-ray ionization rates and FUV intensities.

Concerning the H{\sc i}-to-H$_2$ transition, for a given [O/H] metallicity negative [C/O] ratios do not shift its location and as a result, the molecular mass content does not change. Instead, this transition is more sensitive to the ISM environmental parameters concerning the value of [O/H], the cosmic-ray ionization rate and the FUV intensity as was previously studied \citep[e.g.][]{Bisbas21}. Sub-linear dust-to-gas ratios tend to increase the H{\sc i} abundance, therefore shifting the H{\sc i}-to-H$_2$ transition at higher H-nucleus column densities.

Since carbon is a major coolant in PDRs, in ISM environments where $\rm [C/O]<0$ the average (density-weighted) gas temperature is found to increase. The effect is better seen in the diffuse gas which can increase $T_{\rm gas}$ as much as two times. However, for ISM environments with sub-linear dust-to-gas ratios, FUV photons create more C$^+$ which is an efficient coolant. Although the photoelectric effect operates for a wider range of column densities, the additional C$^+$ decreases the overall gas temperature (e.g. models O$_1$C$_2$D$_{0,1}$-15-2). As discussed earlier, the photoelectric heating also decreases (especially in lower density regions) as a result of the lower $\cal D$ values. In turn, this has consequences in the line emission of all C-cycle coolants. 

Our discussion below is mainly focused on the impact of $\rm [C/O]<0$ on observations of the [C{\sc i}](1-0) emission. The connection of $\rm [C/O]<0$ with `[C{\sc ii}]-deficit' ISM is also examined.

\subsection{Subthermal excitation of [C{\sc i}](1-0)}

Numerical models studying the atomic carbon emission line has been the main focus of various groups in the last few years. \citet{Glover15} examined the [C{\sc i}] fine structure lines of a dynamically evolving giant molecular cloud for typical conditions met in the MW. In their work, they find that carbon is subthermally excited, implying that the observed [C{\sc i}] emission seen in real clouds should underestimate the derived column density of that species. This underestimation should arise from the fact that subthermal excitation reduces the brightness of the line. Observations of [C{\sc i}] lines in high-redshift strongly lensed galaxies by \citet{Harrington21} indicate that they are generally subthermally excited. Similar, results where concluded also by \citet{Papadopoulos22} who studied a sample of 106 galaxies. It should be, therefore, interesting to see how the [C{\sc i}](1-0) emission line behaves in the presented work and whether there is a relationship with the [C/O] ratio. 

To calculate the amount of subthermally excited [C{\sc i}](1-0) emission in our models, an approach similar to \citet{Glover15} is adopted. Through the level populations outputted by {\sc 3d-pdr}, the excitation temperature, $T_{\rm ex}$, is calculated locally in each cell using the expression
\begin{eqnarray}
    T_{\rm ex}=-\frac{h \nu_{10}}{k_{\rm B}}\left[\ln{\left(\frac{s_1}{3s_0}\right)}\right]^{-1},
\end{eqnarray}
where $\nu_{10}$ is the [C{\sc i}](1-0) frequency, $s_0$ and $s_1$ the level populations in the ground and first levels respectively, and the factor 3 results from the statistical weight ratio between these levels. The column of the excitation temperature ($T_{\rm ex,w}$) is then calculated along each line-of-sight, weighted by the local carbon number density. Similarly, the $T_{\rm C,w}$ gas temperature weighted by $n_{\rm C}$ is calculated alongside \citep[see][for more details]{Glover15}. 

\begin{figure}
    \centering
    \includegraphics[width=0.49\textwidth]{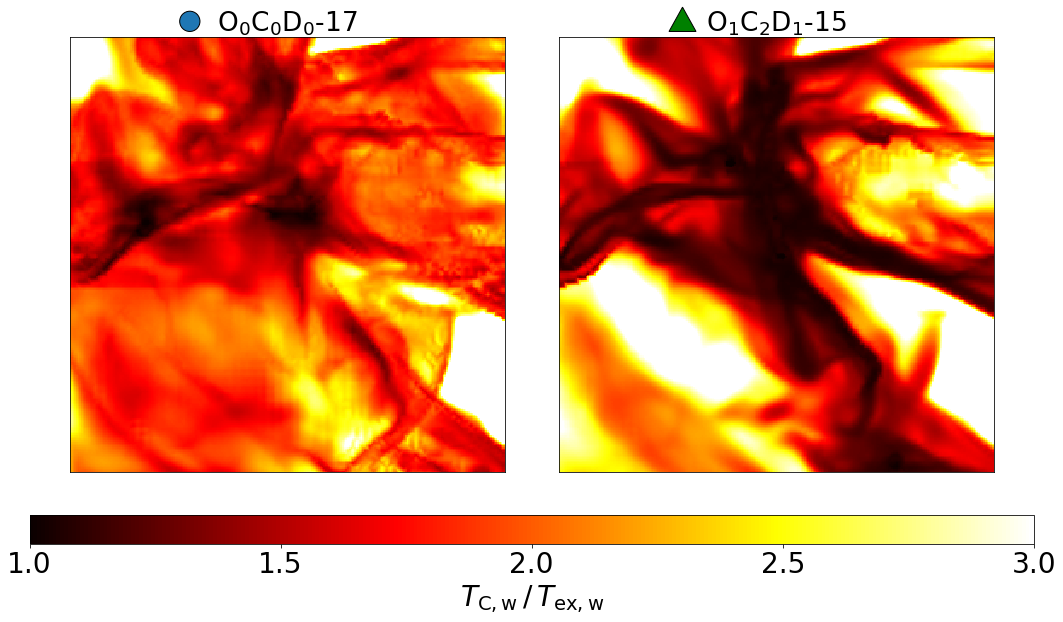}
    \includegraphics[width=0.49\textwidth]{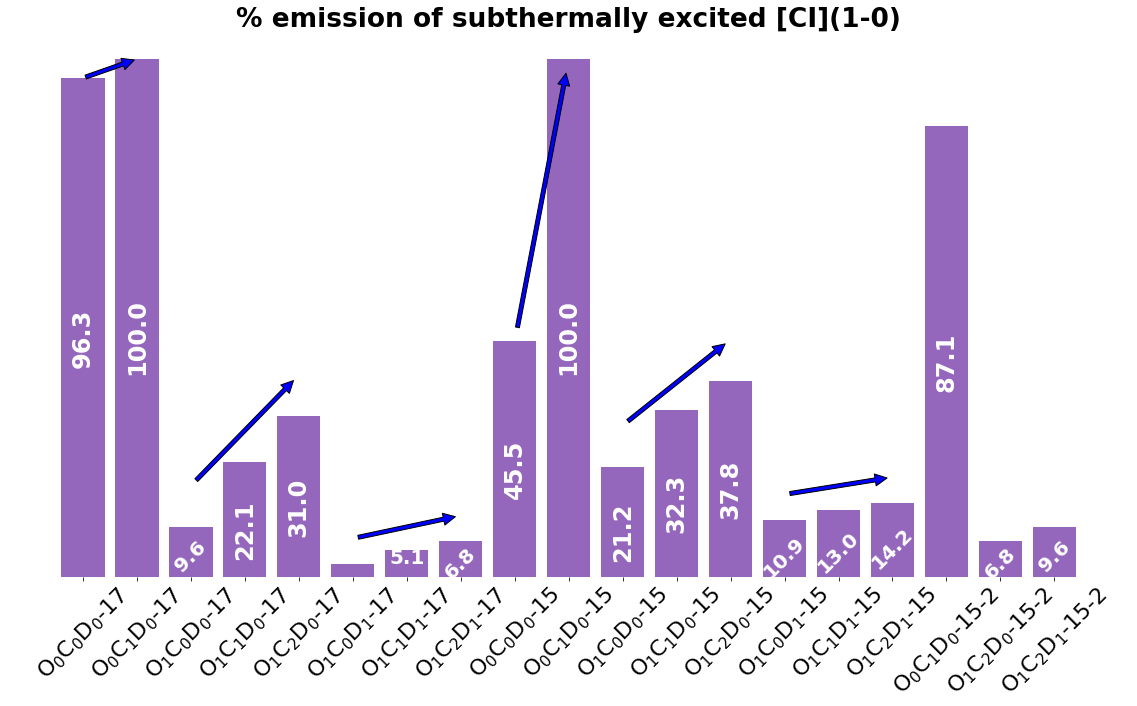}
    \caption{{\it Top:} Maps of the $T_{\rm C,w}/T_{\rm ex,w}$ ratio for the O$_0$C$_0$D$_0$-17 (left) and O$_1$C$_2$D$_1$-15 (right) models. For ratios $\sim\!1$, [C{\sc i}](1-0) is thermalized whereas for $>\!1$ it is sub-thermally excited. {\it Bottom:} Percentage emission of subthermally excited [C{\sc i}](1-0) line in each model. For fixed [O/H], the percentage is increasing with decreasing [C/O]. The arrows point these trends in each group of models.}
    \label{fig:subth}
\end{figure}

The map of the $T_{\rm C,w}/T_{\rm ex,w}$ ratio can be then constructed (see top panels of Fig.~\ref{fig:subth}). For values of $T_{\rm C,w}/T_{\rm ex,w}\simeq1$, the excitation temperature of [C{\sc i}](1-0) is approximately equal to the local gas temperature. In this situation, local thermodynamical equilibrium (LTE) is reached and the line becomes thermalized. This occurs in the parts of the cloud containing densities that are higher than the critical one of [C{\sc i}](1-0) \citep[approximately $10^3\,{\rm cm}^{-3}$ for cold gas, see][]{Papadopoulos04,Glover15}. On the other hand, for values $T_{\rm C,w}/T_{\rm ex,w}>1$, the column of the excitation temperature is lower than the one for the gas temperature implying that radiative de-excitation is more rapid than collisional de-excitation LTE is then not satisfied and the [C{\sc i}](1-0) line is subthermally excited. The top panels of Fig.~\ref{fig:subth} show the aforementioned behavior in two models; the O$_0$C$_0$D$_0$-17 (left panel) and O$_1$C$_2$D$_1$-15 (right panel). In the left one, $T_{\rm C,w}/T_{\rm ex,w}\gtrsim1.5$ in most of the cloud (except for two high-density regions), implying that the majority of [C{\sc i}](1-0) emission is subthermally excited (albeit it is weak for the ISM conditions of O$_0$C$_0$D$_0$-17 as discussed in the previous section). In the right one, corresponding to low-metallicity gas ($\rm [O/H]=-1.21$) with $\rm [C/O]=-0.77$, sub-linear dust-to-gas ratio and high $\zeta_{\rm CR}$ value, the majority of [C{\sc i}](1-0) emission is very thermalized. 

While there is no precise condition defining the transition from thermal-to-subthermal emission, we adopt here the ratio of $1.3$ to define it, meaning a $30\%$ difference between the gas and the excitation temperature. With this criterion, the percentage emission of subthermally excited [C{\sc i}](1-0) is calculated for all models and presented in the bottom panel of Fig.~\ref{fig:subth}. In general, it is found that subthermal emission of [C{\sc i}] occurs only for metal-rich clouds (models with $\rm [O/H]=-0.21$) regardless to the ISM conditions (and except for O$_0$C$_0$D$_0$-15). For low-metallicity conditions, [C{\sc i}](1-0) emission is generally thermalized. There is an interesting pattern seen, however. For fixed [O/H], the percentage of subthermal emission of the carbon line is increasing with decreasing [C/O], albeit not substantially for low-metallicity gas. This trend is marked with blue arrows, separated for each group of models. Considering that the atomic carbon line ratio does not strongly depend on the [C/O] ratio (see Fig.~\ref{fig:ratios}), it is expected that the [C{\sc i}](2-1) line emission will also follow the behavior of [C{\sc i}](1-0) in regards to whether it is subthermally excited or not. 

Therefore, it can be demonstrated that for solar metallicities the [C{\sc i}] emission is generally subthermally excited, in agreement with \citet{Papadopoulos22}. However, in low-metallicity star-forming clouds and galaxies, [C{\sc i}](1-0) emission is generally found to be thermalized. Its intensity is weak and it is a natural consequence of negative [C/O] ratios in the observed ISM rather than being associated with subthermal emission. In turn, this behavior is directly related with the galactic chemical evolution.

\subsection{Three cases of [C{\sc i}]-dark galaxies}

Observations of galaxies that are bright in CO but dark in [C{\sc i}] lines have been reported in the past \citep{Michiyama20,Michiyama21,Harrington21,Dunne22,Lelli23}. Explanations for the low $L_{\rm CI}/L_{\rm CO}$ luminosity ratios include the existence of extended high density regions that are not bright in [C{\sc i}] lines, CI-poor environments, missing fluxes and even specific processes followed during data analysis. However, the possibility of low C/O ratios has not been examined as a potential cause for [C{\sc i}]-dark galaxies. Our attention is drawn for three candidates that may contain $\rm [C/O]<0$ gas: the low-redshift NGC~6052 and NGC~7679 galaxies which are discussed in \citet{Michiyama20,Michiyama21}, and the high-redshift zC-400569 discussed in \citet{Lelli23}.

NGC~6052 (Mrk~297, Arp~209) is a low-redshift ($z=0.01581$) system of colliding galaxies with an $\rm SFR\simeq18\,{\rm M}_{\odot}\,{\rm yr}^{-1}$ \citep{Michiyama20}. The latter group reported nondetection of [C{\sc i}](1-0) and a bright emission of the CO $J=4-3$, resulting in a luminosity ratio of $L_{\rm CI\,1-0}/L_{\rm CO\,4-3}<0.08$. In their analysis, \citet{Michiyama20} used PDR models by assuming MW abundances and they found that such a ratio can be reproduced if the observed number densities are $n_{\rm H}>10^5\,{\rm cm}^{-3}$ which may be reasonably explained from the strong shocks produced due to the merging process. However, our models can reproduce the observed ratio if a low C/O ratio is considered. More specifically, NGC~6052 is observed to be metal rich, with $12+\log({\rm O/H})=8.22-8.85$ \citep{Sage93,James02,Shi05,Rupke08} corresponding to $\rm [O/H]=-0.47$ to $+0.16$, which matches -on average- with O$_0$ models. Models O$_0$C$_1$D$_0$-15 and O$_0$C$_1$D$_0$-15-2 contain a strong emission of CO $J=4-3$, and stronger than the O$_0$C$_0$D$_0$-15 of higher C/O by a factor of $\simeq2$, as indicated in the bottom panel of Fig.~\ref{fig:remission} (see red solid line) and a weak [C{\sc i}](1-0) emission (see red filled circle), which is $\simeq10\times$ fainter than O$_0$C$_0$D$_0$-15. Similarly, the abundance of C is very suppressed for O$_0$C$_1$D$_0$-15 (Fig.~\ref{fig:rabund} bottom panel). The ratio of these two lines for O$_0$C$_2$D$_0$-15 and O$_0$C$_1$D$_0$-15-2 (see Fig.~\ref{fig:ratios}) is close to the observed luminosity ratio\footnote{See \citet{Solomon05} for the equality between the luminosity ratio and the velocity integrated emission ratio.}. The remaining question is whether this galaxy has a [C/O] ratio that is much lower than that of the MW. An insight on that can be taken by \citet{Sage93} who report a high ratio of $^{12}$CO/$^{13}$CO$>\!22$ and \citet{Sage91} who report a high abundance of $^{18}$O in local starburst galaxies. Both these works claim that such a ratio may indicate a higher population of massive stars \citep[such as discussed in][]{Zhang18} than in the MW. If $^{12}$C is partly produced in intermediate and low mass stars (see \S\ref{ssec:Intro-chemicalevolution}), it may in turn produce low C/O ratios. Due to the ongoing collision, an increase in the star-formation rate leads to an increase in the cosmic-ray ionization rate and FUV intensities \citep[e.g.][]{Papadopoulos10a, Kashiyama14, Lahen20}, consistent with the presented models.

NGC~7679 (Mrk 534) is a low-redshift ($z=0.00177$) Seyfert~2 galaxy whose metallicity has not yet been reasonably estimated. Its star-formation rate is $\rm SFR\sim10-21\,{\rm M}\,{\rm yr}^{-1}$ \citep[][]{DeLooze14,Davies16}. Recently, \citet{Michiyama21} reported a low [C{\sc i}](1-0) / CO(4-3) luminosity ratio of $L_{\rm CI}/L_{\rm CO}<0.08$, similar to NGC~6052. They further reported young starbursts, a finding that supports the hypothesis that a larger population of massive stars may exist. NGC~7679 is undergoing interaction, possibly with NGC~7682 \citep{Michiyama21}. Through SNe type II, these conditions may create $\alpha$-enhanced environments resulting in low C/O ratios. Like with NGC~6052 discussed above, our models show that an environment reminiscent to that of O$_0$C$_1$D$_0$-15 and O$_0$C$_1$D$_0$-15-2 models may explain the very weak detection of [C{\sc i}]~(1-0) line.

In studying the kinematics of star-forming galaxies at cosmic-noon, \citet{Lelli23} presented ALMA observations of the zC-400569 galaxy, located at $z=2.24$, in mid-$J$ and [C{\sc i}](1-0) lines. This solar-metallicity galaxy has an elevated $\rm SFR=81\,{\rm M}_{\odot}\,{\rm yr}^{-1}$ and it is, thus, expected it will contain higher cosmic-ray energy densities and possibly also high average FUV radiation field strengths when compared to the MW. While these conditions favor the [C{\sc i}](1-0) emission \citep{Bisbas21}, \citet{Lelli23} reported a weak emission of this line which did not allow to perform kinematic studies. It remains to see whether zC-400569 has $\rm [C/O]<0$, as this would be in accordance to the results of the present work. Should this be true and considering also the relevance with NGC~6052 and NGC~7679, it can be demonstrated that galaxies undergoing starburst may become [C{\sc i}]-dark as a consequence of negative [C/O] ratios resulting from larger populations of massive stars \citep[see, however,][for potential -but localised- differential metal removal through galactic winds]{Sharda23b}.

\subsection{Contribution of $\rm [C/O]<0$ on the [C{\sc ii}]-deficit gas}
\label{ssec:ciideficit}

Observations examining the [C{\sc ii}]/FIR (far-IR) ratio have shown that it has a declining trend with increasing the FIR luminosity \citep{Malhotra01,Luhman03,Casey14,Zhao16}. This trend is known in the literature as ``[C{\sc ii}] deficit''. Many studies have focused on examining its origin \citep{Munoz16,Narayanan17,Smith17,Rybak19,Bisbas22} and have suggested various different physical and chemical processes. The role played by a $\rm [C/O]<0$ gas on the [C{\sc ii}] deficit has not been extensively studied in the past, however some first insight for its contribution was discussed in \citet{Harikane20}, who found that low carbon-to-oxygen ratios are not sufficient to explain the observed trends. It should be therefore interesting to explore whether this claim holds for our set of models.

Assuming optically thin emission for both [C{\sc ii}] and dust (FIR) and following \citet{Bisbas22} for their calculation, we find in our models that low C/O ratios can contribute $\lesssim 3$ times to the [C{\sc ii}] deficit (i.e. to decrease the [C{\sc ii}]/FIR ratio by up to approximately three times). This is a small factor compared to the $\simeq1-2$ orders of magnitude necessary to consider it as a major factor. It is thus demonstrated that low C/O ratios play a minor and hence insignificant role in contributing to the [C{\sc ii}] deficit, in agreement with the findings of \citet{Harikane20}. Although we do not model the ionized phase as mentioned also earlier, it is expected that the above result will also hold even when H{\sc ii} regions are present, since cosmological hydrodynamical simulations of a dwarf galaxy merger \citep{Bisbas22} found that that they may account on average $\sim10\%$ of the total C$^+$ luminosity.

\section{Conclusions}
\label{sec:conclusions}

This work explored the impact of negative [C/O] ratios, implying a sub-solar scaling between carbon and oxygen for low-metallicity gas, on the location of the atomic-to-molecular transition and on the abundances and line emission of the carbon cycle in a star-forming cloud. The {\sc 3d-pdr} astrochemical simulations revealed that consideration of the relative carbon-to-oxygen abundances in both metal-rich and especially metal-poor ISM conditions is of highly importance. \emph{The conclusion is that the [C/O] ratio cannot be neglected from astrochemical calculations concerning the diffuse and dense ISM.} For studies related to (extra-)galactic clouds, the [C/O] ratio constitutes an additional and equally important ISM environmental parameter on top of the three basic and most commonly used (the cosmic-ray ionization rate, the FUV intensity and the metallicity in the form of the [O/H] ratio) in astrochemical modeling. It should be noted that for low-metallicity clouds and galaxies, special attention must be paid also to the dust-to-gas mass ratio. 

The use of [C{\sc i}](1-0) emission as H$_2$ gas mass tracer in galaxies with $\rm [C/O]<0$ is difficult. For such a medium, it appears that the [C{\sc i}] lines may become very weak even when high cosmic-ray energy densities are present. Metal-rich star-forming galaxies of low [C/O] ratios --which are expected to contain enhanced $\zeta_{\rm CR}$ and FUV intensities-- are predicted to appear bright in [C{\sc ii}]~$158\mu$m and CO(1-0) but dark in [C{\sc i}](1-0). Similarly, low-metallicity dwarf galaxies are predicted to be in principle [C{\sc ii}]~$158\mu$m bright with very low emission in CO(1-0) arising from dense clumpy structures only. In these galaxies, [C{\sc i}](1-0) is also expected to be thermalized but very weak. Low C/O abundance ratios contribute insignificantly in making the ISM gas [C{\sc ii}] deficit. It is also expected that O-bearing molecules will be affected both in terms of their relative abundances and line emission. 

The impact of negative [C/O] ratios on the aforementioned abundances and line emission appears to be a complicated and non-linear problem. For a better understanding, especially on the large-scale ISM studies, full hydro-chemical models incorporating dynamical and chemical evolution simultaneously are needed. It is also recommended for observational researchers to report -whenever possible- the C/O abundance ratio, as this is of immense importance for performing robust PDR models.

\section*{Acknowledgements}

The authors thank the anonymous referee for their comments which improved the clarity of the work. The authors thank Piyush Sharda for the discussion and suggestions. Y.Z. is grateful for support from the NSFC (Grant No. 12173079).

\section*{Data Availability}

The data underlying this article will be shared on reasonable request to the corresponding author.



\bibliographystyle{mnras}
\bibliography{alpha_enhanced} 

\begin{thebibliography}{}
\makeatletter
\relax
\def\mn@urlcharsother{\let\do\@makeother \do\$\do\&\do\#\do\^\do\_\do\%\do\~}
\def\mn@doi{\begingroup\mn@urlcharsother \@ifnextchar [ {\mn@doi@}
  {\mn@doi@[]}}
\def\mn@doi@[#1]#2{\def\@tempa{#1}\ifx\@tempa\@empty \href
  {http://dx.doi.org/#2} {doi:#2}\else \href {http://dx.doi.org/#2} {#1}\fi
  \endgroup}
\def\mn@eprint#1#2{\mn@eprint@#1:#2::\@nil}
\def\mn@eprint@arXiv#1{\href {http://arxiv.org/abs/#1} {{\tt arXiv:#1}}}
\def\mn@eprint@dblp#1{\href {http://dblp.uni-trier.de/rec/bibtex/#1.xml}
  {dblp:#1}}
\def\mn@eprint@#1:#2:#3:#4\@nil{\def\@tempa {#1}\def\@tempb {#2}\def\@tempc
  {#3}\ifx \@tempc \@empty \let \@tempc \@tempb \let \@tempb \@tempa \fi \ifx
  \@tempb \@empty \def\@tempb {arXiv}\fi \@ifundefined
  {mn@eprint@\@tempb}{\@tempb:\@tempc}{\expandafter \expandafter \csname
  mn@eprint@\@tempb\endcsname \expandafter{\@tempc}}}

\bibitem[\protect\citeauthoryear{{Accurso}, {Saintonge}, {Bisbas}  \&
  {Viti}}{{Accurso} et~al.}{2017}]{Accurso17}
{Accurso} G.,  {Saintonge} A.,  {Bisbas} T.~G.,   {Viti} S.,  2017, \mn@doi
  [\mnras] {10.1093/mnras/stw2580}, \href
  {https://ui.adsabs.harvard.edu/abs/2017MNRAS.464.3315A} {464, 3315}

\bibitem[\protect\citeauthoryear{{Akerman}, {Carigi}, {Nissen}, {Pettini}  \&
  {Asplund}}{{Akerman} et~al.}{2004}]{Akerman04}
{Akerman} C.~J.,  {Carigi} L.,  {Nissen} P.~E.,  {Pettini} M.,   {Asplund} M.,
  2004, \mn@doi [\aap] {10.1051/0004-6361:20034188}, \href
  {https://ui.adsabs.harvard.edu/abs/2004A&A...414..931A} {414, 931}

\bibitem[\protect\citeauthoryear{{Aller} \& {Greenstein}}{{Aller} \&
  {Greenstein}}{1960}]{Aller60}
{Aller} L.~H.,  {Greenstein} J.~L.,  1960, \mn@doi [\apjs] {10.1086/190054},
  \href {https://ui.adsabs.harvard.edu/abs/1960ApJS....5..139A} {5, 139}

\bibitem[\protect\citeauthoryear{{Amarsi}, {Nissen}  \&
  {Sk{\'u}lad{\'o}ttir}}{{Amarsi} et~al.}{2019}]{Amarsi19c}
{Amarsi} A.~M.,  {Nissen} P.~E.,   {Sk{\'u}lad{\'o}ttir} {\'A}.,  2019, \mn@doi
  [\aap] {10.1051/0004-6361/201936265}, \href
  {https://ui.adsabs.harvard.edu/abs/2019A&A...630A.104A} {630, A104}

\bibitem[\protect\citeauthoryear{{Appleton} et~al.,}{{Appleton}
  et~al.}{2013}]{Appleton13}
{Appleton} P.~N.,  et~al., 2013, \mn@doi [\apj] {10.1088/0004-637X/777/1/66},
  \href {https://ui.adsabs.harvard.edu/abs/2013ApJ...777...66A} {777, 66}

\bibitem[\protect\citeauthoryear{{Arata}, {Yajima}, {Nagamine}, {Abe}  \&
  {Khochfar}}{{Arata} et~al.}{2020}]{Arata20}
{Arata} S.,  {Yajima} H.,  {Nagamine} K.,  {Abe} M.,   {Khochfar} S.,  2020,
  \mn@doi [\mnras] {10.1093/mnras/staa2809}, \href
  {https://ui.adsabs.harvard.edu/abs/2020MNRAS.498.5541A} {498, 5541}

\bibitem[\protect\citeauthoryear{{Arellano-C{\'o}rdova}
  et~al.,}{{Arellano-C{\'o}rdova} et~al.}{2022}]{Arellano23}
{Arellano-C{\'o}rdova} K.~Z.,  et~al., 2022, \mn@doi [\apjl]
  {10.3847/2041-8213/ac9ab2}, \href
  {https://ui.adsabs.harvard.edu/abs/2022ApJ...940L..23A} {940, L23}

\bibitem[\protect\citeauthoryear{{Asplund}, {Grevesse}, {Sauval}  \&
  {Scott}}{{Asplund} et~al.}{2009}]{Asplund2009}
{Asplund} M.,  {Grevesse} N.,  {Sauval} A.~J.,   {Scott} P.,  2009, \mn@doi
  [\araa] {10.1146/annurev.astro.46.060407.145222}, \href
  {https://ui.adsabs.harvard.edu/abs/2009ARA&A..47..481A} {47, 481}

\bibitem[\protect\citeauthoryear{{Asplund}, {Amarsi}  \& {Grevesse}}{{Asplund}
  et~al.}{2021}]{Asplund21}
{Asplund} M.,  {Amarsi} A.~M.,   {Grevesse} N.,  2021, \mn@doi [\aap]
  {10.1051/0004-6361/202140445}, \href
  {https://ui.adsabs.harvard.edu/abs/2021A&A...653A.141A} {653, A141}

\bibitem[\protect\citeauthoryear{{Bate}}{{Bate}}{2014}]{Bate14}
{Bate} M.~R.,  2014, \mn@doi [\mnras] {10.1093/mnras/stu795}, \href
  {https://ui.adsabs.harvard.edu/abs/2014MNRAS.442..285B} {442, 285}

\bibitem[\protect\citeauthoryear{{Bate}}{{Bate}}{2019}]{Bate19}
{Bate} M.~R.,  2019, \mn@doi [\mnras] {10.1093/mnras/stz103}, \href
  {https://ui.adsabs.harvard.edu/abs/2019MNRAS.484.2341B} {484, 2341}

\bibitem[\protect\citeauthoryear{{Bell}, {Viti}  \& {Williams}}{{Bell}
  et~al.}{2007}]{Bell07}
{Bell} T.~A.,  {Viti} S.,   {Williams} D.~A.,  2007, \mn@doi [\mnras]
  {10.1111/j.1365-2966.2007.11830.x}, \href
  {https://ui.adsabs.harvard.edu/abs/2007MNRAS.378..983B} {378, 983}

\bibitem[\protect\citeauthoryear{{Berg}, {Skillman}, {Henry}, {Erb}  \&
  {Carigi}}{{Berg} et~al.}{2016}]{Berg16}
{Berg} D.~A.,  {Skillman} E.~D.,  {Henry} R. B.~C.,  {Erb} D.~K.,   {Carigi}
  L.,  2016, \mn@doi [\apj] {10.3847/0004-637X/827/2/126}, \href
  {https://ui.adsabs.harvard.edu/abs/2016ApJ...827..126B} {827, 126}

\bibitem[\protect\citeauthoryear{{Berg}, {Erb}, {Henry}, {Skillman}  \&
  {McQuinn}}{{Berg} et~al.}{2019}]{Berg19}
{Berg} D.~A.,  {Erb} D.~K.,  {Henry} R. B.~C.,  {Skillman} E.~D.,   {McQuinn}
  K. B.~W.,  2019, \mn@doi [\apj] {10.3847/1538-4357/ab020a}, \href
  {https://ui.adsabs.harvard.edu/abs/2019ApJ...874...93B} {874, 93}

\bibitem[\protect\citeauthoryear{{Bialy}, {Sternberg}, {Lee}, {Le Petit}  \&
  {Roueff}}{{Bialy} et~al.}{2015}]{Bialy15}
{Bialy} S.,  {Sternberg} A.,  {Lee} M.-Y.,  {Le Petit} F.,   {Roueff} E.,
  2015, \mn@doi [\apj] {10.1088/0004-637X/809/2/122}, \href
  {https://ui.adsabs.harvard.edu/abs/2015ApJ...809..122B} {809, 122}

\bibitem[\protect\citeauthoryear{{Bisbas}, {Bell}, {Viti}, {Yates}  \&
  {Barlow}}{{Bisbas} et~al.}{2012}]{Bisbas12}
{Bisbas} T.~G.,  {Bell} T.~A.,  {Viti} S.,  {Yates} J.,   {Barlow} M.~J.,
  2012, \mn@doi [\mnras] {10.1111/j.1365-2966.2012.22077.x}, \href
  {https://ui.adsabs.harvard.edu/abs/2012MNRAS.427.2100B} {427, 2100}

\bibitem[\protect\citeauthoryear{{Bisbas}, {Papadopoulos}  \& {Viti}}{{Bisbas}
  et~al.}{2015}]{Bisbas15}
{Bisbas} T.~G.,  {Papadopoulos} P.~P.,   {Viti} S.,  2015, \mn@doi [\apj]
  {10.1088/0004-637X/803/1/37}, \href
  {https://ui.adsabs.harvard.edu/abs/2015ApJ...803...37B} {803, 37}

\bibitem[\protect\citeauthoryear{{Bisbas}, {van Dishoeck}, {Papadopoulos},
  {Sz{\H{u}}cs}, {Bialy}  \& {Zhang}}{{Bisbas} et~al.}{2017a}]{Bisbas17}
{Bisbas} T.~G.,  {van Dishoeck} E.~F.,  {Papadopoulos} P.~P.,  {Sz{\H{u}}cs}
  L.,  {Bialy} S.,   {Zhang} Z.-Y.,  2017a, \mn@doi [\apj]
  {10.3847/1538-4357/aa696d}, \href
  {https://ui.adsabs.harvard.edu/abs/2017ApJ...839...90B} {839, 90}

\bibitem[\protect\citeauthoryear{{Bisbas}, {Tanaka}, {Tan}, {Wu}  \&
  {Nakamura}}{{Bisbas} et~al.}{2017b}]{Bisbas17b}
{Bisbas} T.~G.,  {Tanaka} K. E.~I.,  {Tan} J.~C.,  {Wu} B.,   {Nakamura} F.,
  2017b, \mn@doi [\apj] {10.3847/1538-4357/aa94c5}, \href
  {https://ui.adsabs.harvard.edu/abs/2017ApJ...850...23B} {850, 23}

\bibitem[\protect\citeauthoryear{{Bisbas} et~al.,}{{Bisbas}
  et~al.}{2018}]{Bisbas18}
{Bisbas} T.~G.,  et~al., 2018, \mn@doi [\mnras] {10.1093/mnrasl/sly039}, \href
  {https://ui.adsabs.harvard.edu/abs/2018MNRAS.478L..54B} {478, L54}

\bibitem[\protect\citeauthoryear{{Bisbas}, {Schruba}  \& {van
  Dishoeck}}{{Bisbas} et~al.}{2019}]{Bisbas19}
{Bisbas} T.~G.,  {Schruba} A.,   {van Dishoeck} E.~F.,  2019, \mn@doi [\mnras]
  {10.1093/mnras/stz405}, \href
  {https://ui.adsabs.harvard.edu/abs/2019MNRAS.485.3097B} {485, 3097}

\bibitem[\protect\citeauthoryear{{Bisbas}, {Tan}  \& {Tanaka}}{{Bisbas}
  et~al.}{2021}]{Bisbas21}
{Bisbas} T.~G.,  {Tan} J.~C.,   {Tanaka} K. E.~I.,  2021, \mn@doi [\mnras]
  {10.1093/mnras/stab121}, \href
  {https://ui.adsabs.harvard.edu/abs/2021MNRAS.502.2701B} {502, 2701}

\bibitem[\protect\citeauthoryear{{Bisbas} et~al.,}{{Bisbas}
  et~al.}{2022}]{Bisbas22}
{Bisbas} T.~G.,  et~al., 2022, \mn@doi [\apj] {10.3847/1538-4357/ac7960}, \href
  {https://ui.adsabs.harvard.edu/abs/2022ApJ...934..115B} {934, 115}

\bibitem[\protect\citeauthoryear{{Bisbas}, {van Dishoeck}, {Hu}  \&
  {Schruba}}{{Bisbas} et~al.}{2023}]{Bisbas23}
{Bisbas} T.~G.,  {van Dishoeck} E.~F.,  {Hu} C.-Y.,   {Schruba} A.,  2023,
  \mn@doi [\mnras] {10.1093/mnras/stac3487}, \href
  {https://ui.adsabs.harvard.edu/abs/2023MNRAS.519..729B} {519, 729}

\bibitem[\protect\citeauthoryear{{Black} \& {Dalgarno}}{{Black} \&
  {Dalgarno}}{1977}]{Black77}
{Black} J.~H.,  {Dalgarno} A.,  1977, \mn@doi [\apjs] {10.1086/190455}, \href
  {https://ui.adsabs.harvard.edu/abs/1977ApJS...34..405B} {34, 405}

\bibitem[\protect\citeauthoryear{{Bolatto}, {Wolfire}  \& {Leroy}}{{Bolatto}
  et~al.}{2013}]{Bolatto13}
{Bolatto} A.~D.,  {Wolfire} M.,   {Leroy} A.~K.,  2013, \mn@doi [\araa]
  {10.1146/annurev-astro-082812-140944}, \href
  {https://ui.adsabs.harvard.edu/abs/2013ARA&A..51..207B} {51, 207}

\bibitem[\protect\citeauthoryear{{Bonifacio} et~al.,}{{Bonifacio}
  et~al.}{2015}]{Bonifacio15}
{Bonifacio} P.,  et~al., 2015, \mn@doi [\aap] {10.1051/0004-6361/201425266},
  \href {https://ui.adsabs.harvard.edu/abs/2015A&A...579A..28B} {579, A28}

\bibitem[\protect\citeauthoryear{{Bothwell} et~al.,}{{Bothwell}
  et~al.}{2017}]{Bothwell17}
{Bothwell} M.~S.,  et~al., 2017, \mn@doi [\mnras] {10.1093/mnras/stw3270},
  \href {https://ui.adsabs.harvard.edu/abs/2017MNRAS.466.2825B} {466, 2825}

\bibitem[\protect\citeauthoryear{{Brauher}, {Dale}  \& {Helou}}{{Brauher}
  et~al.}{2008}]{Brauher08}
{Brauher} J.~R.,  {Dale} D.~A.,   {Helou} G.,  2008, \mn@doi [\apjs]
  {10.1086/590249}, \href
  {https://ui.adsabs.harvard.edu/abs/2008ApJS..178..280B} {178, 280}

\bibitem[\protect\citeauthoryear{{Burbidge}, {Burbidge}, {Fowler}  \&
  {Hoyle}}{{Burbidge} et~al.}{1957}]{B2FH}
{Burbidge} E.~M.,  {Burbidge} G.~R.,  {Fowler} W.~A.,   {Hoyle} F.,  1957,
  \mn@doi [Reviews of Modern Physics] {10.1103/RevModPhys.29.547}, \href
  {https://ui.adsabs.harvard.edu/abs/1957RvMP...29..547B} {29, 547}

\bibitem[\protect\citeauthoryear{{Burkhart}, {Lee}, {Murray}  \&
  {Stanimirovi{\'c}}}{{Burkhart} et~al.}{2015}]{Burkhart15}
{Burkhart} B.,  {Lee} M.-Y.,  {Murray} C.~E.,   {Stanimirovi{\'c}} S.,  2015,
  \mn@doi [\apjl] {10.1088/2041-8205/811/2/L28}, \href
  {https://ui.adsabs.harvard.edu/abs/2015ApJ...811L..28B} {811, L28}

\bibitem[\protect\citeauthoryear{{Cardelli}, {Meyer}, {Jura}  \&
  {Savage}}{{Cardelli} et~al.}{1996}]{Cardelli96}
{Cardelli} J.~A.,  {Meyer} D.~M.,  {Jura} M.,   {Savage} B.~D.,  1996, \mn@doi
  [\apj] {10.1086/177608}, \href
  {https://ui.adsabs.harvard.edu/abs/1996ApJ...467..334C} {467, 334}

\bibitem[\protect\citeauthoryear{{Cartledge}, {Lauroesch}, {Meyer}  \&
  {Sofia}}{{Cartledge} et~al.}{2004}]{Cartledge04}
{Cartledge} S. I.~B.,  {Lauroesch} J.~T.,  {Meyer} D.~M.,   {Sofia} U.~J.,
  2004, \mn@doi [\apj] {10.1086/423270}, \href
  {https://ui.adsabs.harvard.edu/abs/2004ApJ...613.1037C} {613, 1037}

\bibitem[\protect\citeauthoryear{{Casey}, {Narayanan}  \& {Cooray}}{{Casey}
  et~al.}{2014}]{Casey14}
{Casey} C.~M.,  {Narayanan} D.,   {Cooray} A.,  2014, \mn@doi [\physrep]
  {10.1016/j.physrep.2014.02.009}, \href
  {https://ui.adsabs.harvard.edu/abs/2014PhR...541...45C} {541, 45}

\bibitem[\protect\citeauthoryear{{Cazaux} \& {Tielens}}{{Cazaux} \&
  {Tielens}}{2002}]{Cazaux02}
{Cazaux} S.,  {Tielens} A.~G.~G.~M.,  2002, \mn@doi [\apjl] {10.1086/342607},
  \href {https://ui.adsabs.harvard.edu/abs/2002ApJ...575L..29C} {575, L29}

\bibitem[\protect\citeauthoryear{{Cazaux} \& {Tielens}}{{Cazaux} \&
  {Tielens}}{2004}]{Cazaux04}
{Cazaux} S.,  {Tielens} A.~G.~G.~M.,  2004, \mn@doi [\apj] {10.1086/381775},
  \href {https://ui.adsabs.harvard.edu/abs/2004ApJ...604..222C} {604, 222}

\bibitem[\protect\citeauthoryear{{Cazaux} \& {Tielens}}{{Cazaux} \&
  {Tielens}}{2010}]{Cazaux10}
{Cazaux} S.,  {Tielens} A.~G.~G.~M.,  2010, \mn@doi [\apj]
  {10.1088/0004-637X/715/1/698}, \href
  {https://ui.adsabs.harvard.edu/abs/2010ApJ...715..698C} {715, 698}

\bibitem[\protect\citeauthoryear{{Cescutti} \& {Matteucci}}{{Cescutti} \&
  {Matteucci}}{2022}]{Cescutti22}
{Cescutti} G.,  {Matteucci} F.,  2022, \mn@doi [Universe]
  {10.3390/universe8030173}, \href
  {https://ui.adsabs.harvard.edu/abs/2022Univ....8..173C} {8, 173}

\bibitem[\protect\citeauthoryear{{Combes}}{{Combes}}{2018}]{Combes18}
{Combes} F.,  2018, \mn@doi [\aapr] {10.1007/s00159-018-0110-4}, \href
  {https://ui.adsabs.harvard.edu/abs/2018A&ARv..26....5C} {26, 5}

\bibitem[\protect\citeauthoryear{{Cooke}, {Pettini}  \& {Steidel}}{{Cooke}
  et~al.}{2017}]{Cooke17}
{Cooke} R.~J.,  {Pettini} M.,   {Steidel} C.~C.,  2017, \mn@doi [\mnras]
  {10.1093/mnras/stx037}, \href
  {https://ui.adsabs.harvard.edu/abs/2017MNRAS.467..802C} {467, 802}

\bibitem[\protect\citeauthoryear{{Cormier} et~al.,}{{Cormier}
  et~al.}{2019}]{Cormier19}
{Cormier} D.,  et~al., 2019, \mn@doi [\aap] {10.1051/0004-6361/201834457},
  \href {https://ui.adsabs.harvard.edu/abs/2019A&A...626A..23C} {626, A23}

\bibitem[\protect\citeauthoryear{{Cosentino} et~al.,}{{Cosentino}
  et~al.}{2019}]{Cosentino19}
{Cosentino} G.,  et~al., 2019, \mn@doi [\apjl] {10.3847/2041-8213/ab38c5},
  \href {https://ui.adsabs.harvard.edu/abs/2019ApJ...881L..42C} {881, L42}

\bibitem[\protect\citeauthoryear{{Davies} et~al.,}{{Davies}
  et~al.}{2016}]{Davies16}
{Davies} R.~L.,  et~al., 2016, \mn@doi [\mnras] {10.1093/mnras/stw1754}, \href
  {https://ui.adsabs.harvard.edu/abs/2016MNRAS.462.1616D} {462, 1616}

\bibitem[\protect\citeauthoryear{{De Looze}, {Baes}, {Bendo}, {Cortese}  \&
  {Fritz}}{{De Looze} et~al.}{2011}]{DeLooze11}
{De Looze} I.,  {Baes} M.,  {Bendo} G.~J.,  {Cortese} L.,   {Fritz} J.,  2011,
  \mn@doi [\mnras] {10.1111/j.1365-2966.2011.19223.x}, \href
  {https://ui.adsabs.harvard.edu/abs/2011MNRAS.416.2712D} {416, 2712}

\bibitem[\protect\citeauthoryear{{De Looze} et~al.,}{{De Looze}
  et~al.}{2014}]{DeLooze14}
{De Looze} I.,  et~al., 2014, \mn@doi [\aap] {10.1051/0004-6361/201322489},
  \href {https://ui.adsabs.harvard.edu/abs/2014A&A...568A..62D} {568, A62}

\bibitem[\protect\citeauthoryear{{Delgado Mena}, {Adibekyan}, {Santos},
  {Tsantaki}, {Gonz{\'a}lez Hern{\'a}ndez}, {Sousa}  \& {Bertr{\'a}n de
  Lis}}{{Delgado Mena} et~al.}{2021}]{Delgado21}
{Delgado Mena} E.,  {Adibekyan} V.,  {Santos} N.~C.,  {Tsantaki} M.,
  {Gonz{\'a}lez Hern{\'a}ndez} J.~I.,  {Sousa} S.~G.,   {Bertr{\'a}n de Lis}
  S.,  2021, \mn@doi [\aap] {10.1051/0004-6361/202141588}, \href
  {https://ui.adsabs.harvard.edu/abs/2021A&A...655A..99D} {655, A99}

\bibitem[\protect\citeauthoryear{{Draine}}{{Draine}}{1978}]{Draine78}
{Draine} B.~T.,  1978, \mn@doi [\apjs] {10.1086/190513}, \href
  {https://ui.adsabs.harvard.edu/abs/1978ApJS...36..595D} {36, 595}

\bibitem[\protect\citeauthoryear{{Draine}}{{Draine}}{2011}]{Draine11}
{Draine} B.~T.,  2011, {Physics of the Interstellar and Intergalactic Medium}

\bibitem[\protect\citeauthoryear{{Draine} \& {McKee}}{{Draine} \&
  {McKee}}{1993}]{Draine93}
{Draine} B.~T.,  {McKee} C.~F.,  1993, \mn@doi [\araa]
  {10.1146/annurev.aa.31.090193.002105}, \href
  {https://ui.adsabs.harvard.edu/abs/1993ARA&A..31..373D} {31, 373}

\bibitem[\protect\citeauthoryear{{Dufour} \& {Charnley}}{{Dufour} \&
  {Charnley}}{2021}]{Dufour21}
{Dufour} G.,  {Charnley} S.~B.,  2021, \mn@doi [\apj]
  {10.3847/1538-4357/abe1c6}, \href
  {https://ui.adsabs.harvard.edu/abs/2021ApJ...909..171D} {909, 171}

\bibitem[\protect\citeauthoryear{{Dunne}, {Maddox}, {Papadopoulos}, {Ivison}
  \& {Gomez}}{{Dunne} et~al.}{2022}]{Dunne22}
{Dunne} L.,  {Maddox} S.~J.,  {Papadopoulos} P.~P.,  {Ivison} R.~J.,   {Gomez}
  H.~L.,  2022, \mn@doi [\mnras] {10.1093/mnras/stac2098}, \href
  {https://ui.adsabs.harvard.edu/abs/2022MNRAS.517..962D} {517, 962}

\bibitem[\protect\citeauthoryear{{Dwek}}{{Dwek}}{1998}]{Dwek98}
{Dwek} E.,  1998, \mn@doi [\apj] {10.1086/305829}, \href
  {https://ui.adsabs.harvard.edu/abs/1998ApJ...501..643D} {501, 643}

\bibitem[\protect\citeauthoryear{{Esposito}, {Vallini}, {Pozzi}, {Casasola},
  {Mingozzi}, {Vignali}, {Gruppioni}  \& {Salvestrini}}{{Esposito}
  et~al.}{2022}]{Esposito22}
{Esposito} F.,  {Vallini} L.,  {Pozzi} F.,  {Casasola} V.,  {Mingozzi} M.,
  {Vignali} C.,  {Gruppioni} C.,   {Salvestrini} F.,  2022, \mn@doi [\mnras]
  {10.1093/mnras/stac313}, \href
  {https://ui.adsabs.harvard.edu/abs/2022MNRAS.512..686E} {512, 686}

\bibitem[\protect\citeauthoryear{{Esteban}, {M{\'e}ndez-Delgado},
  {Garc{\'\i}a-Rojas}  \& {Arellano-C{\'o}rdova}}{{Esteban}
  et~al.}{2022}]{Esteban22}
{Esteban} C.,  {M{\'e}ndez-Delgado} J.~E.,  {Garc{\'\i}a-Rojas} J.,
  {Arellano-C{\'o}rdova} K.~Z.,  2022, \mn@doi [\apj]
  {10.3847/1538-4357/ac6b38}, \href
  {https://ui.adsabs.harvard.edu/abs/2022ApJ...931...92E} {931, 92}

\bibitem[\protect\citeauthoryear{{Fabbian}, {Nissen}, {Asplund}, {Pettini}  \&
  {Akerman}}{{Fabbian} et~al.}{2009}]{Fabbian09}
{Fabbian} D.,  {Nissen} P.~E.,  {Asplund} M.,  {Pettini} M.,   {Akerman} C.,
  2009, \mn@doi [\aap] {10.1051/0004-6361/200810095}, \href
  {https://ui.adsabs.harvard.edu/abs/2009A&A...500.1143F} {500, 1143}

\bibitem[\protect\citeauthoryear{{Ferland} et~al.,}{{Ferland}
  et~al.}{2017}]{Ferland17}
{Ferland} G.~J.,  et~al., 2017, \mn@doi [\rmxaa] {10.48550/arXiv.1705.10877},
  \href {https://ui.adsabs.harvard.edu/abs/2017RMxAA..53..385F} {53, 385}

\bibitem[\protect\citeauthoryear{{Franeck} et~al.,}{{Franeck}
  et~al.}{2018}]{Franeck18}
{Franeck} A.,  et~al., 2018, \mn@doi [\mnras] {10.1093/mnras/sty2507}, \href
  {https://ui.adsabs.harvard.edu/abs/2018MNRAS.481.4277F} {481, 4277}

\bibitem[\protect\citeauthoryear{{Freer} \& {Fynbo}}{{Freer} \&
  {Fynbo}}{2014}]{Freer14}
{Freer} M.,  {Fynbo} H.~O.~U.,  2014, \mn@doi [Progress in Particle and Nuclear
  Physics] {10.1016/j.ppnp.2014.06.001}, \href
  {https://ui.adsabs.harvard.edu/abs/2014PrPNP..78....1F} {78, 1}

\bibitem[\protect\citeauthoryear{{Frias Castillo} et~al.,}{{Frias Castillo}
  et~al.}{2023}]{FriasCastillo23}
{Frias Castillo} M.,  et~al., 2023, \mn@doi [\apj] {10.3847/1538-4357/acb931},
  \href {https://ui.adsabs.harvard.edu/abs/2023ApJ...945..128F} {945, 128}

\bibitem[\protect\citeauthoryear{Fudamoto et~al.,}{Fudamoto
  et~al.}{2023}]{Fudamoto23}
Fudamoto Y.,  et~al., 2023

\bibitem[\protect\citeauthoryear{{Fuente} et~al.,}{{Fuente}
  et~al.}{2019}]{Fuente19}
{Fuente} A.,  et~al., 2019, \mn@doi [\aap] {10.1051/0004-6361/201834654}, \href
  {https://ui.adsabs.harvard.edu/abs/2019A&A...624A.105F} {624, A105}

\bibitem[\protect\citeauthoryear{{Gaches}, {Offner}  \& {Bisbas}}{{Gaches}
  et~al.}{2019a}]{Gaches19}
{Gaches} B. A.~L.,  {Offner} S. S.~R.,   {Bisbas} T.~G.,  2019a, \mn@doi [\apj]
  {10.3847/1538-4357/ab20c7}, \href
  {https://ui.adsabs.harvard.edu/abs/2019ApJ...878..105G} {878, 105}

\bibitem[\protect\citeauthoryear{{Gaches}, {Offner}  \& {Bisbas}}{{Gaches}
  et~al.}{2019b}]{Gaches19b}
{Gaches} B. A.~L.,  {Offner} S. S.~R.,   {Bisbas} T.~G.,  2019b, \mn@doi [\apj]
  {10.3847/1538-4357/ab3c5c}, \href
  {https://ui.adsabs.harvard.edu/abs/2019ApJ...883..190G} {883, 190}

\bibitem[\protect\citeauthoryear{{Gaches}, {Bisbas}  \& {Bialy}}{{Gaches}
  et~al.}{2022a}]{Gaches22}
{Gaches} B. A.~L.,  {Bisbas} T.~G.,   {Bialy} S.,  2022a, \mn@doi [\aap]
  {10.1051/0004-6361/202142411}, \href
  {https://ui.adsabs.harvard.edu/abs/2022A&A...658A.151G} {658, A151}

\bibitem[\protect\citeauthoryear{{Gaches}, {Bialy}, {Bisbas}, {Padovani},
  {Seifried}  \& {Walch}}{{Gaches} et~al.}{2022b}]{Gaches22b}
{Gaches} B. A.~L.,  {Bialy} S.,  {Bisbas} T.~G.,  {Padovani} M.,  {Seifried}
  D.,   {Walch} S.,  2022b, \mn@doi [\aap] {10.1051/0004-6361/202244090}, \href
  {https://ui.adsabs.harvard.edu/abs/2022A&A...664A.150G} {664, A150}

\bibitem[\protect\citeauthoryear{{Galametz}, {Madden}, {Galliano}, {Hony},
  {Bendo}  \& {Sauvage}}{{Galametz} et~al.}{2011}]{Galametz11}
{Galametz} M.,  {Madden} S.~C.,  {Galliano} F.,  {Hony} S.,  {Bendo} G.~J.,
  {Sauvage} M.,  2011, \mn@doi [\aap] {10.1051/0004-6361/201014904}, \href
  {https://ui.adsabs.harvard.edu/abs/2011A&A...532A..56G} {532, A56}

\bibitem[\protect\citeauthoryear{{Gallino}, {Arlandini}, {Busso}, {Lugaro},
  {Travaglio}, {Straniero}, {Chieffi}  \& {Limongi}}{{Gallino}
  et~al.}{1998}]{Gallino98}
{Gallino} R.,  {Arlandini} C.,  {Busso} M.,  {Lugaro} M.,  {Travaglio} C.,
  {Straniero} O.,  {Chieffi} A.,   {Limongi} M.,  1998, \mn@doi [\apj]
  {10.1086/305437}, \href
  {https://ui.adsabs.harvard.edu/abs/1998ApJ...497..388G} {497, 388}

\bibitem[\protect\citeauthoryear{{Garnett}, {Skillman}, {Dufour}, {Peimbert},
  {Torres-Peimbert}, {Terlevich}, {Terlevich}  \& {Shields}}{{Garnett}
  et~al.}{1995}]{Garnett95}
{Garnett} D.~R.,  {Skillman} E.~D.,  {Dufour} R.~J.,  {Peimbert} M.,
  {Torres-Peimbert} S.,  {Terlevich} R.,  {Terlevich} E.,   {Shields} G.~A.,
  1995, \mn@doi [\apj] {10.1086/175503}, \href
  {https://ui.adsabs.harvard.edu/abs/1995ApJ...443...64G} {443, 64}

\bibitem[\protect\citeauthoryear{{Genzel} et~al.,}{{Genzel}
  et~al.}{2012}]{Genzel12}
{Genzel} R.,  et~al., 2012, \mn@doi [\apj] {10.1088/0004-637X/746/1/69}, \href
  {https://ui.adsabs.harvard.edu/abs/2012ApJ...746...69G} {746, 69}

\bibitem[\protect\citeauthoryear{{Girichidis} et~al.,}{{Girichidis}
  et~al.}{2016}]{Girichidis16}
{Girichidis} P.,  et~al., 2016, \mn@doi [\mnras] {10.1093/mnras/stv2742}, \href
  {https://ui.adsabs.harvard.edu/abs/2016MNRAS.456.3432G} {456, 3432}

\bibitem[\protect\citeauthoryear{{Glover} \& {Clark}}{{Glover} \&
  {Clark}}{2016}]{Glover16}
{Glover} S. C.~O.,  {Clark} P.~C.,  2016, \mn@doi [\mnras]
  {10.1093/mnras/stv2863}, \href
  {https://ui.adsabs.harvard.edu/abs/2016MNRAS.456.3596G} {456, 3596}

\bibitem[\protect\citeauthoryear{{Glover}, {Federrath}, {Mac Low}  \&
  {Klessen}}{{Glover} et~al.}{2010}]{Glover10}
{Glover} S.~C.~O.,  {Federrath} C.,  {Mac Low} M.~M.,   {Klessen} R.~S.,  2010,
  \mn@doi [\mnras] {10.1111/j.1365-2966.2009.15718.x}, \href
  {https://ui.adsabs.harvard.edu/abs/2010MNRAS.404....2G} {404, 2}

\bibitem[\protect\citeauthoryear{{Glover}, {Clark}, {Micic}  \&
  {Molina}}{{Glover} et~al.}{2015}]{Glover15}
{Glover} S. C.~O.,  {Clark} P.~C.,  {Micic} M.,   {Molina} F.,  2015, \mn@doi
  [\mnras] {10.1093/mnras/stu2699}, \href
  {https://ui.adsabs.harvard.edu/abs/2015MNRAS.448.1607G} {448, 1607}

\bibitem[\protect\citeauthoryear{{Goldsmith}, {Langer}, {Pineda}  \&
  {Velusamy}}{{Goldsmith} et~al.}{2012}]{Goldsmith12}
{Goldsmith} P.~F.,  {Langer} W.~D.,  {Pineda} J.~L.,   {Velusamy} T.,  2012,
  \mn@doi [\apjs] {10.1088/0067-0049/203/1/13}, \href
  {https://ui.adsabs.harvard.edu/abs/2012ApJS..203...13G} {203, 13}

\bibitem[\protect\citeauthoryear{{Gong}, {Ostriker}  \& {Wolfire}}{{Gong}
  et~al.}{2017}]{Gong17}
{Gong} M.,  {Ostriker} E.~C.,   {Wolfire} M.~G.,  2017, \mn@doi [\apj]
  {10.3847/1538-4357/aa7561}, \href
  {https://ui.adsabs.harvard.edu/abs/2017ApJ...843...38G} {843, 38}

\bibitem[\protect\citeauthoryear{{Gong}, {Ostriker}, {Kim}  \& {Kim}}{{Gong}
  et~al.}{2020}]{Gong20}
{Gong} M.,  {Ostriker} E.~C.,  {Kim} C.-G.,   {Kim} J.-G.,  2020, \mn@doi
  [\apj] {10.3847/1538-4357/abbdab}, \href
  {https://ui.adsabs.harvard.edu/abs/2020ApJ...903..142G} {903, 142}

\bibitem[\protect\citeauthoryear{{Grenier}, {Black}  \& {Strong}}{{Grenier}
  et~al.}{2015}]{Grenier15}
{Grenier} I.~A.,  {Black} J.~H.,   {Strong} A.~W.,  2015, \mn@doi [\araa]
  {10.1146/annurev-astro-082214-122457}, \href
  {https://ui.adsabs.harvard.edu/abs/2015ARA&A..53..199G} {53, 199}

\bibitem[\protect\citeauthoryear{{Gullberg} et~al.,}{{Gullberg}
  et~al.}{2016}]{Gullberg16}
{Gullberg} B.,  et~al., 2016, \mn@doi [\aap] {10.1051/0004-6361/201527647},
  \href {https://ui.adsabs.harvard.edu/abs/2016A&A...591A..73G} {591, A73}

\bibitem[\protect\citeauthoryear{{Harikane} et~al.,}{{Harikane}
  et~al.}{2020}]{Harikane20}
{Harikane} Y.,  et~al., 2020, \mn@doi [\apj] {10.3847/1538-4357/ab94bd}, \href
  {https://ui.adsabs.harvard.edu/abs/2020ApJ...896...93H} {896, 93}

\bibitem[\protect\citeauthoryear{{Harrington} et~al.,}{{Harrington}
  et~al.}{2021}]{Harrington21}
{Harrington} K.~C.,  et~al., 2021, \mn@doi [\apj] {10.3847/1538-4357/abcc01},
  \href {https://ui.adsabs.harvard.edu/abs/2021ApJ...908...95H} {908, 95}

\bibitem[\protect\citeauthoryear{{Haworth}, {Glover}, {Koepferl}, {Bisbas}  \&
  {Dale}}{{Haworth} et~al.}{2018}]{Haworth18}
{Haworth} T.~J.,  {Glover} S. C.~O.,  {Koepferl} C.~M.,  {Bisbas} T.~G.,
  {Dale} J.~E.,  2018, \mn@doi [\nar] {10.1016/j.newar.2018.06.001}, \href
  {https://ui.adsabs.harvard.edu/abs/2018NewAR..82....1H} {82, 1}

\bibitem[\protect\citeauthoryear{{Herrera-Camus} et~al.,}{{Herrera-Camus}
  et~al.}{2012}]{Herrera12}
{Herrera-Camus} R.,  et~al., 2012, \mn@doi [\apj]
  {10.1088/0004-637X/752/2/112}, \href
  {https://ui.adsabs.harvard.edu/abs/2012ApJ...752..112H} {752, 112}

\bibitem[\protect\citeauthoryear{{Herrera-Camus} et~al.,}{{Herrera-Camus}
  et~al.}{2015}]{Herrera15}
{Herrera-Camus} R.,  et~al., 2015, \mn@doi [\apj] {10.1088/0004-637X/800/1/1},
  \href {https://ui.adsabs.harvard.edu/abs/2015ApJ...800....1H} {800, 1}

\bibitem[\protect\citeauthoryear{{Herrera-Camus} et~al.,}{{Herrera-Camus}
  et~al.}{2018}]{Herrera18}
{Herrera-Camus} R.,  et~al., 2018, \mn@doi [\apj] {10.3847/1538-4357/aac0f6},
  \href {https://ui.adsabs.harvard.edu/abs/2018ApJ...861...94H} {861, 94}

\bibitem[\protect\citeauthoryear{{Hollenbach} \& {Tielens}}{{Hollenbach} \&
  {Tielens}}{1999}]{Hollenbach99}
{Hollenbach} D.~J.,  {Tielens} A.~G.~G.~M.,  1999, \mn@doi [Reviews of Modern
  Physics] {10.1103/RevModPhys.71.173}, \href
  {https://ui.adsabs.harvard.edu/abs/1999RvMP...71..173H} {71, 173}

\bibitem[\protect\citeauthoryear{{Hoyle}}{{Hoyle}}{1954}]{Hoyle54}
{Hoyle} F.,  1954, \mn@doi [\apjs] {10.1086/190005}, \href
  {https://ui.adsabs.harvard.edu/abs/1954ApJS....1..121H} {1, 121}

\bibitem[\protect\citeauthoryear{{Hu}, {Sternberg}  \& {van Dishoeck}}{{Hu}
  et~al.}{2021}]{Hu21}
{Hu} C.-Y.,  {Sternberg} A.,   {van Dishoeck} E.~F.,  2021, \mn@doi [\apj]
  {10.3847/1538-4357/ac0dbd}, \href
  {https://ui.adsabs.harvard.edu/abs/2021ApJ...920...44H} {920, 44}

\bibitem[\protect\citeauthoryear{{Indriolo}}{{Indriolo}}{2023}]{Indriolo23}
{Indriolo} N.,  2023, arXiv e-prints, \href
  {https://ui.adsabs.harvard.edu/abs/2023arXiv230313689I} {p. arXiv:2303.13689}

\bibitem[\protect\citeauthoryear{{Indriolo}, {Bergin}, {Falgarone}, {Godard},
  {Zwaan}, {Neufeld}  \& {Wolfire}}{{Indriolo} et~al.}{2018}]{Indriolo18}
{Indriolo} N.,  {Bergin} E.~A.,  {Falgarone} E.,  {Godard} B.,  {Zwaan} M.~A.,
  {Neufeld} D.~A.,   {Wolfire} M.~G.,  2018, \mn@doi [\apj]
  {10.3847/1538-4357/aad7b3}, \href
  {https://ui.adsabs.harvard.edu/abs/2018ApJ...865..127I} {865, 127}

\bibitem[\protect\citeauthoryear{{Isobe} et~al.,}{{Isobe}
  et~al.}{2023}]{Isobe23}
{Isobe} Y.,  et~al., 2023, \mn@doi [arXiv e-prints]
  {10.48550/arXiv.2307.00710}, \href
  {https://ui.adsabs.harvard.edu/abs/2023arXiv230700710I} {p. arXiv:2307.00710}

\bibitem[\protect\citeauthoryear{{Israel} \& {Baas}}{{Israel} \&
  {Baas}}{2002}]{Israel02}
{Israel} F.~P.,  {Baas} F.,  2002, \mn@doi [\aap] {10.1051/0004-6361:20011736},
  \href {https://ui.adsabs.harvard.edu/abs/2002A&A...383...82I} {383, 82}

\bibitem[\protect\citeauthoryear{{Izotov}, {Schaerer}, {Worseck}, {Berg},
  {Chisholm}, {Ravindranath}  \& {Thuan}}{{Izotov} et~al.}{2023}]{Izotov23}
{Izotov} Y.~I.,  {Schaerer} D.,  {Worseck} G.,  {Berg} D.,  {Chisholm} J.,
  {Ravindranath} S.,   {Thuan} T.~X.,  2023, \mn@doi [\mnras]
  {10.1093/mnras/stad1036}, \href
  {https://ui.adsabs.harvard.edu/abs/2023MNRAS.tmp..946I} {}

\bibitem[\protect\citeauthoryear{{James}, {Dunne}, {Eales}  \&
  {Edmunds}}{{James} et~al.}{2002}]{James02}
{James} A.,  {Dunne} L.,  {Eales} S.,   {Edmunds} M.~G.,  2002, \mn@doi
  [\mnras] {10.1046/j.1365-8711.2002.05660.x}, \href
  {https://ui.adsabs.harvard.edu/abs/2002MNRAS.335..753J} {335, 753}

\bibitem[\protect\citeauthoryear{{James}, {Viti}, {Holdship}  \&
  {Jim{\'e}nez-Serra}}{{James} et~al.}{2020}]{James20}
{James} T.~A.,  {Viti} S.,  {Holdship} J.,   {Jim{\'e}nez-Serra} I.,  2020,
  \mn@doi [\aap] {10.1051/0004-6361/201936536}, \href
  {https://ui.adsabs.harvard.edu/abs/2020A&A...634A..17J} {634, A17}

\bibitem[\protect\citeauthoryear{{Jiao}, {Zhao}, {Zhu}, {Lu}, {Gao}  \&
  {Zhang}}{{Jiao} et~al.}{2017}]{Jiao17}
{Jiao} Q.,  {Zhao} Y.,  {Zhu} M.,  {Lu} N.,  {Gao} Y.,   {Zhang} Z.-Y.,  2017,
  \mn@doi [\apjl] {10.3847/2041-8213/aa6f0f}, \href
  {https://ui.adsabs.harvard.edu/abs/2017ApJ...840L..18J} {840, L18}

\bibitem[\protect\citeauthoryear{{Jiao} et~al.,}{{Jiao} et~al.}{2019}]{Jiao19}
{Jiao} Q.,  et~al., 2019, \mn@doi [\apj] {10.3847/1538-4357/ab29ed}, \href
  {https://ui.adsabs.harvard.edu/abs/2019ApJ...880..133J} {880, 133}

\bibitem[\protect\citeauthoryear{{Jones} et~al.,}{{Jones}
  et~al.}{2023}]{Jones23}
{Jones} T.,  et~al., 2023, \mn@doi [\apjl] {10.3847/2041-8213/acd938}, \href
  {https://ui.adsabs.harvard.edu/abs/2023ApJ...951L..17J} {951, L17}

\bibitem[\protect\citeauthoryear{{Jura}}{{Jura}}{1974}]{Jura74}
{Jura} M.,  1974, \mn@doi [\apj] {10.1086/152975}, \href
  {https://ui.adsabs.harvard.edu/abs/1974ApJ...191..375J} {191, 375}

\bibitem[\protect\citeauthoryear{{Kajino}, {Aoki}, {Balantekin}, {Diehl},
  {Famiano}  \& {Mathews}}{{Kajino} et~al.}{2019}]{Kajino19}
{Kajino} T.,  {Aoki} W.,  {Balantekin} A.~B.,  {Diehl} R.,  {Famiano} M.~A.,
  {Mathews} G.~J.,  2019, \mn@doi [Progress in Particle and Nuclear Physics]
  {10.1016/j.ppnp.2019.02.008}, \href
  {https://ui.adsabs.harvard.edu/abs/2019PrPNP.107..109K} {107, 109}

\bibitem[\protect\citeauthoryear{{Kasen}, {Metzger}, {Barnes}, {Quataert}  \&
  {Ramirez-Ruiz}}{{Kasen} et~al.}{2017}]{Kasen17}
{Kasen} D.,  {Metzger} B.,  {Barnes} J.,  {Quataert} E.,   {Ramirez-Ruiz} E.,
  2017, \mn@doi [\nat] {10.1038/nature24453}, \href
  {https://ui.adsabs.harvard.edu/abs/2017Natur.551...80K} {551, 80}

\bibitem[\protect\citeauthoryear{{Kashiyama} \& {M{\'e}sz{\'a}ros}}{{Kashiyama}
  \& {M{\'e}sz{\'a}ros}}{2014}]{Kashiyama14}
{Kashiyama} K.,  {M{\'e}sz{\'a}ros} P.,  2014, \mn@doi [\apjl]
  {10.1088/2041-8205/790/1/L14}, \href
  {https://ui.adsabs.harvard.edu/abs/2014ApJ...790L..14K} {790, L14}

\bibitem[\protect\citeauthoryear{{Katz} et~al.,}{{Katz} et~al.}{2022}]{Katz22}
{Katz} H.,  et~al., 2022, \mn@doi [\mnras] {10.1093/mnras/stac028}, \href
  {https://ui.adsabs.harvard.edu/abs/2022MNRAS.510.5603K} {510, 5603}

\bibitem[\protect\citeauthoryear{{Kaufman}, {Wolfire}, {Hollenbach}  \&
  {Luhman}}{{Kaufman} et~al.}{1999}]{Kaufman99}
{Kaufman} M.~J.,  {Wolfire} M.~G.,  {Hollenbach} D.~J.,   {Luhman} M.~L.,
  1999, \mn@doi [\apj] {10.1086/308102}, \href
  {https://ui.adsabs.harvard.edu/abs/1999ApJ...527..795K} {527, 795}

\bibitem[\protect\citeauthoryear{{Kelly}, {Viti}, {Garc{\'\i}a-Burillo},
  {Fuente}, {Usero}, {Krips}  \& {Neri}}{{Kelly} et~al.}{2017}]{Kelly17}
{Kelly} G.,  {Viti} S.,  {Garc{\'\i}a-Burillo} S.,  {Fuente} A.,  {Usero} A.,
  {Krips} M.,   {Neri} R.,  2017, \mn@doi [\aap] {10.1051/0004-6361/201628946},
  \href {https://ui.adsabs.harvard.edu/abs/2017A&A...597A..11K} {597, A11}

\bibitem[\protect\citeauthoryear{{Klitsch} et~al.,}{{Klitsch}
  et~al.}{2022}]{Klitsch22}
{Klitsch} A.,  et~al., 2022, \mn@doi [\mnras] {10.1093/mnras/stac1190}, \href
  {https://ui.adsabs.harvard.edu/abs/2022MNRAS.514.2346K} {514, 2346}

\bibitem[\protect\citeauthoryear{{Kreckel} et~al.,}{{Kreckel}
  et~al.}{2019}]{Kreckel19}
{Kreckel} K.,  et~al., 2019, \mn@doi [\apj] {10.3847/1538-4357/ab5115}, \href
  {https://ui.adsabs.harvard.edu/abs/2019ApJ...887...80K} {887, 80}

\bibitem[\protect\citeauthoryear{{Krips} et~al.,}{{Krips}
  et~al.}{2016}]{Krips16}
{Krips} M.,  et~al., 2016, \mn@doi [\aap] {10.1051/0004-6361/201628882}, \href
  {https://ui.adsabs.harvard.edu/abs/2016A&A...592L...3K} {592, L3}

\bibitem[\protect\citeauthoryear{{Lah{\'e}n}, {Naab}, {Johansson}, {Elmegreen},
  {Hu}, {Walch}, {Steinwandel}  \& {Moster}}{{Lah{\'e}n}
  et~al.}{2020}]{Lahen20}
{Lah{\'e}n} N.,  {Naab} T.,  {Johansson} P.~H.,  {Elmegreen} B.,  {Hu} C.-Y.,
  {Walch} S.,  {Steinwandel} U.~P.,   {Moster} B.~P.,  2020, \mn@doi [\apj]
  {10.3847/1538-4357/ab7190}, \href
  {https://ui.adsabs.harvard.edu/abs/2020ApJ...891....2L} {891, 2}

\bibitem[\protect\citeauthoryear{{Le Bourlot}, {Pineau des Forets}, {Roueff}
  \& {Schilke}}{{Le Bourlot} et~al.}{1993}]{LeBourlot93}
{Le Bourlot} J.,  {Pineau des Forets} G.,  {Roueff} E.,   {Schilke} P.,  1993,
  \mn@doi [\apjl] {10.1086/187077}, \href
  {https://ui.adsabs.harvard.edu/abs/1993ApJ...416L..87L} {416, L87}

\bibitem[\protect\citeauthoryear{{Le Petit}, {Nehm{\'e}}, {Le Bourlot}  \&
  {Roueff}}{{Le Petit} et~al.}{2006}]{LePetit06}
{Le Petit} F.,  {Nehm{\'e}} C.,  {Le Bourlot} J.,   {Roueff} E.,  2006, \mn@doi
  [\apjs] {10.1086/503252}, \href
  {https://ui.adsabs.harvard.edu/abs/2006ApJS..164..506L} {164, 506}

\bibitem[\protect\citeauthoryear{{Lee}, {Herbst}, {Pineau des Forets}, {Roueff}
   \& {Le Bourlot}}{{Lee} et~al.}{1996}]{Lee96}
{Lee} H.~H.,  {Herbst} E.,  {Pineau des Forets} G.,  {Roueff} E.,   {Le
  Bourlot} J.,  1996, \aap, \href
  {https://ui.adsabs.harvard.edu/abs/1996A&A...311..690L} {311, 690}

\bibitem[\protect\citeauthoryear{{Lelli} et~al.,}{{Lelli}
  et~al.}{2023}]{Lelli23}
{Lelli} F.,  et~al., 2023, \mn@doi [\aap] {10.1051/0004-6361/202245105}, \href
  {https://ui.adsabs.harvard.edu/abs/2023A&A...672A.106L} {672, A106}

\bibitem[\protect\citeauthoryear{{Li}, {Millar}, {Heays}, {Walsh}, {van
  Dishoeck}  \& {Cherchneff}}{{Li} et~al.}{2016}]{Li16}
{Li} X.,  {Millar} T.~J.,  {Heays} A.~N.,  {Walsh} C.,  {van Dishoeck} E.~F.,
  {Cherchneff} I.,  2016, \mn@doi [\aap] {10.1051/0004-6361/201525739}, \href
  {https://ui.adsabs.harvard.edu/abs/2016A&A...588A...4L} {588, A4}

\bibitem[\protect\citeauthoryear{{Liang} et~al.,}{{Liang}
  et~al.}{2023}]{Liang23}
{Liang} L.,  et~al., 2023, \mn@doi [arXiv e-prints]
  {10.48550/arXiv.2301.04149}, \href
  {https://ui.adsabs.harvard.edu/abs/2023arXiv230104149L} {p. arXiv:2301.04149}

\bibitem[\protect\citeauthoryear{{Limongi} \& {Chieffi}}{{Limongi} \&
  {Chieffi}}{2018}]{Limongi18}
{Limongi} M.,  {Chieffi} A.,  2018, \mn@doi [\apjs] {10.3847/1538-4365/aacb24},
  \href {https://ui.adsabs.harvard.edu/abs/2018ApJS..237...13L} {237, 13}

\bibitem[\protect\citeauthoryear{{Lique}, {Werfelli}, {Halvick}, {Stoecklin},
  {Faure}, {Wiesenfeld}  \& {Dagdigian}}{{Lique} et~al.}{2013}]{Lique13}
{Lique} F.,  {Werfelli} G.,  {Halvick} P.,  {Stoecklin} T.,  {Faure} A.,
  {Wiesenfeld} L.,   {Dagdigian} P.~J.,  2013, \mn@doi [\jcp]
  {10.1063/1.4807311}, \href
  {https://ui.adsabs.harvard.edu/abs/2013JChPh.138t4314L} {138, 204314}

\bibitem[\protect\citeauthoryear{{Lo} et~al.,}{{Lo} et~al.}{2014}]{Lo14}
{Lo} N.,  et~al., 2014, \mn@doi [\apjl] {10.1088/2041-8205/797/2/L17}, \href
  {https://ui.adsabs.harvard.edu/abs/2014ApJ...797L..17L} {797, L17}

\bibitem[\protect\citeauthoryear{{Luhman}, {Satyapal}, {Fischer}, {Wolfire},
  {Sturm}, {Dudley}, {Lutz}  \& {Genzel}}{{Luhman} et~al.}{2003}]{Luhman03}
{Luhman} M.~L.,  {Satyapal} S.,  {Fischer} J.,  {Wolfire} M.~G.,  {Sturm} E.,
  {Dudley} C.~C.,  {Lutz} D.,   {Genzel} R.,  2003, \mn@doi [\apj]
  {10.1086/376965}, \href
  {https://ui.adsabs.harvard.edu/abs/2003ApJ...594..758L} {594, 758}

\bibitem[\protect\citeauthoryear{{Luo} et~al.,}{{Luo} et~al.}{2020}]{Luo20}
{Luo} G.,  et~al., 2020, \mn@doi [\apjl] {10.3847/2041-8213/ab6337}, \href
  {https://ui.adsabs.harvard.edu/abs/2020ApJ...889L...4L} {889, L4}

\bibitem[\protect\citeauthoryear{{Luo} et~al.,}{{Luo} et~al.}{2023}]{Luo23}
{Luo} G.,  et~al., 2023, \mn@doi [\apj] {10.3847/1538-4357/aca657}, \href
  {https://ui.adsabs.harvard.edu/abs/2023ApJ...942..101L} {942, 101}

\bibitem[\protect\citeauthoryear{{Mackey}, {Walch}, {Seifried}, {Glover},
  {W{\"u}nsch}  \& {Aharonian}}{{Mackey} et~al.}{2019}]{Mackey19}
{Mackey} J.,  {Walch} S.,  {Seifried} D.,  {Glover} S. C.~O.,  {W{\"u}nsch} R.,
    {Aharonian} F.,  2019, \mn@doi [\mnras] {10.1093/mnras/stz902}, \href
  {https://ui.adsabs.harvard.edu/abs/2019MNRAS.486.1094M} {486, 1094}

\bibitem[\protect\citeauthoryear{{Madden} et~al.,}{{Madden}
  et~al.}{2020}]{Madden20}
{Madden} S.~C.,  et~al., 2020, \mn@doi [\aap] {10.1051/0004-6361/202038860},
  \href {https://ui.adsabs.harvard.edu/abs/2020A&A...643A.141M} {643, A141}

\bibitem[\protect\citeauthoryear{{Maiolino} \& {Mannucci}}{{Maiolino} \&
  {Mannucci}}{2019}]{Maiolino19}
{Maiolino} R.,  {Mannucci} F.,  2019, \mn@doi [\aapr]
  {10.1007/s00159-018-0112-2}, \href
  {https://ui.adsabs.harvard.edu/abs/2019A&ARv..27....3M} {27, 3}

\bibitem[\protect\citeauthoryear{{Malhotra} et~al.,}{{Malhotra}
  et~al.}{2001}]{Malhotra01}
{Malhotra} S.,  et~al., 2001, \mn@doi [\apj] {10.1086/323046}, \href
  {https://ui.adsabs.harvard.edu/abs/2001ApJ...561..766M} {561, 766}

\bibitem[\protect\citeauthoryear{{Maloney}, {Hollenbach}  \&
  {Tielens}}{{Maloney} et~al.}{1996}]{Maloney96}
{Maloney} P.~R.,  {Hollenbach} D.~J.,   {Tielens} A.~G.~G.~M.,  1996, \mn@doi
  [\apj] {10.1086/177532}, \href
  {https://ui.adsabs.harvard.edu/abs/1996ApJ...466..561M} {466, 561}

\bibitem[\protect\citeauthoryear{{Mashian} et~al.,}{{Mashian}
  et~al.}{2015}]{Mashian15}
{Mashian} N.,  et~al., 2015, \mn@doi [\apj] {10.1088/0004-637X/802/2/81}, \href
  {https://ui.adsabs.harvard.edu/abs/2015ApJ...802...81M} {802, 81}

\bibitem[\protect\citeauthoryear{{Mathis}}{{Mathis}}{1990}]{Mathis90}
{Mathis} J.~S.,  1990, \mn@doi [\araa] {10.1146/annurev.aa.28.090190.000345},
  \href {https://ui.adsabs.harvard.edu/abs/1990ARA&A..28...37M} {28, 37}

\bibitem[\protect\citeauthoryear{{Mathis}, {Rumpl}  \& {Nordsieck}}{{Mathis}
  et~al.}{1977}]{Mathis97}
{Mathis} J.~S.,  {Rumpl} W.,   {Nordsieck} K.~H.,  1977, \mn@doi [\apj]
  {10.1086/155591}, \href
  {https://ui.adsabs.harvard.edu/abs/1977ApJ...217..425M} {217, 425}

\bibitem[\protect\citeauthoryear{{Matsunaga} et~al.,}{{Matsunaga}
  et~al.}{2023}]{Matsunaga23}
{Matsunaga} N.,  et~al., 2023, arXiv e-prints, \href
  {https://ui.adsabs.harvard.edu/abs/2023arXiv230802853M} {p. arXiv:2308.02853}

\bibitem[\protect\citeauthoryear{{McElroy}, {Walsh}, {Markwick}, {Cordiner},
  {Smith}  \& {Millar}}{{McElroy} et~al.}{2013}]{McElroy13}
{McElroy} D.,  {Walsh} C.,  {Markwick} A.~J.,  {Cordiner} M.~A.,  {Smith} K.,
  {Millar} T.~J.,  2013, \mn@doi [\aap] {10.1051/0004-6361/201220465}, \href
  {https://ui.adsabs.harvard.edu/abs/2013A&A...550A..36M} {550, A36}

\bibitem[\protect\citeauthoryear{{Meijerink}, {Spaans}  \&
  {Israel}}{{Meijerink} et~al.}{2006}]{Meijerink06}
{Meijerink} R.,  {Spaans} M.,   {Israel} F.~P.,  2006, \mn@doi [\apjl]
  {10.1086/508938}, \href
  {https://ui.adsabs.harvard.edu/abs/2006ApJ...650L.103M} {650, L103}

\bibitem[\protect\citeauthoryear{{Meijerink}, {Spaans}  \&
  {Israel}}{{Meijerink} et~al.}{2007}]{Meijerink07}
{Meijerink} R.,  {Spaans} M.,   {Israel} F.~P.,  2007, \mn@doi [\aap]
  {10.1051/0004-6361:20066130}, \href
  {https://ui.adsabs.harvard.edu/abs/2007A&A...461..793M} {461, 793}

\bibitem[\protect\citeauthoryear{{Meijerink}, {Spaans}, {Loenen}  \& {van der
  Werf}}{{Meijerink} et~al.}{2011}]{Meijeirink11}
{Meijerink} R.,  {Spaans} M.,  {Loenen} A.~F.,   {van der Werf} P.~P.,  2011,
  \mn@doi [\aap] {10.1051/0004-6361/201015136}, \href
  {https://ui.adsabs.harvard.edu/abs/2011A&A...525A.119M} {525, A119}

\bibitem[\protect\citeauthoryear{{Michiyama} et~al.,}{{Michiyama}
  et~al.}{2020}]{Michiyama20}
{Michiyama} T.,  et~al., 2020, \mn@doi [\apjl] {10.3847/2041-8213/ab9d28},
  \href {https://ui.adsabs.harvard.edu/abs/2020ApJ...897L..19M} {897, L19}

\bibitem[\protect\citeauthoryear{{Michiyama} et~al.,}{{Michiyama}
  et~al.}{2021}]{Michiyama21}
{Michiyama} T.,  et~al., 2021, \mn@doi [\apjs] {10.3847/1538-4365/ac16df},
  \href {https://ui.adsabs.harvard.edu/abs/2021ApJS..257...28M} {257, 28}

\bibitem[\protect\citeauthoryear{{Montoya Arroyave} et~al.,}{{Montoya Arroyave}
  et~al.}{2023}]{MontoyaArroyave23}
{Montoya Arroyave} I.,  et~al., 2023, \mn@doi [\aap]
  {10.1051/0004-6361/202245046}, \href
  {https://ui.adsabs.harvard.edu/abs/2023A&A...673A..13M} {673, A13}

\bibitem[\protect\citeauthoryear{{Mu{\~n}oz} \& {Oh}}{{Mu{\~n}oz} \&
  {Oh}}{2016}]{Munoz16}
{Mu{\~n}oz} J.~A.,  {Oh} S.~P.,  2016, \mn@doi [\mnras]
  {10.1093/mnras/stw2102}, \href
  {https://ui.adsabs.harvard.edu/abs/2016MNRAS.463.2085M} {463, 2085}

\bibitem[\protect\citeauthoryear{{Narayanan} \& {Krumholz}}{{Narayanan} \&
  {Krumholz}}{2014}]{Narayanan14}
{Narayanan} D.,  {Krumholz} M.~R.,  2014, \mn@doi [\mnras]
  {10.1093/mnras/stu834}, \href
  {https://ui.adsabs.harvard.edu/abs/2014MNRAS.442.1411N} {442, 1411}

\bibitem[\protect\citeauthoryear{{Narayanan} \& {Krumholz}}{{Narayanan} \&
  {Krumholz}}{2017}]{Narayanan17}
{Narayanan} D.,  {Krumholz} M.~R.,  2017, \mn@doi [\mnras]
  {10.1093/mnras/stw3218}, \href
  {https://ui.adsabs.harvard.edu/abs/2017MNRAS.467...50N} {467, 50}

\bibitem[\protect\citeauthoryear{{Narayanan}, {Krumholz}, {Ostriker}  \&
  {Hernquist}}{{Narayanan} et~al.}{2012}]{Narayanan12}
{Narayanan} D.,  {Krumholz} M.~R.,  {Ostriker} E.~C.,   {Hernquist} L.,  2012,
  \mn@doi [\mnras] {10.1111/j.1365-2966.2012.20536.x}, \href
  {https://ui.adsabs.harvard.edu/abs/2012MNRAS.421.3127N} {421, 3127}

\bibitem[\protect\citeauthoryear{{Nicholls}, {Sutherland}, {Dopita}, {Kewley}
  \& {Groves}}{{Nicholls} et~al.}{2017}]{Nicholls2017}
{Nicholls} D.~C.,  {Sutherland} R.~S.,  {Dopita} M.~A.,  {Kewley} L.~J.,
  {Groves} B.~A.,  2017, \mn@doi [\mnras] {10.1093/mnras/stw3235}, \href
  {https://ui.adsabs.harvard.edu/abs/2017MNRAS.466.4403N} {466, 4403}

\bibitem[\protect\citeauthoryear{{{\"O}berg}, {Murray-Clay}  \&
  {Bergin}}{{{\"O}berg} et~al.}{2011}]{Oberg11}
{{\"O}berg} K.~I.,  {Murray-Clay} R.,   {Bergin} E.~A.,  2011, \mn@doi [\apjl]
  {10.1088/2041-8205/743/1/L16}, \href
  {https://ui.adsabs.harvard.edu/abs/2011ApJ...743L..16O} {743, L16}

\bibitem[\protect\citeauthoryear{{Oesch} et~al.,}{{Oesch}
  et~al.}{2016}]{Oesch16}
{Oesch} P.~A.,  et~al., 2016, \mn@doi [\apj] {10.3847/0004-637X/819/2/129},
  \href {https://ui.adsabs.harvard.edu/abs/2016ApJ...819..129O} {819, 129}

\bibitem[\protect\citeauthoryear{{Offner}, {Bisbas}, {Viti}  \&
  {Bell}}{{Offner} et~al.}{2013}]{Offner13}
{Offner} S. S.~R.,  {Bisbas} T.~G.,  {Viti} S.,   {Bell} T.~A.,  2013, \mn@doi
  [\apj] {10.1088/0004-637X/770/1/49}, \href
  {https://ui.adsabs.harvard.edu/abs/2013ApJ...770...49O} {770, 49}

\bibitem[\protect\citeauthoryear{{Offner}, {Bisbas}, {Bell}  \&
  {Viti}}{{Offner} et~al.}{2014}]{Offner14}
{Offner} S.~S.~R.,  {Bisbas} T.~G.,  {Bell} T.~A.,   {Viti} S.,  2014, \mn@doi
  [\mnras] {10.1093/mnrasl/slu013}, \href
  {https://ui.adsabs.harvard.edu/abs/2014MNRAS.440L..81O} {440, L81}

\bibitem[\protect\citeauthoryear{{Pagel}}{{Pagel}}{2009}]{Pagel09}
{Pagel} B. E.~J.,  2009, {Nucleosynthesis and Chemical Evolution of Galaxies}

\bibitem[\protect\citeauthoryear{{Papadopoulos}}{{Papadopoulos}}{2010}]{Papadopoulos10a}
{Papadopoulos} P.~P.,  2010, \mn@doi [\apj] {10.1088/0004-637X/720/1/226},
  \href {https://ui.adsabs.harvard.edu/abs/2010ApJ...720..226P} {720, 226}

\bibitem[\protect\citeauthoryear{{Papadopoulos}, {Thi}  \&
  {Viti}}{{Papadopoulos} et~al.}{2004}]{Papadopoulos04}
{Papadopoulos} P.~P.,  {Thi} W.~F.,   {Viti} S.,  2004, \mn@doi [\mnras]
  {10.1111/j.1365-2966.2004.07762.x}, \href
  {https://ui.adsabs.harvard.edu/abs/2004MNRAS.351..147P} {351, 147}

\bibitem[\protect\citeauthoryear{{Papadopoulos}, {van der Werf}, {Isaak}  \&
  {Xilouris}}{{Papadopoulos} et~al.}{2010}]{Papadopoulos10b}
{Papadopoulos} P.~P.,  {van der Werf} P.,  {Isaak} K.,   {Xilouris} E.~M.,
  2010, \mn@doi [\apj] {10.1088/0004-637X/715/2/775}, \href
  {https://ui.adsabs.harvard.edu/abs/2010ApJ...715..775P} {715, 775}

\bibitem[\protect\citeauthoryear{{Papadopoulos}, {Bisbas}  \&
  {Zhang}}{{Papadopoulos} et~al.}{2018}]{Papadopoulos18}
{Papadopoulos} P.~P.,  {Bisbas} T.~G.,   {Zhang} Z.-Y.,  2018, \mn@doi [\mnras]
  {10.1093/mnras/sty1077}, \href
  {https://ui.adsabs.harvard.edu/abs/2018MNRAS.478.1716P} {478, 1716}

\bibitem[\protect\citeauthoryear{{Papadopoulos}, {Dunne}  \&
  {Maddox}}{{Papadopoulos} et~al.}{2022}]{Papadopoulos22}
{Papadopoulos} P.,  {Dunne} L.,   {Maddox} S.,  2022, \mn@doi [\mnras]
  {10.1093/mnras/stab3194}, \href
  {https://ui.adsabs.harvard.edu/abs/2022MNRAS.510..725P} {510, 725}

\bibitem[\protect\citeauthoryear{Pietrow, Hoppe, Bergemann  \& Calvo}{Pietrow
  et~al.}{2023}]{Pietrow23}
Pietrow A. G.~M.,  Hoppe R.,  Bergemann M.,   Calvo F.,  2023

\bibitem[\protect\citeauthoryear{{Pineda}, {Langer}  \& {Goldsmith}}{{Pineda}
  et~al.}{2014}]{Pineda14}
{Pineda} J.~L.,  {Langer} W.~D.,   {Goldsmith} P.~F.,  2014, \mn@doi [\aap]
  {10.1051/0004-6361/201424054}, \href
  {https://ui.adsabs.harvard.edu/abs/2014A&A...570A.121P} {570, A121}

\bibitem[\protect\citeauthoryear{{Ravindranath}, {Monroe}, {Jaskot}, {Ferguson}
   \& {Tumlinson}}{{Ravindranath} et~al.}{2020}]{Ravindranath20}
{Ravindranath} S.,  {Monroe} T.,  {Jaskot} A.,  {Ferguson} H.~C.,   {Tumlinson}
  J.,  2020, \mn@doi [\apj] {10.3847/1538-4357/ab91a5}, \href
  {https://ui.adsabs.harvard.edu/abs/2020ApJ...896..170R} {896, 170}

\bibitem[\protect\citeauthoryear{{R{\'e}my-Ruyer} et~al.,}{{R{\'e}my-Ruyer}
  et~al.}{2013}]{RemyRuyer13}
{R{\'e}my-Ruyer} A.,  et~al., 2013, \mn@doi [\aap]
  {10.1051/0004-6361/201321602}, \href
  {https://ui.adsabs.harvard.edu/abs/2013A&A...557A..95R} {557, A95}

\bibitem[\protect\citeauthoryear{{R{\'e}my-Ruyer} et~al.,}{{R{\'e}my-Ruyer}
  et~al.}{2014}]{RemyRuyer14}
{R{\'e}my-Ruyer} A.,  et~al., 2014, \mn@doi [\aap]
  {10.1051/0004-6361/201322803}, \href
  {https://ui.adsabs.harvard.edu/abs/2014A&A...563A..31R} {563, A31}

\bibitem[\protect\citeauthoryear{{Richings} \& {Schaye}}{{Richings} \&
  {Schaye}}{2016}]{Richings16}
{Richings} A.~J.,  {Schaye} J.,  2016, \mn@doi [\mnras] {10.1093/mnras/stw327},
  \href {https://ui.adsabs.harvard.edu/abs/2016MNRAS.458..270R} {458, 270}

\bibitem[\protect\citeauthoryear{{R{\"o}llig} \&
  {Ossenkopf-Okada}}{{R{\"o}llig} \& {Ossenkopf-Okada}}{2022}]{Roellig22}
{R{\"o}llig} M.,  {Ossenkopf-Okada} V.,  2022, \mn@doi [\aap]
  {10.1051/0004-6361/202141854}, \href
  {https://ui.adsabs.harvard.edu/abs/2022A&A...664A..67R} {664, A67}

\bibitem[\protect\citeauthoryear{{R{\"o}llig} et~al.,}{{R{\"o}llig}
  et~al.}{2007}]{Roellig07}
{R{\"o}llig} M.,  et~al., 2007, \mn@doi [\aap] {10.1051/0004-6361:20065918},
  \href {https://ui.adsabs.harvard.edu/abs/2007A&A...467..187R} {467, 187}

\bibitem[\protect\citeauthoryear{{Romano}}{{Romano}}{2022}]{Romano22}
{Romano} D.,  2022, \mn@doi [\aapr] {10.1007/s00159-022-00144-z}, \href
  {https://ui.adsabs.harvard.edu/abs/2022A&ARv..30....7R} {30, 7}

\bibitem[\protect\citeauthoryear{{Romano}, {Franchini}, {Grisoni}, {Spitoni},
  {Matteucci}  \& {Morossi}}{{Romano} et~al.}{2020}]{Romano20}
{Romano} D.,  {Franchini} M.,  {Grisoni} V.,  {Spitoni} E.,  {Matteucci} F.,
  {Morossi} C.,  2020, \mn@doi [\aap] {10.1051/0004-6361/202037972}, \href
  {https://ui.adsabs.harvard.edu/abs/2020A&A...639A..37R} {639, A37}

\bibitem[\protect\citeauthoryear{{Rosenberg} et~al.,}{{Rosenberg}
  et~al.}{2015}]{Rosenberg15}
{Rosenberg} M.~J.~F.,  et~al., 2015, \mn@doi [\apj]
  {10.1088/0004-637X/801/2/72}, \href
  {https://ui.adsabs.harvard.edu/abs/2015ApJ...801...72R} {801, 72}

\bibitem[\protect\citeauthoryear{{Roueff} \& {Le Bourlot}}{{Roueff} \& {Le
  Bourlot}}{2020}]{Roueff20}
{Roueff} E.,  {Le Bourlot} J.,  2020, \mn@doi [\aap]
  {10.1051/0004-6361/202039085}, \href
  {https://ui.adsabs.harvard.edu/abs/2020A&A...643A.121R} {643, A121}

\bibitem[\protect\citeauthoryear{{Rupke}, {Veilleux}  \& {Baker}}{{Rupke}
  et~al.}{2008}]{Rupke08}
{Rupke} D. S.~N.,  {Veilleux} S.,   {Baker} A.~J.,  2008, \mn@doi [\apj]
  {10.1086/522363}, \href
  {https://ui.adsabs.harvard.edu/abs/2008ApJ...674..172R} {674, 172}

\bibitem[\protect\citeauthoryear{{Rybak} et~al.,}{{Rybak}
  et~al.}{2019}]{Rybak19}
{Rybak} M.,  et~al., 2019, \mn@doi [\apj] {10.3847/1538-4357/ab0e0f}, \href
  {https://ui.adsabs.harvard.edu/abs/2019ApJ...876..112R} {876, 112}

\bibitem[\protect\citeauthoryear{{Sage}, {Mauersberger}  \& {Henkel}}{{Sage}
  et~al.}{1991}]{Sage91}
{Sage} L.~J.,  {Mauersberger} R.,   {Henkel} C.,  1991, \aap, \href
  {https://ui.adsabs.harvard.edu/abs/1991A&A...249...31S} {249, 31}

\bibitem[\protect\citeauthoryear{{Sage}, {Loose}  \& {Salzer}}{{Sage}
  et~al.}{1993}]{Sage93}
{Sage} L.~J.,  {Loose} H.~H.,   {Salzer} J.~J.,  1993, \aap, \href
  {https://ui.adsabs.harvard.edu/abs/1993A&A...273....6S} {273, 6}

\bibitem[\protect\citeauthoryear{{Sandstrom} et~al.,}{{Sandstrom}
  et~al.}{2013}]{Sandstrom13}
{Sandstrom} K.~M.,  et~al., 2013, \mn@doi [\apj] {10.1088/0004-637X/777/1/5},
  \href {https://ui.adsabs.harvard.edu/abs/2013ApJ...777....5S} {777, 5}

\bibitem[\protect\citeauthoryear{{Schneider} et~al.,}{{Schneider}
  et~al.}{2023}]{Schneider23}
{Schneider} N.,  et~al., 2023, \mn@doi [Nature Astronomy]
  {10.1038/s41550-023-01901-5}, \href
  {https://ui.adsabs.harvard.edu/abs/2023NatAs...7..546S} {7, 546}

\bibitem[\protect\citeauthoryear{{Seifried}, {Haid}, {Walch}, {Borchert}  \&
  {Bisbas}}{{Seifried} et~al.}{2020}]{Seifried20}
{Seifried} D.,  {Haid} S.,  {Walch} S.,  {Borchert} E.~M.~A.,   {Bisbas} T.~G.,
   2020, \mn@doi [\mnras] {10.1093/mnras/stz3563}, \href
  {https://ui.adsabs.harvard.edu/abs/2020MNRAS.492.1465S} {492, 1465}

\bibitem[\protect\citeauthoryear{{Sharda} \& {Krumholz}}{{Sharda} \&
  {Krumholz}}{2022}]{Sharda22}
{Sharda} P.,  {Krumholz} M.~R.,  2022, \mn@doi [\mnras]
  {10.1093/mnras/stab2921}, \href
  {https://ui.adsabs.harvard.edu/abs/2022MNRAS.509.1959S} {509, 1959}

\bibitem[\protect\citeauthoryear{{Sharda}, {Ginzburg}, {Krumholz}, {Forbes},
  {Wisnioski}, {Mingozzi}, {Zovaro}  \& {Dekel}}{{Sharda}
  et~al.}{2023a}]{Sharda23b}
{Sharda} P.,  {Ginzburg} O.,  {Krumholz} M.~R.,  {Forbes} J.~C.,  {Wisnioski}
  E.,  {Mingozzi} M.,  {Zovaro} H. R.~M.,   {Dekel} A.,  2023a, \mn@doi [arXiv
  e-prints] {10.48550/arXiv.2303.15853}, \href
  {https://ui.adsabs.harvard.edu/abs/2023arXiv230315853S} {p. arXiv:2303.15853}

\bibitem[\protect\citeauthoryear{{Sharda}, {Amarsi}, {Grasha}, {Krumholz},
  {Yong}, {Chiaki}, {Roy}  \& {Nordlander}}{{Sharda} et~al.}{2023b}]{Sharda23}
{Sharda} P.,  {Amarsi} A.~M.,  {Grasha} K.,  {Krumholz} M.~R.,  {Yong} D.,
  {Chiaki} G.,  {Roy} A.,   {Nordlander} T.,  2023b, \mn@doi [\mnras]
  {10.1093/mnras/stac3315}, \href
  {https://ui.adsabs.harvard.edu/abs/2023MNRAS.518.3985S} {518, 3985}

\bibitem[\protect\citeauthoryear{{Sharda}, {Amarsi}, {Grasha}, {Krumholz},
  {Yong}, {Chiaki}, {Roy}  \& {Nordlander}}{{Sharda}
  et~al.}{2023c}]{Sharda23cor}
{Sharda} P.,  {Amarsi} A.~M.,  {Grasha} K.,  {Krumholz} M.~R.,  {Yong} D.,
  {Chiaki} G.,  {Roy} A.,   {Nordlander} T.,  2023c, \mn@doi [\mnras]
  {10.1093/mnras/stad2534}, \href
  {https://ui.adsabs.harvard.edu/abs/2023MNRAS.525.3316S} {525, 3316}

\bibitem[\protect\citeauthoryear{{Shi}, {Kong}, {Li}  \& {Cheng}}{{Shi}
  et~al.}{2005}]{Shi05}
{Shi} F.,  {Kong} X.,  {Li} C.,   {Cheng} F.~Z.,  2005, \mn@doi [\aap]
  {10.1051/0004-6361:20041945}, \href
  {https://ui.adsabs.harvard.edu/abs/2005A&A...437..849S} {437, 849}

\bibitem[\protect\citeauthoryear{{Berg}, {Erb}, {Henry}, {Skillman}  \&
  {McQuinn}}{Ski}{1998}]{Skillman1998}
 1998, {Stellar astrophysics for the local group : VIII Canary Islands Winter
  School of Astrophysics}

\bibitem[\protect\citeauthoryear{{Smith} et~al.,}{{Smith}
  et~al.}{2017}]{Smith17}
{Smith} J.~D.~T.,  et~al., 2017, \mn@doi [\apj] {10.3847/1538-4357/834/1/5},
  \href {https://ui.adsabs.harvard.edu/abs/2017ApJ...834....5S} {834, 5}

\bibitem[\protect\citeauthoryear{{Smith} et~al.,}{{Smith}
  et~al.}{2020}]{Smith20}
{Smith} R.~J.,  et~al., 2020, \mn@doi [\mnras] {10.1093/mnras/stz3328}, \href
  {https://ui.adsabs.harvard.edu/abs/2020MNRAS.492.1594S} {492, 1594}

\bibitem[\protect\citeauthoryear{{Solomon} \& {Vanden Bout}}{{Solomon} \&
  {Vanden Bout}}{2005}]{Solomon05}
{Solomon} P.~M.,  {Vanden Bout} P.~A.,  2005, \mn@doi [\araa]
  {10.1146/annurev.astro.43.051804.102221}, \href
  {https://ui.adsabs.harvard.edu/abs/2005ARA&A..43..677S} {43, 677}

\bibitem[\protect\citeauthoryear{{Spitoni}, {Silva Aguirre}, {Matteucci},
  {Calura}  \& {Grisoni}}{{Spitoni} et~al.}{2019}]{Spitoni19}
{Spitoni} E.,  {Silva Aguirre} V.,  {Matteucci} F.,  {Calura} F.,   {Grisoni}
  V.,  2019, \mn@doi [\aap] {10.1051/0004-6361/201834188}, \href
  {https://ui.adsabs.harvard.edu/abs/2019A&A...623A..60S} {623, A60}

\bibitem[\protect\citeauthoryear{{Stacey}, {Geis}, {Genzel}, {Lugten},
  {Poglitsch}, {Sternberg}  \& {Townes}}{{Stacey} et~al.}{1991}]{Stacey91}
{Stacey} G.~J.,  {Geis} N.,  {Genzel} R.,  {Lugten} J.~B.,  {Poglitsch} A.,
  {Sternberg} A.,   {Townes} C.~H.,  1991, \mn@doi [\apj] {10.1086/170062},
  \href {https://ui.adsabs.harvard.edu/abs/1991ApJ...373..423S} {373, 423}

\bibitem[\protect\citeauthoryear{{Stanley} et~al.,}{{Stanley}
  et~al.}{2023}]{Stanley23}
{Stanley} F.,  et~al., 2023, \mn@doi [\apj] {10.3847/1538-4357/acb6f7}, \href
  {https://ui.adsabs.harvard.edu/abs/2023ApJ...945...24S} {945, 24}

\bibitem[\protect\citeauthoryear{{Sternberg}, {Le Petit}, {Roueff}  \& {Le
  Bourlot}}{{Sternberg} et~al.}{2014}]{Sternberg14}
{Sternberg} A.,  {Le Petit} F.,  {Roueff} E.,   {Le Bourlot} J.,  2014, \mn@doi
  [\apj] {10.1088/0004-637X/790/1/10}, \href
  {https://ui.adsabs.harvard.edu/abs/2014ApJ...790...10S} {790, 10}

\bibitem[\protect\citeauthoryear{{Strong}, {Moskalenko}  \& {Ptuskin}}{{Strong}
  et~al.}{2007}]{Strong07}
{Strong} A.~W.,  {Moskalenko} I.~V.,   {Ptuskin} V.~S.,  2007, \mn@doi [Annual
  Review of Nuclear and Particle Science]
  {10.1146/annurev.nucl.57.090506.123011}, \href
  {https://ui.adsabs.harvard.edu/abs/2007ARNPS..57..285S} {57, 285}

\bibitem[\protect\citeauthoryear{{Su{\'a}rez-Andr{\'e}s}, {Israelian},
  {Gonz{\'a}lez Hern{\'a}ndez}, {Adibekyan}, {Delgado Mena}, {Santos}  \&
  {Sousa}}{{Su{\'a}rez-Andr{\'e}s} et~al.}{2018}]{Surez2018}
{Su{\'a}rez-Andr{\'e}s} L.,  {Israelian} G.,  {Gonz{\'a}lez Hern{\'a}ndez}
  J.~I.,  {Adibekyan} V.~Z.,  {Delgado Mena} E.,  {Santos} N.~C.,   {Sousa}
  S.~G.,  2018, \mn@doi [\aap] {10.1051/0004-6361/201730743}, \href
  {https://ui.adsabs.harvard.edu/abs/2018A&A...614A..84S} {614, A84}

\bibitem[\protect\citeauthoryear{{Sutter} et~al.,}{{Sutter}
  et~al.}{2019}]{Sutter19}
{Sutter} J.,  et~al., 2019, \mn@doi [\apj] {10.3847/1538-4357/ab4da5}, \href
  {https://ui.adsabs.harvard.edu/abs/2019ApJ...886...60S} {886, 60}

\bibitem[\protect\citeauthoryear{{Tacconi} et~al.,}{{Tacconi}
  et~al.}{2008}]{Tacconi08}
{Tacconi} L.~J.,  et~al., 2008, \mn@doi [\apj] {10.1086/587168}, \href
  {https://ui.adsabs.harvard.edu/abs/2008ApJ...680..246T} {680, 246}

\bibitem[\protect\citeauthoryear{{Tanaka} \& {Omukai}}{{Tanaka} \&
  {Omukai}}{2014}]{Tanaka14}
{Tanaka} K. E.~I.,  {Omukai} K.,  2014, \mn@doi [\mnras]
  {10.1093/mnras/stu069}, \href
  {https://ui.adsabs.harvard.edu/abs/2014MNRAS.439.1884T} {439, 1884}

\bibitem[\protect\citeauthoryear{{Tanaka}, {Tan}, {Zhang}  \&
  {Hosokawa}}{{Tanaka} et~al.}{2018}]{Tanaka18}
{Tanaka} K. E.~I.,  {Tan} J.~C.,  {Zhang} Y.,   {Hosokawa} T.,  2018, \mn@doi
  [\apj] {10.3847/1538-4357/aac892}, \href
  {https://ui.adsabs.harvard.edu/abs/2018ApJ...861...68T} {861, 68}

\bibitem[\protect\citeauthoryear{{Tielens}}{{Tielens}}{2005}]{Tielens05}
{Tielens} A.~G.~G.~M.,  2005, {The Physics and Chemistry of the Interstellar
  Medium}

\bibitem[\protect\citeauthoryear{{Trainor}, {Strom}, {Steidel}  \&
  {Rudie}}{{Trainor} et~al.}{2016}]{Trainor16}
{Trainor} R.~F.,  {Strom} A.~L.,  {Steidel} C.~C.,   {Rudie} G.~C.,  2016,
  \mn@doi [\apj] {10.3847/0004-637X/832/2/171}, \href
  {https://ui.adsabs.harvard.edu/abs/2016ApJ...832..171T} {832, 171}

\bibitem[\protect\citeauthoryear{{Valentino} et~al.,}{{Valentino}
  et~al.}{2018}]{Valentino18}
{Valentino} F.,  et~al., 2018, \mn@doi [\apj] {10.3847/1538-4357/aaeb88}, \href
  {https://ui.adsabs.harvard.edu/abs/2018ApJ...869...27V} {869, 27}

\bibitem[\protect\citeauthoryear{{Valentino} et~al.,}{{Valentino}
  et~al.}{2020}]{Valentino20}
{Valentino} F.,  et~al., 2020, \mn@doi [\apj] {10.3847/1538-4357/ab6603}, \href
  {https://ui.adsabs.harvard.edu/abs/2020ApJ...890...24V} {890, 24}

\bibitem[\protect\citeauthoryear{{Vallini}, {Pallottini}, {Ferrara},
  {Gallerani}, {Sobacchi}  \& {Behrens}}{{Vallini} et~al.}{2018}]{Vallini18}
{Vallini} L.,  {Pallottini} A.,  {Ferrara} A.,  {Gallerani} S.,  {Sobacchi} E.,
    {Behrens} C.,  2018, \mn@doi [\mnras] {10.1093/mnras/stx2376}, \href
  {https://ui.adsabs.harvard.edu/abs/2018MNRAS.473..271V} {473, 271}

\bibitem[\protect\citeauthoryear{{Vallini}, {Tielens}, {Pallottini},
  {Gallerani}, {Gruppioni}, {Carniani}, {Pozzi}  \& {Talia}}{{Vallini}
  et~al.}{2019}]{Vallini19}
{Vallini} L.,  {Tielens} A.~G.~G.~M.,  {Pallottini} A.,  {Gallerani} S.,
  {Gruppioni} C.,  {Carniani} S.,  {Pozzi} F.,   {Talia} M.,  2019, \mn@doi
  [\mnras] {10.1093/mnras/stz2837}, \href
  {https://ui.adsabs.harvard.edu/abs/2019MNRAS.490.4502V} {490, 4502}

\bibitem[\protect\citeauthoryear{{Viti}, {Roueff}, {Hartquist}, {Pineau des
  For{\^e}ts}  \& {Williams}}{{Viti} et~al.}{2001}]{Viti01}
{Viti} S.,  {Roueff} E.,  {Hartquist} T.~W.,  {Pineau des For{\^e}ts} G.,
  {Williams} D.~A.,  2001, \mn@doi [\aap] {10.1051/0004-6361:20010246}, \href
  {https://ui.adsabs.harvard.edu/abs/2001A&A...370..557V} {370, 557}

\bibitem[\protect\citeauthoryear{{Walch} et~al.,}{{Walch}
  et~al.}{2015}]{Walch15}
{Walch} S.,  et~al., 2015, \mn@doi [\mnras] {10.1093/mnras/stv1975}, \href
  {https://ui.adsabs.harvard.edu/abs/2015MNRAS.454..238W} {454, 238}

\bibitem[\protect\citeauthoryear{{Wolfire}, {McKee}, {Hollenbach}  \&
  {Tielens}}{{Wolfire} et~al.}{2003}]{Wolfire03}
{Wolfire} M.~G.,  {McKee} C.~F.,  {Hollenbach} D.,   {Tielens} A.~G.~G.~M.,
  2003, \mn@doi [\apj] {10.1086/368016}, \href
  {https://ui.adsabs.harvard.edu/abs/2003ApJ...587..278W} {587, 278}

\bibitem[\protect\citeauthoryear{{Wolfire}, {Tielens}, {Hollenbach}  \&
  {Kaufman}}{{Wolfire} et~al.}{2008}]{Wolfire08}
{Wolfire} M.~G.,  {Tielens} A.~G.~G.~M.,  {Hollenbach} D.,   {Kaufman} M.~J.,
  2008, \mn@doi [\apj] {10.1086/587688}, \href
  {https://ui.adsabs.harvard.edu/abs/2008ApJ...680..384W} {680, 384}

\bibitem[\protect\citeauthoryear{{Wolfire}, {Vallini}  \& {Chevance}}{{Wolfire}
  et~al.}{2022}]{Wolfire22}
{Wolfire} M.~G.,  {Vallini} L.,   {Chevance} M.,  2022, \mn@doi [\araa]
  {10.1146/annurev-astro-052920-010254}, \href
  {https://ui.adsabs.harvard.edu/abs/2022ARA&A..60..247W} {60, 247}

\bibitem[\protect\citeauthoryear{{Wu}, {Tan}, {Nakamura}, {Van Loo}, {Christie}
   \& {Collins}}{{Wu} et~al.}{2017}]{Wu17}
{Wu} B.,  {Tan} J.~C.,  {Nakamura} F.,  {Van Loo} S.,  {Christie} D.,
  {Collins} D.,  2017, \mn@doi [\apj] {10.3847/1538-4357/835/2/137}, \href
  {https://ui.adsabs.harvard.edu/abs/2017ApJ...835..137W} {835, 137}

\bibitem[\protect\citeauthoryear{{Wuyts} et~al.,}{{Wuyts}
  et~al.}{2016}]{Wuyts16}
{Wuyts} E.,  et~al., 2016, \mn@doi [\apj] {10.3847/0004-637X/827/1/74}, \href
  {https://ui.adsabs.harvard.edu/abs/2016ApJ...827...74W} {827, 74}

\bibitem[\protect\citeauthoryear{{Xie}, {Allen}  \& {Langer}}{{Xie}
  et~al.}{1995}]{Xie95}
{Xie} T.,  {Allen} M.,   {Langer} W.~D.,  1995, \mn@doi [\apj]
  {10.1086/175305}, \href
  {https://ui.adsabs.harvard.edu/abs/1995ApJ...440..674X} {440, 674}

\bibitem[\protect\citeauthoryear{{Yang} et~al.,}{{Yang} et~al.}{2023}]{Yang23}
{Yang} C.,  et~al., 2023, \mn@doi [arXiv e-prints] {10.48550/arXiv.2308.07368},
  \href {https://ui.adsabs.harvard.edu/abs/2023arXiv230807368Y} {p.
  arXiv:2308.07368}

\bibitem[\protect\citeauthoryear{{Zanella} et~al.,}{{Zanella}
  et~al.}{2018}]{Zanella18}
{Zanella} A.,  et~al., 2018, \mn@doi [\mnras] {10.1093/mnras/sty2394}, \href
  {https://ui.adsabs.harvard.edu/abs/2018MNRAS.481.1976Z} {481, 1976}

\bibitem[\protect\citeauthoryear{{Zhang} et~al.,}{{Zhang}
  et~al.}{2014}]{Zhang14}
{Zhang} Z.-Y.,  et~al., 2014, \mn@doi [\aap] {10.1051/0004-6361/201322639},
  \href {https://ui.adsabs.harvard.edu/abs/2014A&A...568A.122Z} {568, A122}

\bibitem[\protect\citeauthoryear{{Zhang}, {Romano}, {Ivison}, {Papadopoulos}
  \& {Matteucci}}{{Zhang} et~al.}{2018}]{Zhang18}
{Zhang} Z.-Y.,  {Romano} D.,  {Ivison} R.~J.,  {Papadopoulos} P.~P.,
  {Matteucci} F.,  2018, \mn@doi [\nat] {10.1038/s41586-018-0196-x}, \href
  {https://ui.adsabs.harvard.edu/abs/2018Natur.558..260Z} {558, 260}

\bibitem[\protect\citeauthoryear{{Zhao}, {Yan}  \& {Tsai}}{{Zhao}
  et~al.}{2016}]{Zhao16}
{Zhao} Y.,  {Yan} L.,   {Tsai} C.-W.,  2016, \mn@doi [\apj]
  {10.3847/0004-637X/824/2/146}, \href
  {https://ui.adsabs.harvard.edu/abs/2016ApJ...824..146Z} {824, 146}

\bibitem[\protect\citeauthoryear{{Zuo}, {Li}, {Peek}, {Chang}, {Zhang},
  {Chapman}, {Goldsmith}  \& {Zhang}}{{Zuo} et~al.}{2018}]{Zuo2018}
{Zuo} P.,  {Li} D.,  {Peek} J.~E.~G.,  {Chang} Q.,  {Zhang} X.,  {Chapman} N.,
  {Goldsmith} P.~F.,   {Zhang} Z.-Y.,  2018, \mn@doi [\apj]
  {10.3847/1538-4357/aad571}, \href
  {https://ui.adsabs.harvard.edu/abs/2018ApJ...867...13Z} {867, 13}

\bibitem[\protect\citeauthoryear{{van Dishoeck} \& {Black}}{{van Dishoeck} \&
  {Black}}{1986}]{vDishoeck86}
{van Dishoeck} E.~F.,  {Black} J.~H.,  1986, \mn@doi [\apjs] {10.1086/191135},
  \href {https://ui.adsabs.harvard.edu/abs/1986ApJS...62..109V} {62, 109}

\bibitem[\protect\citeauthoryear{{van Dishoeck} \& {Black}}{{van Dishoeck} \&
  {Black}}{1988}]{vDishoeck88}
{van Dishoeck} E.~F.,  {Black} J.~H.,  1988, \mn@doi [\apj] {10.1086/166877},
  \href {https://ui.adsabs.harvard.edu/abs/1988ApJ...334..771V} {334, 771}

\bibitem[\protect\citeauthoryear{{van Dishoeck} et~al.,}{{van Dishoeck}
  et~al.}{2023}]{vDishoeck23}
{van Dishoeck} E.~F.,  et~al., 2023, \mn@doi [arXiv e-prints]
  {10.48550/arXiv.2307.11817}, \href
  {https://ui.adsabs.harvard.edu/abs/2023arXiv230711817V} {p. arXiv:2307.11817}

\makeatother
\end{thebibliography}



\appendix

\section{Line ratios}
\label{app:ratios}

\begin{figure}
    \centering
    \includegraphics[width=0.46\textwidth]{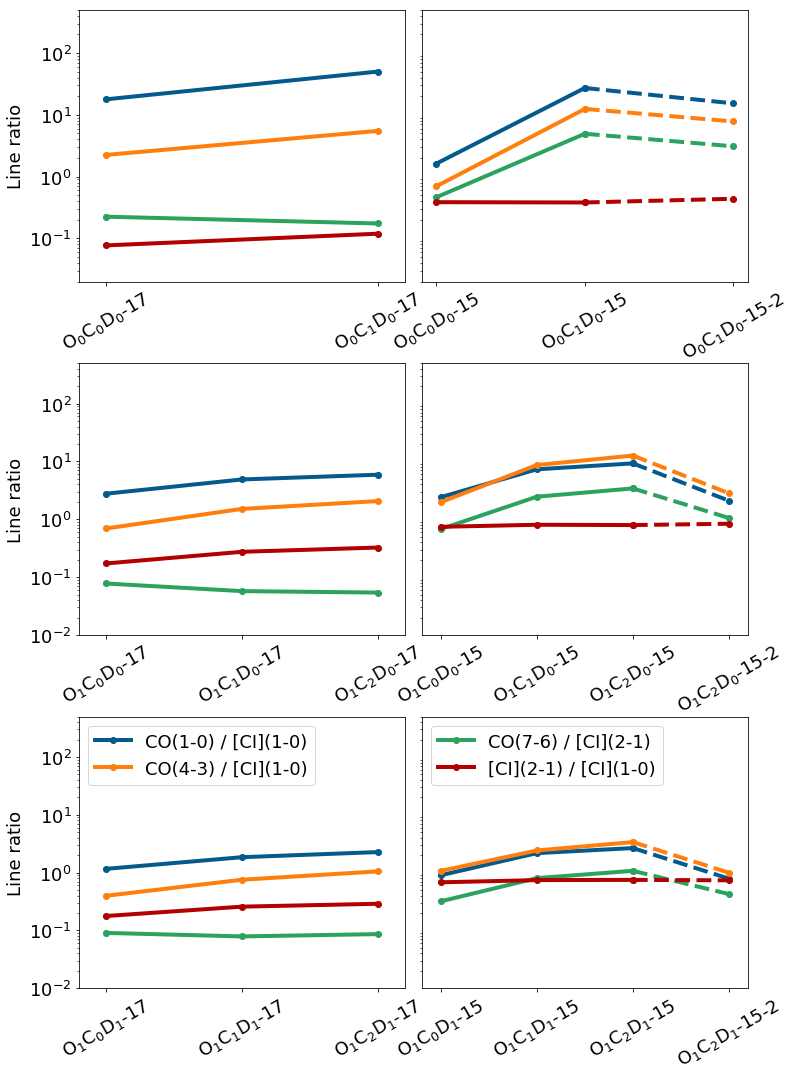}
    \caption{Emission line ratios between CO and [C{\sc i}] for all explored simulation cases and that are commonly used in observations.}
    \label{fig:ratios}
\end{figure}

Line ratios are frequently used in observations to study the ISM conditions and constrain the environmental parameters in MW clouds and distant extragalactic objects \citep[e.g.][]{Israel02,Gullberg16,Krips16,Valentino18,Valentino20,Papadopoulos22}. Here, we explore the ratios of CO(1-0)/[C{\sc i}](1-0) the lines of which are both important coolants of the dense gas, CO(4-3)/[C{\sc i}](1-0) and CO(7-6)/[C{\sc i}](2-1) as the frequency separation between these lines is small and are often observed together especially with ALMA, and the atomic carbon line ratio, [C{\sc i}](1-0)/[C{\sc i}](2-1). Figure~\ref{fig:ratios} shows the emission line ratios of the aforementioned combinations of CO and [C{\sc i}] for the environmental parameters explored. 

It is found that for $\zeta_{\rm CR}=10^{-17}\,{\rm s}^{-1}$, all above line ratios remain approximately constant as a function of [C/O], except for the CO(1-0)/[C{\sc i}](1-0) and CO(4-3)/[C{\sc i}](1-0) for the O$_0$C$_1$D$_0$-17 model in which they show a slight increase when compared to the O$_0$C$_0$D$_0$-17. The most interesting features, however, occur for $\zeta_{\rm CR}=10^{-15}\,{\rm s}^{-1}$. Here, it is found that all line ratios, except for the atomic carbon line ratio of [C{\sc i}](1-0)/[C{\sc i}](2-1), are increased with decreasing [C/O] for fixed [O/H]. The increase in FUV radiation dissociates CO and ionizes C, therefore affecting their emission. The imprint of that can be seen in the second column of Fig.~\ref{fig:ratios} (dashed lines), where a decreasing trend is observed between the high and low FUV models. 

The atomic carbon line ratio remains overall remarkably constant as a function of [C/O] in all cases, even when a higher FUV radiation field is applied. However, this ratio increases with the increasing $\zeta_{\rm CR}$ as can be seen when comparing it between the left and the right columns of the above figure. This is in accordance with the findings of \citet{Bisbas21,Bisbas23}, who suggest that the atomic carbon line ratio can be used to infer the cosmic-ray ionization rate, particularly in extragalactic star-forming systems, as it is not strongly affected by other changes in the ISM environmental parameters. 

\section{Mass distribution of C$^+$, C and CO with molecular gas defined as $\chi_{\rm H2}\ge0.45$}
\label{app:origin}

\begin{figure}
    \centering
    \includegraphics[width=0.49\textwidth]{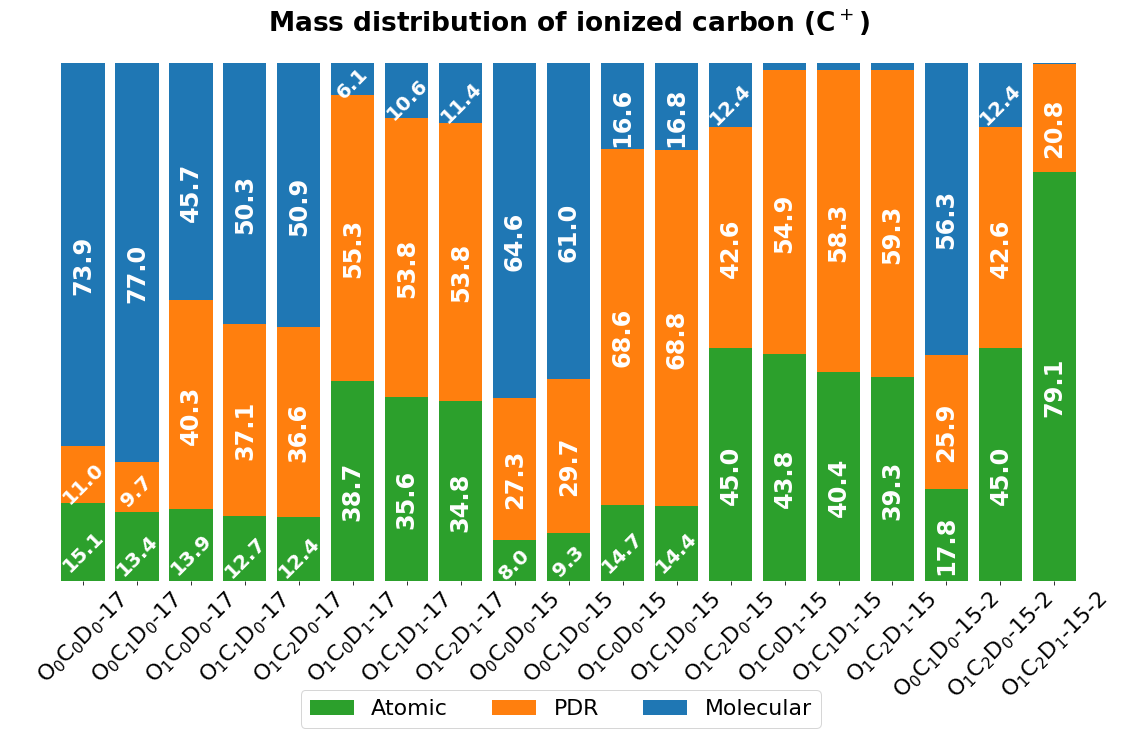}
    \includegraphics[width=0.49\textwidth]{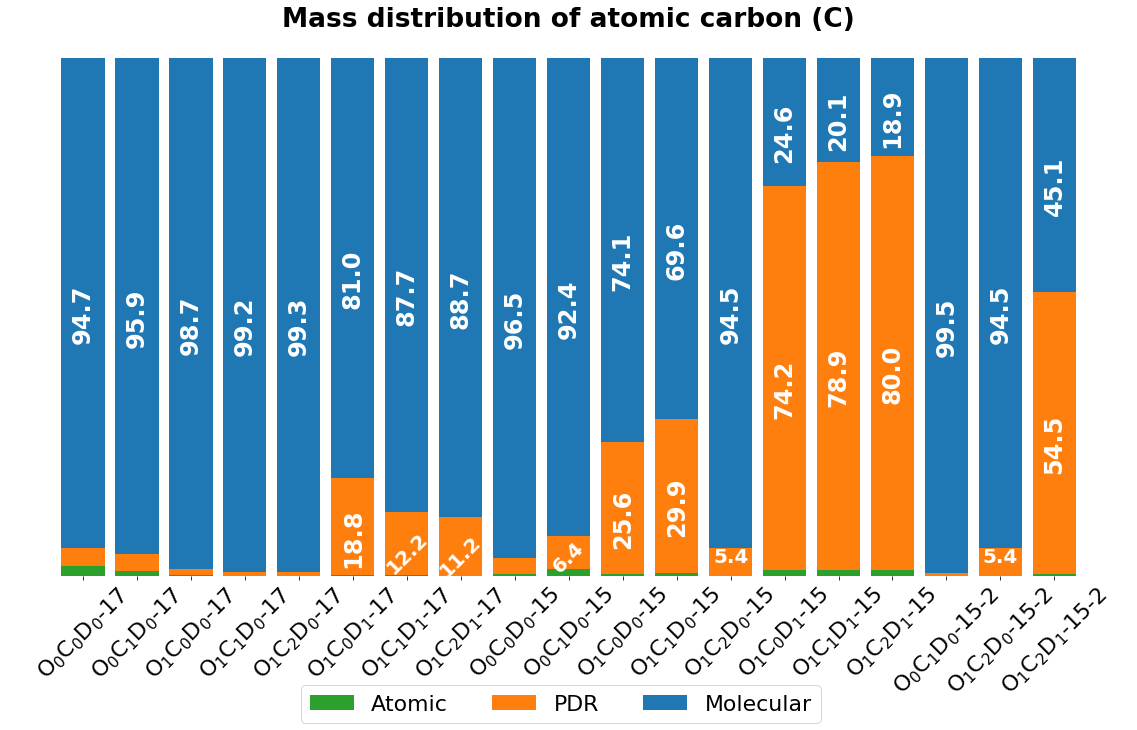}
    \includegraphics[width=0.49\textwidth]{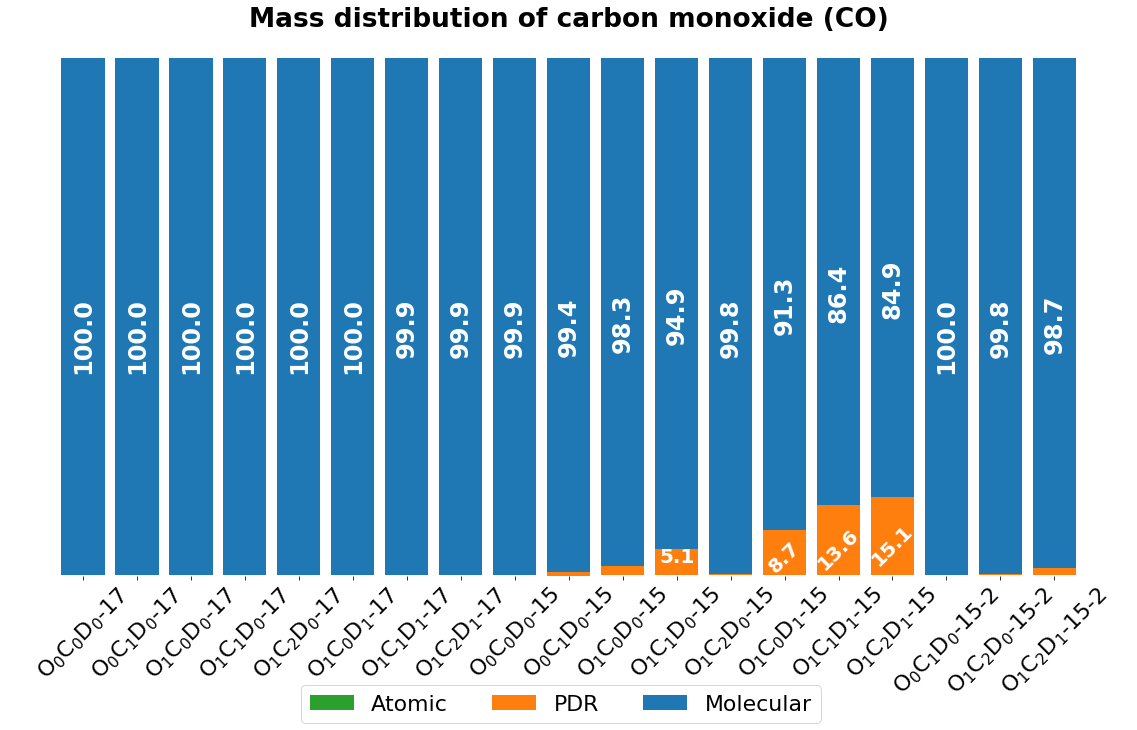}
    \caption{As in Figure~\ref{fig:origin2pcent} but with the molecular gas defined as the gas with $x_{\rm HI}\ge0.45$, and the PDR with $x_{\rm HI}<0.9$ and $x_{\rm H2}<0.45$, accordingly.}
    \label{fig:origin10pcent}
\end{figure}

As discussed, the $\chi_{\rm H2}\ge0.49$ condition for defining the molecular phase was based on observations of Galactic infrared dark clouds \citep{Zuo2018}. It would be interesting, however, to explore how the above results change if this condition is relaxed to $\chi_{\rm H2}\ge0.45$. Figure~\ref{fig:origin10pcent}  shows how $M_{\rm C^+}$, $M_{\rm C}$ and $M_{\rm CO}$ are now associated with the PDR and molecular gas phases. For C$^+$ and C, it is found that the contribution from the molecular gas phase is increased. The effect is more prominent, however, for the C species, as in all environmental parameters explored and except for the lower $\cal D$ at $\zeta_{\rm CR}=10^{-15}\,{\rm s}^{-1}$, it is associated with $>\!60\%$ or even $>\!90\%$ with the H$_2$-rich gas. For CO, the bottom panel of Fig.~\ref{fig:origin10pcent} indicates that CO is always associated with the molecular region as defined with $x_{\rm H2}\ge0.45$.
It can be, therefore, argued that the transition layer between C and CO is sharp and very close to the border of H$_2$-rich gas. A similar picture occurs for the C$^+$ to C transition, albeit at lower column densities.

\bsp	
\label{lastpage}
\end{document}